\documentclass[12pt]{article}             
\usepackage{feynmp}                       

\newcommand{\comment}[1]{}
\newcommand{\note}[1]{}
\newcommand{\bib}[1]{\bibitem{#1}}

\newcommand{\3}{\ss}
\newcommand{\ia}{{\"{\i}}}   
\newcommand{\absatz}{\vspace{2ex}\noindent}

\newcommand{\journal}[4]{{#1} {\bf{#2}}, #3 (#4)}
\newcommand{\JMathP}{\emph{J.\ Math.\ Phys.\ }}
\newcommand{\NP}{\emph{Nucl.\ Phys.\ }}
\newcommand{\NPB}{\emph{Nucl.\ Phys.\ }{B}}
\newcommand{\PLB}{\emph{Phys.\ Lett.\ }{B}}
\newcommand{\PRD}{\emph{Phys.\ Rev.\ }{D}}
\newcommand{\PRL}{\emph{Phys.\ Rev.\ Lett.\ }}
\newcommand{\book}[3]{\emph{#1}; #2 (#3)}
\newcommand{\preprint}[4][to be published]{\emph{#2}; #3, #4 (#1)}

\newcommand{\dis}{\displaystyle}
\newcommand{\non}{\nonumber}

\newcommand{\half}{\frac{1}{2}}
\newcommand{\e}{\mathrm{e}}
\newcommand{\ii}{\mathrm{i}}
\newcommand{\dd}{\mathrm{d}}

\newcommand{\xs}{\vec{X}_{\mathrm{s}}}
\newcommand{\ts}{T_{\mathrm{s}}}

\newcommand{\xu}{\vec{X}_{\mathrm{u}}}
\newcommand{\tu}{T_{\mathrm{u}}}
\newcommand{\Qs}{Q_{\mathrm{s}}}
\newcommand{\Qp}{Q_{\mathrm{p}}}
\newcommand{\Qu}{Q_{\mathrm{u}}}
\newcommand{\calQs}{{\calQ_{{\rm s}}}}
\newcommand{\calQp}{{\calQ_{{\rm p}}}}
\newcommand{\calQu}{{\calQ_{{\rm u}}}}

\newcommand{\Amu}{{A^\mu}}
\newcommand{\Asmu}{{A_{{\rm s}}^\mu}}
\newcommand{\Apmu}{{A_{{\rm p}}^\mu}}
\newcommand{\Aumu}{{A_{{\rm u}}^\mu}}
\newcommand{\Av}{\vec{A}}

\newcommand{\Avp}{{\vec{A}_{{\rm p}}}}

\newcommand{\calAsmu}{{\calA_{{\rm s}}^\mu}}
\newcommand{\calApmu}{{\calA_{{\rm p}}^\mu}}
\newcommand{\calAumu}{{\calA_{{\rm u}}^\mu}}

\newcommand{\Xvs}{\vec{X}_{\mathrm{s}}}

\newcommand{\Ts}{T_{\mathrm{s}}}

\newcommand{\dedreiXs}{\dd^{3}\:\!\! X_{\mathrm{s}}\;}
\newcommand{\dedreiXu}{\dd^{3}\:\!\! X_{\mathrm{u}}\;}

\newcommand{\deTs}{\dd T_{\mathrm{s}}\;}
\newcommand{\deTu}{\dd T_{\mathrm{u}}\;}

\newcommand{\deint}[2]{\dd^{#1}\! #2\;}
\newcommand{\deintdim}[2]{\frac{\dd^{#1}\! #2}{(2\pi)^{#1}}\;}

\newcommand{\kv}{\vec{k}}
\newcommand{\pv}{\vec{\,\!p}\!\:{}}
\newcommand{\qv}{\vec{\,\!q}\!\:{}}

\newcommand{\xv}{\vec{x}}
\newcommand{\yv}{\vec{y}}

\newcommand{\dedreiy}{\dd^{3}\:\!\! y\;}
\newcommand{\de}{\partial}
\newcommand{\dev}{\vec{\de}}

\newcommand{\calA}{\mathcal{A}}\newcommand{\calC}{\mathcal{C}}
\newcommand{\calG}{\mathcal{G}}
\newcommand{\calL}{\mathcal{L}}\newcommand{\calO}{\mathcal{O}}
\newcommand{\calQ}{\mathcal{Q}}

\newcommand{\pc}{\mathrm{P.C.}}

\begin{document}

\begin{fmffile}{nrqfeyn}
\fmfset{curly_len}{2mm}
\fmfset{wiggly_len}{3mm}
\newcommand{\feynbox}[2]{\mbox{\parbox{#1}{#2}}}
\newcommand{\fs}{\scriptstyle} 
\newcommand{\hq}{\hspace{0.5em}}
\fmfcmd{%
  style_def double_dashsolid_lr expr p =
  save oldpen;
  pen oldpen;
  oldpen := currentpen;
  pickup oldpen scaled 2;
  cdraw p shifted (thick*(0,-0.5));
  pickup oldpen;
  draw_dashes p shifted (thick*(0,0.5));
  enddef;}                     
\fmfcmd{%
  style_def double_dashsolid_ud expr p =
  save oldpen;
  pen oldpen;
  oldpen := currentpen;
  pickup oldpen scaled 2;
  cdraw p shifted (thick*(-0.5,0));
  pickup oldpen;
  draw_dashes p shifted (thick*(0.5,0));
  enddef;}                     

%

\begin{titlepage}
\begin{flushright}
  hep-ph/9810235\\ NT@UW-98-22 \\ 1st October 1998 \\ Revised 14th April 1999\\
\end{flushright}
\vspace*{1.5cm}
\begin{center}
  
  \LARGE{\textbf{Power Counting and $\beta$ Function in NRQCD}}

\end{center}
\vspace*{1.0cm}
\begin{center}
  \textbf{Harald W.\ Grie\3hammer\footnote{Email: hgrie@phys.washington.edu}}
  
  \vspace*{0.2cm}
  
 \emph{Nuclear Theory Group, Department of Physics, University of Washington,\\
    Box 351 560, Seattle, WA 98195-1560, USA} \vspace*{0.2cm}

\end{center}

\vspace*{2.0cm}

\begin{abstract}
  A computation of the NRQCD $\beta$ function both in the Lorentz gauge family
  and in the Coulomb gauge to one loop order endorses a velocity power counting
  scheme for dimensionally regularised NRQCD. In addition to the ultrasoft
  scale represented by bremsstrahlung gluons and the potential scale with
  Coulomb gluons and on-shell quarks, a soft r\'egime is identified in which
  energies and momenta are of order $Mv$, gluons are on shell and the quark
  propagator becomes static. The instantaneous gluon propagator has a non-zero
  vacuum polarisation only because of contributions from this r\'egime,
  irrespective of the gauge chosen. Rules are derived which allow one to read
  up from a given graph whether it is zero because of the homogene{\ia}ty of
  dimensional regularisation. They also apply to threshold expansion and are
  used to prove that ultrasoft quarks with energy and momentum of order $Mv^2$
  decouple from the theory.
\end{abstract}
\vskip 1.0cm
\noindent
\begin{tabular}{rl}
Suggested PACS numbers:& 12.38.Bx, 12.39.Hg, 12.39.Jh, 11.10.Gh.\\[1ex]
Suggested Keywords: &\begin{minipage}[t]{10cm}
                    Non-Relativistic QCD, Heavy Quark Effective Theory,
                    effective field theory, threshold expansion,
                    renormalisation, $\beta$ function, dimensional
                    regularisation.
                    \end{minipage}
\end{tabular}
\end{titlepage}

\setcounter{page}{2} \setcounter{footnote}{0} \newpage

%

\section{Introduction}
\label{intro}
\setcounter{equation}{0}

Non-Relativistic QCD~\cite{CaswellLepage,BBL} takes advantage of the fact that
when a heavy quark is nearly on shell and its energy is dominated by its mass
$M$, the resulting non-relativistic system exhibits two small expansion
parameters: the coupling constant $g$ and the particle velocity $v$. The
Coulomb interaction rules the level spacing in Charmonium and Bottomium
because $\alpha_\mathrm{s}$ is small enough for perturbative calculations and
the relative velocity of the quarks is $v\sim \alpha_\mathrm{s}(Mv)$ by virtue
of the virial theorem, where the scale of $g$ is set by the inverse Bohr
radius $Mv$ of the system. Albeit $\alpha_\mathrm{s}$ increases with
decreasing $Q^2\sim (Mv)^2$, a window between the relativistic perturbative
and the confinement r\'egime remains in which both $\alpha_\mathrm{s}$ and $v$
is small (e.g.\ for Bottomium, $\alpha_\mathrm{s}(M_\mathrm{b}v)\approx
0.35$). Therefore, wave functions and potentials obtained by re-summation of
ladder diagrams involving relevant couplings as $v\to 0$ may be used to
account for bound state physics, and calculations of production cross
sections, hyperfine splittings, lifetimes, threshold properties etc.\ are much
facilitated. The Abelian analogue, NRQED, to which all considerations in this
article apply equally well, has also simplified precision calculations in
Positronium. An incomplete list of recent references may
include~\cite{KinoshitaNio}-\cite{BrambillaVairo}.

The NRQCD Lagrangean in terms of the heavy quark (anti-quark) bi-spinors $Q$
($\bar{Q}$) and gluons ($D_\mu=\de_\mu+\ii gA_\mu$) reads
\begin{eqnarray}
  \lefteqn{\calL_\mathrm{NRQCD}=
        Q^\dagger\Big(\ii\de_0-gA_0\Big)Q \;+\;
        \frac{{c_1}}{2M}\;Q^\dagger \vec{D}^2 Q\;+
        \;\frac{{c_2}}{8M^3}\; Q^\dagger\vec{D}^4 Q\;+\dots}\non\\
      \label{nrqcdlagr}
    &&+ \;\frac{g{c_3}}{2M}\; Q^\dagger\vec{\sigma}\cdot\vec{B}Q\;+\dots
        \;-\;\frac{g^2{d_1}}{4M^2}\;\Big(\calC\bar{Q}^\dagger\vec{\sigma}Q\Big)
        \cdot\Big(Q^\dagger \vec{\sigma}\calC\bar{Q}\Big)\;+\dots\\
    &&-\;\frac{{e_1}}{4}\; F_{\mu\nu}^a F^{\mu\nu,\,a}+
        \;\frac{g^3{e_2}}{480 \pi^2 M^2}\;F_{\mu\nu}^a  D^2
        F^{\mu\nu,\,a}+\;\dots+\;\calL_\mathrm{GFix}\;\;,\non
\end{eqnarray}
where the coefficients $c_i,d_i,e_i,\dots$ encode the ultraviolet physics: All
excitations with four-momenta of the order of $M$ and higher are integrated out
and give rise e.g.\ to four-point interactions between quarks ($d_1\not=0$).
The perturbative part of the coefficients can be determined by matching NRQCD
matrix elements to their QCD counterparts in the r\'egime where both theories
are perturbative in the coupling constant. This has been performed to
$\calO(M^{-3})$~\cite{Manohar}. At tree level, a Foldy-Wouthuysen
transformation gives $c_i=1,e_1=1$, and loop corrections are down by powers of
$g$, the most famous example being the coefficient for the Fermi term related
to the anomalous magnetic moment of the electron,
$c_3=1+\frac{\alpha_\mathrm{s}}{2\pi}+\dots$. Further coefficients to enter at
one loop level are $d_1=e_2=1$. Lorentz invariance demands $c_1=c_2=1$ to all
orders.

The Lagrangean (\ref{nrqcdlagr}) consists of infinitely many terms constrained
only by the symmetries of the theory and is non-renormalisable. Predictive
power is nonetheless established when only a finite number of terms contribute
to a given order in the two expansion parameters\footnote{For clarity, the two
  will be distinguished in the following, although $v\sim\alpha_\mathrm{s}$,
  as noted above.}, $g$ and $v$. Besides the heavy quark mass $M$, the typical
binding energy and momentum scales in NRQCD are the non-relativistic kinetic
energy $Mv^2$ and the momentum $Mv$ of the quark, which in Quarkonia appear as
the energy and inverse size of the bound state~\cite{CaswellLepage,BBL}. Since
for the smallest of the three scales $\alpha_\mathrm{s}(Mv^2)\not<1$ in
Bottomium and Charmonium, an expansion in $g$ is justified only for the
interactions taking place on scales $Mv$ and higher in the real world, but one
can imagine a world in which $Mv^2\gg\Lambda_\mathrm{QCD}$. Toponium fulfils
this requirement but decays mainly weakly and very fast so that QCD effects do
not dominate.

Because the effective Lagrangean does not exhibit the non-relativistic
expansion parameter $v$ explicitly, a power counting scheme has to be
established which determines uniquely which terms in the Lagrangean must be
taken into account to render consistent calculations and predictive power to a
given accuracy in $v$. It is at this point that NRQCD can serve as a ``toy
model'' for nuclear physics~\cite{LukeManohar} (although this grossly
understates its value\footnote{The most striking difference to nuclear physics
  is that NRQCD has no exceptionally large scattering length to accommodate.}):
It will establish what the relevant kinematic r\'egimes and infrared variables
are in a theory with three (or more) separate scales, and it will demonstrate
how to count powers of the non-relativistic expansion parameter, $v$. Recently,
velocity power counting rules were established for a toy model~\cite{hgpub3}
following Beneke and Smirnov's threshold expansion~\cite{BenekeSmirnov}, which
has become very useful for understanding non-relativistic effective theories.
The relevant energy and momentum r\'egimes and low energy degrees of freedom
were found for dimensionally regularised non-relativistic theories. This
article presents the extension to NRQCD and as an application the calculation
of the NRQCD $\beta$ function. It will also be argued that the diagrammatic
approach advocated here sheds a new light on threshold
expansion~\cite{BenekeSmirnov} and helps simplify calculations.

Both from a conceptual and a technical point of view the calculation of the
NRQCD $\beta$ function proves interesting. NRQCD is a well-defined field theory
of quarks and gluons. One can therefore investigate its renormalisation group
equations under the assumption that perturbation theory is applicable at all
scales. For example, the running and mixing of the dimension six operators has
been investigated to one loop order by Bauer and Manohar~\cite{BauerManohar}.
On the other hand, as NRQCD is the low energy limit of QCD, it must reproduce
the QCD $\beta$ function with $N_\mathrm{F}$ light quarks below the scale $M$,
but this has not been demonstrated so far. The problem with the power counting
developed until now~\cite{LukeManohar,LukeSavage,GrinsteinRothstein} was that
gluons mediating the Coulomb interaction (``potential gluons'') seemed to have
a vanishing vacuum polarisation and hence a $\beta$ function which was neither
gauge invariant nor agreed with the QCD result, while the coupling of gluons on
a much lower scale which mediate bremsstrahlung processes ran as expected. The
apparent impasse which results is resolved in this article.

Straightforward as the computation may thus seem with its outcome already
anticipated, the prime goal of this article is not the result for the NRQCD
$\beta$ function; the objective is rather to shed more light on power counting
in NRQCD and to provide a comparison between calculations in the effective and
the full theory. It will show that the Lorentz gauge family (and indeed any
standard gauge) is a legitimate gauge choice in NRQCD, not only the Coulomb
gauge. It will prove that the recently discussed soft
r\'egime~\cite{hgpub3,BenekeSmirnov,hgpub4} with quarks and gluons of
four-momenta of order $Mv$ is indispensable for NRQCD to describe the correct
infrared limit of QCD. It will also test the validity of the power counting
proposed. In the Coulomb gauge, it will demonstrate that the renormalisation of
the quark propagator and of the leading order quark gluon vertex are trivial,
so that for the $\beta$ function in this gauge, only a computation of the gluon
vacuum polarisation is necessary. From a technical point of view, dimensional
regularisation for non-covariant loop integrals proves to be a simple regulator
choice. It allows to develop a set of simple diagrammatic rules to determine
which graphs yield non-zero contributions in perturbation theory. This reduces
the computational effort considerably as compared to other regularisation
schemes in which power divergences are present like e.g.\ when using a lattice
cut-off. Because of the association of threshold expansion and NRQCD, such
rules can be applied to the former, too. Finally, the calculation of the NRQCD
$\beta$ function presented here is -- especially in the Feynman gauge --
simpler than its QCD counterpart and shows some pedagogically intriguing
aspects.

\absatz The article is organised as follows: In Sect.\ 
\ref{sec:velocitypowercounting}, the velocity power counting of NRQCD is
presented. The relevant r\'egimes of NRQCD are identified (Sect.\ 
\ref{sec:regimes}), extending the formalism of Luke and
Savage~\cite{LukeSavage} by the soft r\'egime of Beneke and
Smirnov~\cite{BenekeSmirnov}. The section then proposes the rescaling rules
necessary for a Lagrangean with manifest velocity power counting (Sect.\ 
\ref{sec:props}) and gives the vertex (Sect.\ \ref{sec:vertex}) and loop
velocity power counting rules (Sect.\ \ref{sec:loop}). Gauge invariance of the
power counting is addressed in Sect.\ \ref{sec:gaugeinvariance}. The
calculation of the NRQCD $\beta$ function in the Lorentz gauges in Sect.\ 
\ref{sec:betafunction} starts by highlighting the intimate relation between
threshold expansion and the proposed NRQCD power counting, which will
additionally be summarised in the Conclusions. Diagrammatic rules developed in
Sect.\ \ref{sec:diagrammar} help to facilitate the calculation considerably and
are used in Sect.\ \ref{sec:theorems} to prove the absence of ultrasoft heavy
quarks. As a tilly, the $\beta$ function is computed in the Coulomb gauge in
Sect.\ \ref{sec:coulombbeta}. Summary and outlook conclude the article,
together with an appendix containing exemplary calculations in non-covariant
dimensional regularisation.

\section{Velocity Power Counting}
\label{sec:velocitypowercounting}
\setcounter{equation}{0}

\subsection{R\'egimes of NRQCD}
\label{sec:regimes}

The first power counting rules for NRQCD were derived by Lepage et al.\ 
\cite{LMNMH} in the Coulomb gauge using the consistency of the equations of
motion and a momentum cut-off. Only Coulomb gluons with typical momenta of
order $Mv$ were considered, taking into account retardation, but neglecting
bremsstrahlung effects. The power counting also held to all orders only after
all power divergences were subtracted.

Simpler counting rules were proposed by Luke and Manohar for Coulomb
interactions~\cite{LukeManohar}, and by Grinstein and Rothstein for
bremsstrahlung processes~\cite{GrinsteinRothstein}. Luke and
Savage~\cite{LukeSavage} united these schemes using dimensional regularisation,
so that the tree level power counting is automatically preserved to all orders
in perturbation theory up to logarithmic corrections. They also extended the
formalism to include the Lorentz gauges. Labelle~\cite{Labelle} proposed a
similar scheme in time ordered perturbation theory. Beneke and
Smirnov~\cite{BenekeSmirnov} observed that the collinear divergence of the two
gluon exchange contribution to Coulomb scattering between non-relativistic
particles near threshold is not reproduced in this version of NRQCD because the
soft low energy r\'egime is not taken into account. A recent
article~\cite{hgpub3} has shown that an extension of the work
in~\cite{LukeManohar,LukeSavage,GrinsteinRothstein} using dimensional
regularisation, as motivated from threshold expansion~\cite{BenekeSmirnov},
resolves this conflict in a toy model. The complete power counting scheme of
NRQCD is presented briefly in the following (see also~\cite{hgpub4}). In
Sect.\ \ref{sec:nrqcdandthreshold}, the intimate relation between NRQCD as
developed in recent
years~\cite{LukeManohar,hgpub3,LukeSavage,GrinsteinRothstein} and threshold
expansion is addressed in more detail.

The NRQCD propagators are read up from the Lagrangean (\ref{nrqcdlagr}) as
\begin{equation}
  \label{nonrelprop}
  Q\;:\;\frac{\ii\;\mathrm{Num}}{T-\frac{\pv^2}{2M}+\ii\epsilon}\;\;,\;\;
   \Amu\;:\;\frac{\ii\;\mathrm{Num}}{k^2+\ii\epsilon}\;\;,
\end{equation}
where $T=p_0-M=\frac{\pv^2}{2M}+\dots$ is the kinetic energy of the quark.
``$\mathrm{Num}$'' are numerators containing the appropriate colour, Dirac and
flavour indices, all of which are unimportant for the considerations in this
section and will be suppressed throughout this article. Gauge dependence of the
gluon propagator for the power counting is addressed in Sect.\ 
\ref{sec:gaugeinvariance}.

Cuts and poles in scattering amplitudes close to threshold stem from saddle
points of the loop integrand, corresponding to bound states and on-shell
propagation of particles in intermediate states. They give rise to infrared
divergences and in general dominate contributions to scattering amplitudes.
With the two low energy scales at hand, and energies and momenta being of
either scale, three r\'egimes are identified in which either the quark or the
gluon in (\ref{nonrelprop}) is on shell, as noted by Beneke and
Smirnov~\cite{BenekeSmirnov}:
\begin{eqnarray}
   \mbox{soft r\'egime: }&\Asmu:&\;k_0\sim |\kv|\sim Mv\;\;,\non\\
  \label{regimes}
   \mbox{potential r\'egime: }&Q_\mathrm{p}:&\;T\sim Mv^2\;,\; |\pv|\sim
   Mv\;\;,\\
   \mbox{ultrasoft r\'egime: }&\Aumu:&\;k_0\sim |\kv|\sim Mv^2\non
\end{eqnarray}
Note that the scale $M$ has been integrated out when deriving the
non-relativistic Lagrangean (\ref{nrqcdlagr}), so that a ``hard'' r\'egime
with on shell gluons and quarks of four-momentum of order $M$ cannot be
considered.

What is the particle content of each of the three r\'egimes? Ultrasoft gluons
$\Aumu$ are emitted as bremsstrahlung or from excited states in the bound
system, and hence are physical. Soft gluons $\Asmu$ do not describe
bremsstrahlung: Because in- and outgoing quarks $Q_\mathrm{p}$ are close to
their mass shell, they have an energy of order $Mv^2$. Therefore, overall
energy conservation forbids all processes with outgoing soft gluons but without
ingoing ones, and vice versa, as their energy is of order $Mv$. Finally, gluons
which change the quark momenta but keep them close to their mass shell relate
the (instantaneous) Coulomb interaction:
\begin{equation}
  \label{pgluon}
  \Apmu\;\;:\;\;k_0\sim Mv^2\;,\;|\kv|\sim Mv
\end{equation}
When a soft gluon $\Asmu$ couples to a potential quark $ Q_\mathrm{p}$, the
outgoing quark is far off its mass shell and carries energy and momentum of
order $Mv$. Therefore, consistency requires the existence of quarks in the
soft r\'egime as well,
\begin{equation}
  \label{squark}
  Q_\mathrm{s}\;\;:\;\;T\sim |\pv|\sim Mv\;\;.
\end{equation}
With these five fields $\Qs,\Qp,\Asmu,\Apmu,\Aumu$ representing quarks and
gluons in the three different non-relativistic r\'egimes, soft, potential and
ultrasoft, NRQCD becomes self-consistent. These fields are the
infrared-relevant degrees of freedom representing one and the same
non-relativistic particle in the respective kinematic r\'egimes and came
naturally by identifying all possible particle poles in the non-relativistic
propagators. Their interactions are fixed by the non-relativistic Lagrangean
(\ref{nrqcdlagr}). Section \ref{sec:theorems} will prove that a hypothetical
ultrasoft quark, created e.g.\ by the radiation of a potential gluon off a
potential quark, decouples completely from the theory. On the other hand,
Sect.\ \ref{sec:betafunction} will demonstrate that in order to obtain the
correct result for the NRQCD $\beta$ function, all fields listed above in the
three r\'egimes have to be accounted for. Therefore, the particle content
presented above is not only consistent but both minimal and complete.

In order to guarantee that there is no overlap between interactions and
particles in different r\'egimes, the regularisation scheme must finally be
chosen such that the expansion around one saddle point in the loop integral as
performed above does not obtain any contribution from other r\'egimes,
represented by other saddle points in the loop integrals. One might use an
energy and momentum cut-off separating the soft from the potential, and the
potential from the ultrasoft r\'egime, but the integrals encountered can in
general not be performed analytically and in closed form. Furthermore, by
introducing another, artificial scale, cut-off regularisation jeopardises power
counting and symmetries (gauge, chiral, \dots). Power divergences occur when
the (un-physical) cut-off is removed in intermediate steps but not in the
final, physical result, complicating the computation. In contradistinction,
using dimensional regularisation \emph{after} the saddle point expansion
preserves power counting because its homogeneity guarantees that contributions
from different saddle points and r\'egimes do not overlap. (A simple example is
given in Ref.~\cite{hgpub3}.) Homogene{\ia}ty will also be essential when
developing diagrammatic rules to classify graphs as zero in Sect.\ 
\ref{sec:diagrammar} and when showing the decoupling of the ultrasoft quark in
Sect.\ \ref{sec:theorems}.

\subsection{Rescaling Rules and Propagators}
\label{sec:props}

In order to establish explicit velocity power counting in the NRQCD
Lagrangean, one rescales the space-time coordinates such that typical momenta
in either r\'egime are dimensionless, as proposed by Luke and
Manohar~\cite{LukeManohar} for the potential r\'egime and by Grinstein and
Rothstein~\cite{GrinsteinRothstein} for the ultrasoft one:
\begin{equation}
  \begin{array}{rll}
    \mbox{soft: }\;\;& t=(Mv)^{-1}\;\ts\;\;,&\xv=(Mv)^{-1}\;\xs\;\;,\\
    \label{xtscaling}
    \mbox{potential: }\;\;& t=(Mv^2)^{-1}\;\tu\;\;,&\xv=(Mv)^{-1}\;\xs\;\;,\\
    \mbox{ultrasoft: }\;\;& t=(Mv^2)^{-1}\;\tu\;\;,&\xv=(Mv^2)^{-1}\;\xu\;\;.
  \end{array}
\end{equation}
For the propagator terms in the NRQCD Lagrangean to be normalised as order
$v^0$, one sets for the representatives of the gluons in the three
r\'egimes~\cite{hgpub3,hgpub4}
\begin{eqnarray}
   \mbox{soft:} && \Asmu(\xv,t) = (Mv)\; \calAsmu(\xs,\ts)\;\;,\non\\
   \label{gluonscaling}
   \mbox{potential:} && \Apmu(\xv,t) = (Mv^{\frac{3}{2}})\;
                                           \calApmu(\xs,\tu)\;\;,\\
   \mbox{ultrasoft:} && \Aumu(\xv,t) = (Mv^2)\; \calAumu(\xu,\tu)\;\;,\non
\end{eqnarray}
and for the quark representatives
\begin{eqnarray}
  \label{quarkscaling}
  \mbox{soft:} && Q_\mathrm{s}(\xv,t) = (Mv)^{\frac{3}{2}}\;
  \calQs(\xs,\ts)\;\;,\\
  \mbox{potential:} &&Q_\mathrm{p}(\xv,t) = (Mv)^{\frac{3}{2}}\;
  \calQp(\xs,\tu)\;\;.\non
\end{eqnarray}
The rescaled free quark Lagrangean reads then
\begin{eqnarray}
  \label{qsfreelagr}
   \mbox{soft:} &&\dedreiXs\deTs\calQs^\dagger\Big(\ii\de_0+
   \frac{{v}}{2} \;\vec{\de}^2 \Big)\calQs\;\;,\\
  \label{qpfreelagr}
   \mbox{potential:} &&\dedreiXs\deTu
   \calQp^\dagger\Big(\ii\de_0+\frac{1}{2}\; \vec{\de}^2 \Big)\calQp\;\;.
\end{eqnarray}
Here, as in the following, the positions of the fields have been left out
whenever they coincide with the rescaled variables of the volume element.
Derivatives are to be taken with respect to the rescaled variables of the
volume element unless otherwise stated. In order to maintain velocity power
counting, corrections of order $v$ or higher must be treated as insertions, so
that one reads up the (un-rescaled) quark propagators and insertion as
\begin{eqnarray}
  \label{qsprop}
       \mbox{soft:} && \Qs:\;
       \feynbox{60\unitlength}{
       \begin{fmfgraph*}(60,30)
         \fmfleft{i}
         \fmfright{o}
         \fmf{heavy,foreground=red,width=thin,label=$\fs(T,,\pv)$,
           label.side=left}{i,o}
       \end{fmfgraph*}}
       \;:\;\frac{\ii}{T+\ii\epsilon}\;\;,\;\;
      \feynbox{60\unitlength}{
      \begin{fmfgraph*}(60,30)
        \fmfleft{i}
        \fmfright{o}
        \fmf{double,foreground=red,width=thin}{i,v,o}
        \fmfv{decor.shape=cross,foreground=red,label=$\fs(T,,\pv)$,
          label.angle=90,
          label.dist=0.15w}{v}
      \end{fmfgraph*}}
      \;:\;-\ii\;\frac{\pv^2}{2M}\;=\;\pc(v^1)\;\;,\;\;\;\;\;{}{}\\
   \label{qpprop}
    \mbox{potential:} && \Qp:\;
    \feynbox{60\unitlength}{
    \begin{fmfgraph*}(60,30)
      \fmfleft{i}
      \fmfright{o}
      \fmf{fermion,width=thick,label=$\fs(T,,\pv)$,label.side=left}{i,o}
    \end{fmfgraph*}}
    \;:\;\frac{\ii}{T-\frac{\pv^2}{2M}+\ii\epsilon}\;\;.
\end{eqnarray}
The soft quark becomes static because $T\sim Mv \gg \frac{\pv^2}{2M}\sim Mv^2$
allows one to expand the non-relativistic propagator (\ref{nonrelprop}) and to
treat the momentum term as an insertion.

Throughout this article, the symbol $\pc(v^n)$ denotes at which order in the
velocity power counting a certain term or diagram contributes. In the
approach presented, each line in a loop diagram counts as $\pc(v^0)$ because
the strength of the propagator has been set to unity in the rescaled
Lagrangean (\ref{qsfreelagr}/\ref{qpfreelagr}). The rescaled Lagrangean
measures the strength of the insertion relative to the propagator. Therefore,
the soft quark insertion (\ref{qsprop}) does not count as $\frac{\pv^2}{2M}$
which scales like $v^2$, but as $\frac{\pv^2}{2M}/T=\pc(v)$. In
contradistinction to the approach in~\cite{BenekeSmirnov}, one will therefore
not obtain the absolute order in $v$ at which a given graph contributes, but
the relative order in $v$ between graphs or vertices is read off more easily
and asserted correctly.

\absatz The gauge fixing term was included in the NRQCD Lagrangean
(\ref{nrqcdlagr}) since the decomposition of the Lagrangean into a free and an
interacting part is gauge dependent. Because of the difference between
canonical and physical momentum, it is important to specify the gauge
\emph{before} identifying to which order in $v$ a certain r\'egime in the
Lagrangean contributes, as will be seen shortly. In the following, the Feynman
rules for the Lorentz gauges and for the Coulomb gauge are derived explicitly.
Sect.\ \ref{sec:gaugeinvariance} will comment on the gauge invariance of the
procedure.

The rescaled free gluon Lagrangean in the Lorentz gauges reads
\begin{eqnarray}
  \label{gsfreelagrlorentz}
  \mbox{soft:} &&\dedreiXs\deTs
  \frac{1}{2}\;\calAsmu \Big[\de^2 g_{\mu\nu}-(1-\frac{1}{\alpha})\de_\mu
  \de_\nu\Big]\calA_\mathrm{s}^{\nu}\;\;,\\
  \label{gpfreelagrlorentz}
  \mbox{potential:} &&\dedreiXs\deTu
  \frac{1}{2}\;\calA_{\mathrm{p}}^\mu \Big[ g_{\mu\nu} (v^2\de_0^2-\dev^2)-\\
  &&\phantom{\dedreiXs\deTu\frac{1}{2}\;}-
  (1-\frac{1}{\alpha}) (v \delta_{\mu 0}\de_0+\delta_{\mu i}\de_i)
  (v \delta_{\nu 0}\de_0+\delta_{\nu i}\de_i)\Big]\calA_\mathrm{u}^{\nu}
  \;\;,\non\\
  \label{gufreelagrlorentz}
  \mbox{ultrasoft:} && \dedreiXu\deTu
  \frac{1}{2}\;\calAumu \Big[\de^2 g_{\mu\nu}-(1-\frac{1}{\alpha})\de_\mu
  \de_\nu\Big]\calA_\mathrm{u}^{\nu}\;\;,
\end{eqnarray}
while in the Coulomb gauge
\begin{eqnarray}
  \label{gsfreelagrcoulomb}
  \mbox{soft:} &&\dedreiXs\deTs
  \half\;{\calA_{i,\mathrm{s}}}
        \Big[(\dev^2-\de_0^2)\delta_{ij}-\de_i\de_j\Big]
      {\calA_{j,\mathrm{s}}}\;\;,\\
  \label{gpfreelagrcoulomb}
  \mbox{potential:} &&\dedreiXs\deTu
   \half\;\Big[{\calA_{0,\mathrm{p}}}\dev^2{\calA_{0,\mathrm{p}}}+
     {\calA_{i,\mathrm{p}}}
        (\dev^2\delta_{ij}-\de_i\de_j-{v^2}\de_0^2\delta_{ij})
      {\calA_{j,\mathrm{p}}}\Big]\;,\\
  \label{gufreelagrcoulomb}
  \mbox{ultrasoft:} &&\dedreiXs\deTs
  \half\;{\calA_{i,\mathrm{u}}}\Big[(\dev^2-\de_0^2)\delta_{ij}-\de_i\de_j\Big]
      {\calA_{j,\mathrm{u}}}\;\;.
\end{eqnarray}
The (un-rescaled) Coulomb and Lorentz gauge propagators are therefore given as
($\delta_\mathrm{tr}^{ij}(\kv)=\delta^{ij}-\frac{k^i k^j}{\kv^2}$,
$\calG_{\mu\nu}(k)=-(g_{\mu\nu}-(1-\alpha)\frac{k_\mu
  k_\nu}{k^2+\ii\epsilon})$)
\begin{equation}
\begin{array}{rrcc}  
  &&\mbox{Coulomb gauge}&\mbox{Lorentz gauges}\non\\[2ex]
  \label{gsprop}
       \mbox{soft:}\;\;&\Asmu:\;
    \feynbox{60\unitlength}{
      \begin{fmfgraph*}(60,30)
        \fmfleft{i}
        \fmfv{label=$\fs \mu$,label.angle=90,label.dist=0.1w}{i}
        \fmfright{o}
        \fmfv{label=$\fs \nu$,label.angle=90,label.dist=0.1w}{o}
        \fmf{zigzag,foreground=red,width=thin,label=$\fs k$,
          label.side=left}{i,o}
      \end{fmfgraph*}}
    :&\dis\frac{\ii\;\delta_\mathrm{tr}^{ij}(\kv)}{k^2+\ii\epsilon}\;\;&
    \dis\frac{\ii\;\calG_{\mu\nu}(k)}{k^2+\ii\epsilon}\;\;,
    \\[2ex]
  \label{gpprop}
    \mbox{potential:} \;\;&
    A_{\mathrm{p},0}:\;
      \feynbox{60\unitlength}{
        \begin{fmfgraph*}(60,30)
          \fmfleft{i}
          \fmfright{o}
          \fmfv{label=$\fs 0$,label.angle=90,label.dist=0.1w}{i}
          \fmfv{label=$\fs 0$,label.angle=90,label.dist=0.1w}{o}
          \fmf{dashes,width=thin,label=${\fs k,,A_0}$,label.side=left}{i,o}
        \end{fmfgraph*}}
      :&\dis\frac{-\ii}{-\kv^2+\ii\epsilon}\;\;&
      \dis\frac{-\ii}{-\kv^2+\ii\epsilon}\;\;,\\[2ex]
    &\Avp:\;
      \feynbox{60\unitlength}{
        \begin{fmfgraph*}(60,30)
          \fmfleft{i}
          \fmfv{label=$\fs i$,label.angle=90,label.dist=0.1w}{i}
          \fmfright{o}
          \fmfv{label=$\fs j$,label.angle=90,label.dist=0.1w}{o}
          \fmf{dashes,width=thin,label=${\fs k,,\Av}$,label.side=left}{i,o}
        \end{fmfgraph*}}
      :&\dis\frac{\ii\;\delta_\mathrm{tr}^{ij}(\kv)}{-\kv^2+\ii\epsilon}\;\;&
      \dis\frac{\ii\;\big[\delta^{ij}+(1-\alpha)
         \frac{k_ik_j}{-\kv^2+\ii\epsilon}\big]}{-\kv^2+\ii\epsilon}\;\;,
       \non\\[2ex]
    \label{guprop} \mbox{ultrasoft:}\;\;&
      \Aumu:\;
      \feynbox{60\unitlength}{
        \begin{fmfgraph*}(60,30)
          \fmfleft{i}
          \fmfv{label=$\fs \mu$,label.angle=90,label.dist=0.1w}{i}
          \fmfright{o}
          \fmfv{label=$\fs \nu$,label.angle=90,label.dist=0.1w}{o}
          \fmf{photon,foreground=blue,width=thin,label=$\fs k$,
            label.side=left}{i,o}
        \end{fmfgraph*}}
      :&\dis\frac{\ii\;\delta_\mathrm{tr}^{ij}(\kv)}{k^2+\ii\epsilon}\;\;&
      \dis\frac{\ii\;\calG_{\mu\nu}(k)}{k^2+\ii\epsilon}\;\;.
\end{array}
\end{equation}
The potential gluon becomes instantaneous in both gauges as expected for the
particle mediating the Coulomb interaction. Insertions are necessary only in
the potential r\'egime:
\begin{equation}
\begin{array}{rcc}
  &\;\;\mbox{Coulomb gauge}\;\;&\;\;\mbox{Lorentz gauges}\;\;\\[1.5ex]
     \feynbox{60\unitlength}{
       \begin{fmfgraph*}(60,30)
         \fmfleft{i}
         \fmfright{o}
         \fmfv{label=$\fs 0$,label.angle=90,label.dist=0.1w}{i}
         \fmfv{label=$\fs 0$,label.angle=90,label.dist=0.1w}{o}
         \fmf{dashes,width=thin}{i,v,o}
         \fmfv{decor.shape=cross,label=${\fs k,,A_0}$,label.angle=90,
           label.dist=0.15w}{v}
       \end{fmfgraph*}}
     :&\dis \mbox{none}&\dis-\ii\;\frac{k_0^2}{\alpha}=\pc(v^2)\;\;,\\[1.5ex]
       \label{ginsertions}
       \feynbox{60\unitlength}{
         \begin{fmfgraph*}(60,30)
           \fmfleft{i}
           \fmfright{o}
           \fmfv{label=$\fs i$,label.angle=90,label.dist=0.1w}{i}
           \fmfv{label=$\fs j$,label.angle=90,label.dist=0.1w}{o}
           \fmf{dashes,width=thin}{i,v,o}
           \fmfv{decor.shape=cross,label=${\fs
            k,,\Av}$,label.angle=90,label.dist=0.15w}{v}
      \end{fmfgraph*}}
    :&\;\;\dis \ii\; k_0^2\;\delta_{ij}\;=\;\pc(v^2)\;\;&
      \non\ii\; k_0^2\;\delta_{ij}\;=\;\pc(v^2)\;\;,\\[1.5ex]
      \feynbox{60\unitlength}{
        \begin{fmfgraph*}(60,30)
          \fmfleft{i}
          \fmfright{o}
          \fmfv{label=$\fs i$,label.angle=90,label.dist=0.1w}{i}
          \fmfv{label=$\fs 0$,label.angle=90,label.dist=0.1w}{o}
          \fmf{dashes,width=thin}{i,v,o}
          \fmfv{decor.shape=cross,label=${\fs
            k,,\Amu}$,label.angle=90,label.dist=0.15w}{v}
        \end{fmfgraph*}} 
      :&\dis\mbox{none}&\;\;\dis-\ii\;(1-\frac{1}{\alpha})\;k_i
                 k_0=\pc(v)
\end{array}
\end{equation}
The Lorentz gauge propagators and insertions for the potential and ultrasoft
r\'egimes were first given in~\cite{LukeSavage}. Especially for the Feynman
gauge $\alpha=1$ ($\calG_{\mu\nu}(k)=g_{\mu\nu}$), Lorentz and Coulomb gauges
in the potential r\'egime differ only by insertions, i.e.\ at higher order in
$v$.

As seen from (\ref{gsfreelagrcoulomb}--\ref{gufreelagrcoulomb}), the choice of
the Coulomb gauge makes $A_0$ instantaneous to all orders in $v$, and hence it
contributes in the potential r\'egime, only. Since in this gauge, $A_0$ solely
mediates the instantaneous Coulomb potential (physical fields are transverse by
virtue of Gau\3' law), this result was to be expected. The field $\Avp$ is
associated with retardation effects like spin-orbit coupling and the Darwin
term in (\ref{nrqcdlagr}). In the Coulomb gauge, the advantages of $A_0$ only
contributing in the potential r\'egime and of its propagator having no
insertions or admixtures with the vector components of the gauge field, are
balanced by the fact that a more lengthy and cumbersome renormalisation seems
necessary. The calculation of the NRQCD $\beta$ function will prove that the
Lorentz gauges are a legitimate gauge choice in NRQCD, although the number of
diagrams appears to be larger because of a larger number of vertices, as will
be seen now.

\subsection{Vertex Rules}
\label{sec:vertex}

By experience, particles in the various r\'egimes couple: On-shell (potential)
quarks radiate bremsstrahlung (ultrasoft) gluons. In general, one must allow
all couplings between the various r\'egimes which obey ``scale conservation'':
Both energies and momenta must be conserved within each r\'egime to the order
in $v$ one works. This will exclude for example the coupling of two potential
quarks ($T\sim Mv^2$) to one soft gluon ($q_0\sim Mv$), but not to two soft
gluons via the $Q^\dagger \vec{A}\cdot\vec{A}Q$ term of the Lagrangean
(\ref{nrqcdlagr}). For a more rigorous derivation of this rule see Sect.\ 
\ref{sec:theorems}.

As an example, consider a bremsstrahlung-like process: the radiation of a soft
scalar gluon $A_{\mathrm{s}\,,0}$ off a soft quark, resulting in a potential
quark with four momentum $(T^\prime_\mathrm{p},\pv^\prime)$. The rescaled
interaction Lagrangean with its Hermitean conjugate reads
\begin{equation}
  \label{ssplagr}
  \dedreiXs\deTs \Big[- g \;{v^0}\; \calQs^\dagger(\Xvs,\Ts)\,
   \calA_{\mathrm{s},\,0}(\Xvs,\Ts) \, \calQp (\Xvs,{v}\Ts)\;+\mbox{ H.c. }
   \Big]\;\;.
\end{equation}
Note that the scaling r\'egime of the volume element is set by the particle
with the highest momentum and energy. Therefore, one obtains the power counting
of a given vertex in a specific r\'egime. As the external quarks are on shell,
i.e.\ in the potential r\'egime, the power counting rules of subgraphs whose
scaling rules have been determined in the soft r\'egime have to be transferred
to the potential r\'egime. This step is postponed to the next section. The
strength of the example vertex above is read up as $v^0$ \emph{in the soft
  r\'egime}. This is clearly also the strength of the Hermitean conjugate
vertex.

The interaction Lagrangean is non-local in the rescaled variables, and in
order to maintain velocity power counting, $\calQp (\Xvs,{v}\Ts)$ has to be
expanded about $\calQp (\Xvs,0)$ in powers of $v$. The Feynman rule for this
vertex is after inverting the rescaling
\begin{equation}
  \rule{0pt}{30pt}
  \label{sspvertex}
    \feynbox{75\unitlength}{
      \begin{fmfgraph*}(75,30)
        \fmfstraight
        \fmftop{i,o}
        \fmfbottom{u}
        \fmf{heavy,foreground=red,width=thin,label=$\fs(T,,\pv)$,
          label.side=left,label.dist=0.1w}{i,v}
        \fmf{fermion,width=thick,label=$\fs(T^\prime_\mathrm{p},,\pv^\prime)$,
          label.side=left,label.dist=0.1w}{v,o}
        \fmffreeze
        \fmf{zigzag,foreground=red,width=thin,label=$\uparrow\fs q,,A_0$}{u,v}
      \end{fmfgraph*}}
    \begin{array}{l}
          :\;-\ii g \;(2\pi)^4\;
          \delta^{(3)}(\pv-\pv^\prime+\qv)\times \\
          \;\;\;\;\;\times\Big[\exp\Big({-T^\prime_\mathrm{p}\;
            \frac{\partial}{\partial
          (T+q_0)}}\Big)\;\delta(T+q_0)\Big] \;=\;\pc(\e^v)\;\;.
      \end{array}
\end{equation}
The Hermitean conjugate vertex is built analogously. As the potential quark
has a much smaller energy than either of the soft particles, it can -- by the
uncertainty relation -- not resolve the precise time at which the soft quark
emits or absorbs the soft gluon. This ``temporal'' multipole expansion comes
technically from the different scaling of $\xv$ and $t$ in the three
r\'egimes. The multipole expansion symbolised in the power counting by $\e^v$
corresponds term by term to an expansion in $v$ and should be truncated at the
desired order. In general, the coupling between particles of different
r\'egimes will not be point-like but contain multipole expansions for the
particle belonging to the weaker kinematic r\'egime. That the coupling of
potential quarks to ultrasoft gluons requires a momentum multipole expansion
as in atomic physics has been observed by Grinstein and
Rothstein~\cite{GrinsteinRothstein}, and by Labelle~\cite{Labelle}.

Amongst the fields introduced, six scale conserving interactions are allowed
within and between the various r\'egimes for any coupling of one gluon to two
quarks. Because only in the Coulomb gauge, $A_0$ is a field in the potential
r\'egime only, its propagator always being instantaneous, only the first two
interactions exist for the scalar coupling $-g Q^\dagger A_0 Q$ of table
\ref{scalarvertex} in this gauge choice. The $v$ counting for the lowest order
quark-gluon interactions from this vertex is presented in table
\ref{scalarvertex}.
\begin{table}[!htb]
  \caption{\label{scalarvertex} \sl Velocity power counting and vertices for
    the interaction Lagrangean \protect$-g Q^\dagger A_0 Q$. In the Coulomb
    gauge, only the first two diagrams exist. The last line indicates the
    field for which an energy or momentum multipole expansion has to be
    performed.} 
\begin{center}
  \footnotesize
\begin{tabular}{|r|c|c||c|c|c|c|}
  \hline
  Vertex\rule[-20pt]{0pt}{48pt}&
      \feynbox{50\unitlength}{
      \begin{fmfgraph*}(50,25)
        \fmfstraight
        \fmftop{i,o}
        \fmfbottom{u}
        \fmf{vanilla,width=thick}{i,v}
        \fmf{vanilla,width=thick}{o,v}
        \fmffreeze
        \fmf{dashes,width=thin,label=$\fs A_0$}{u,v}
      \end{fmfgraph*}}&
      \feynbox{50\unitlength}{
        \begin{fmfgraph*}(50,25)
          \fmfstraight
          \fmftop{i,o}
          \fmfbottom{u}
          \fmf{double,foreground=red,width=thin}{i,v}
          \fmf{double,foreground=red,width=thin}{o,v}
          \fmffreeze
          \fmf{dashes,width=thin}{u,v}
      \end{fmfgraph*}}&
      \feynbox{50\unitlength}{
      \begin{fmfgraph*}(50,25)
        \fmfstraight
        \fmftop{i,o}
        \fmfbottom{u}
        \fmf{vanilla,width=thick}{i,v}
        \fmf{vanilla,width=thick}{o,v}
        \fmffreeze
        \fmf{photon,foreground=blue,width=thin}{u,v}
      \end{fmfgraph*}}&
      \feynbox{50\unitlength}{
      \begin{fmfgraph*}(50,25)
        \fmfstraight
        \fmftop{i,o}
        \fmfbottom{u}
        \fmf{double,foreground=red,width=thin}{i,v}
        \fmf{double,foreground=red,width=thin}{o,v}
        \fmffreeze
        \fmf{zigzag,foreground=red,width=thin}{u,v}
      \end{fmfgraph*}}&
      \feynbox{50\unitlength}{
      \begin{fmfgraph*}(50,25)
        \fmfstraight
        \fmftop{i,o}
        \fmfbottom{u}
        \fmf{double,foreground=red,width=thin}{i,v}
        \fmf{vanilla,width=thick}{o,v}
        \fmffreeze
        \fmf{zigzag,foreground=red,width=thin}{u,v}
      \end{fmfgraph*}}&
      \feynbox{50\unitlength}{
      \begin{fmfgraph*}(50,25)
        \fmfstraight
        \fmftop{i,o}
        \fmfbottom{u}
        \fmf{double,foreground=red,width=thin}{i,v}
        \fmf{double,foreground=red,width=thin}{o,v}
        \fmffreeze
        \fmf{photon,foreground=blue,width=thin}{u,v}
      \end{fmfgraph*}}
  \\[2ex]
  \hline
  \protect$\pc$\rule[-12pt]{0pt}{30pt}&
  \protect$\frac{1}{\sqrt{v}}$&
  \protect$v^\frac{1}{2}$&
  \protect$v^0$&
  \protect$v^0$&
  \protect$v^0$&
  \protect$v$
  \\
  \hline
  multipole \rule[-12pt]{0pt}{30pt}&
  none&
  \protect${A^{0}_{{\rm p}}}(\xv)$&
  \protect${A^{0}_{{\rm u}}}(\xv)$&
  none&
  \protect$\Qp(t)$&
  \protect${A^{0}_{{\rm u}}}(\xv,t)$ 
  \\
  \hline
\end{tabular}
\end{center}
\end{table}

Note that -- although both describing interactions with physical gluons -- soft
and ultrasoft couplings occur at different orders in $v$. On the level of the
vertex rules, an overlap of different r\'egimes resulting in double counting is
prevented by the fact that in addition to most of the propagators, all vertices
are distinct because of different multipole expansion rules.

This approach may be compared to power counting in loop diagrams as proposed in
Beneke and Smirnov's threshold expansion~\cite{BenekeSmirnov} where the
strength of all scalar gluon interactions is $v^0$ because the scalar vertex
$Q^\dagger A_0 Q$ of table \ref{scalarvertex} does not contain derivatives.
Here, one can easily see that the Coulomb interaction is the only relevant
coupling as $v\to 0$ and that it scales like $v^{-1/2}$. This follows
immediately from the rescaling rules proposed. Also, the suppression of
bremsstrahlung processes relative to the Coulomb interaction is evident from
table \ref{scalarvertex}. In threshold expansion, these features are
established by considering scattering processes with an arbitrary number of
loops.

Velocity power counting for other vertices is again obtained by rescaling and
multipole expansion. For later reference, the counting rules for the coupling
of one and two gluons, either minimally or via the Fermi term, and the rules
for the three gluon interaction are displayed in tables \ref{vectorvertex},
\ref{threegluonvertex} and \ref{twogluetwoquarkvertex}. In the second,
couplings between on shell gluons in the same r\'egime are all of order
$v^0=c=1$ as expected for relativistic particles.

In the minimal coupling term $-\frac{\ii g}{M}\;Q^\dagger\dev\cdot\Av Q$, the
derivative acts on both the gluon and the quark field. Because the quark is
either soft or potential, its derivative from $Q^\dagger\Av\cdot\dev Q$ is
rescaled as $\dev\to(Mv)\,\dev_\mathrm{s}$. The same holds when the derivative
acts on a soft or potential gluon. But both for the one gluon part of the Fermi
interaction $Q^\dagger\vec{\sigma}\cdot\vec{B}Q$ and for the term
$Q^\dagger(\dev\cdot\Av)Q$ of the minimal coupling, the derivative acts on an
ultrasoft gluon field and must hence be rescaled as
$\dev\to(Mv^2)\,\dev_\mathrm{u}$. That this part of the vertex coupling is one
power in $v$ weaker than the part where the derivative acts on the quark, is
recorded in parentheses in table \ref{vectorvertex}. The $v$ rules for the
coupling of one gluon to the quark via the Fermi term are identical to those
sub-dominant contributions to the minimal coupling vertex for ultrasoft gluons,
and to the dominant contributions in the other cases. A similar provision is
made in table \ref{threegluonvertex}, but is not necessary in table
\ref{twogluetwoquarkvertex}.
\begin{table}[!htb]
  \caption{\label{vectorvertex} \sl 
    Velocity power counting and vertices for the interaction
    \protect$-\frac{\ii g}{M}\;Q^\dagger\dev\cdot\Av Q$. When the spatial
    derivative acts on an ultrasoft gluon field, the power counting in
    parentheses is the sub-dominant contribution as explained in the text.
    It coincides with the power counting for the one gluon component of the
    Fermi vertex.}
\begin{center}
  \footnotesize
\begin{tabular}{|r|c|c|c|c|c|c|}
  \hline
  Vertex\rule[-20pt]{0pt}{48pt}&
        \feynbox{50\unitlength}{
        \begin{fmfgraph*}(50,25)
          \fmfstraight
          \fmftop{i,o}
          \fmfbottom{u}
          \fmf{vanilla,width=thick}{i,v}
          \fmf{vanilla,width=thick}{o,v}
          \fmffreeze \fmf{dashes,width=thin,label=$\fs\Av$}{u,v}
        \end{fmfgraph*}}&
      \feynbox{50\unitlength}{
      \begin{fmfgraph*}(50,25)
        \fmfstraight
        \fmftop{i,o}
        \fmfbottom{u}
        \fmf{vanilla,width=thick}{i,v}
        \fmf{vanilla,width=thick}{o,v}
        \fmffreeze
        \fmf{photon,foreground=blue,width=thin}{u,v}
      \end{fmfgraph*}}&
      \feynbox{50\unitlength}{
      \begin{fmfgraph*}(50,25)
        \fmfstraight
        \fmftop{i,o}
        \fmfbottom{u}
        \fmf{double,foreground=red,width=thin}{i,v}
        \fmf{double,foreground=red,width=thin}{o,v}
        \fmffreeze
        \fmf{zigzag,foreground=red,width=thin}{u,v}
      \end{fmfgraph*}}&
      \feynbox{50\unitlength}{
      \begin{fmfgraph*}(50,25)
        \fmfstraight
        \fmftop{i,o}
        \fmfbottom{u}
        \fmf{double,foreground=red,width=thin}{i,v}
        \fmf{vanilla,width=thick}{o,v}
        \fmffreeze
        \fmf{zigzag,foreground=red,width=thin}{u,v}
      \end{fmfgraph*}}&
      \feynbox{50\unitlength}{
      \begin{fmfgraph*}(50,25)
        \fmfstraight
        \fmftop{i,o}
        \fmfbottom{u}
        \fmf{double,foreground=red,width=thin}{i,v}
        \fmf{double,foreground=red,width=thin}{o,v}
        \fmffreeze
        \fmf{dashes,width=thin}{u,v}
      \end{fmfgraph*}}&
      \feynbox{50\unitlength}{
      \begin{fmfgraph*}(50,25)
        \fmfstraight
        \fmftop{i,o}
        \fmfbottom{u}
        \fmf{double,foreground=red,width=thin}{i,v}
        \fmf{double,foreground=red,width=thin}{o,v}
        \fmffreeze
        \fmf{photon,foreground=blue,width=thin}{u,v}
      \end{fmfgraph*}}
  \\[2ex]
  \hline
  \protect$\pc$\rule[-12pt]{0pt}{30pt}&
  \protect$v^\frac{1}{2}$&
  \protect$v$ ($v^2$)&
  \protect$v$&
  \protect$v$&
  \protect$v^{\frac{3}{2}}$&
  \protect$v^2$ ($v^3$)
  \\
  \hline
\end{tabular}
\end{center}
\end{table}
\begin{table}[!htb]
  \caption{\label{threegluonvertex} \sl Velocity power counting and vertices
    for the gluonic interaction Lagrangean \protect$\frac{g}{2} f^{abc}
    (\de_\mu A_\nu^a-\de_\nu A_\mu^a)\,A^{b,\,\mu}\,A^{c,\,\nu}$. Sub-dominant
    contributions in parentheses.} 
\begin{center}
  \footnotesize
\begin{tabular}{|r|c|c|c|c|c|c|}
  \hline
  Vertex\rule[-20pt]{0pt}{48pt}&
      \feynbox{40\unitlength}{
        \begin{fmfgraph*}(40,30)
          \fmfstraight
          \fmftop{i,o}
          \fmfbottom{u}
          \fmf{dashes}{i,v}
          \fmf{dashes}{o,v} 
          \fmf{dashes}{u,v}
        \end{fmfgraph*}}&
      \feynbox{40\unitlength}{
        \begin{fmfgraph*}(40,30)
          \fmfstraight
          \fmftop{i,o}
          \fmfbottom{u}
          \fmf{dashes}{i,v}
          \fmf{photon,foreground=blue}{o,v} 
          \fmf{dashes}{u,v}
        \end{fmfgraph*}}&
      \feynbox{40\unitlength}{
        \begin{fmfgraph*}(40,30)
          \fmfstraight
          \fmftop{i,o}
          \fmfbottom{u}
          \fmf{zigzag,foreground=red}{i,v}
          \fmf{zigzag,foreground=red}{o,v} 
          \fmf{zigzag,foreground=red}{u,v}
        \end{fmfgraph*}}
      &
      \feynbox{40\unitlength}{
        \begin{fmfgraph*}(40,30)
          \fmfstraight
          \fmftop{i,o}
          \fmfbottom{u}
          \fmf{zigzag,foreground=red}{i,v}
          \fmf{zigzag,foreground=red}{o,v} 
          \fmf{dashes}{u,v}
        \end{fmfgraph*}}
      &
      \feynbox{40\unitlength}{
        \begin{fmfgraph*}(40,30)
          \fmfstraight
          \fmftop{i,o}
          \fmfbottom{u}
          \fmf{zigzag,foreground=red}{i,v}
          \fmf{zigzag,foreground=red}{o,v} 
          \fmf{photon,foreground=blue}{u,v}
        \end{fmfgraph*}}
      &
      \feynbox{40\unitlength}{
        \begin{fmfgraph*}(40,30)
          \fmfstraight
          \fmftop{i,o}
          \fmfbottom{u}
          \fmf{photon,foreground=blue}{i,v}
          \fmf{photon,foreground=blue}{o,v} 
          \fmf{photon,foreground=blue}{u,v}
        \end{fmfgraph*}}
  \\[2ex]
  \hline
  \protect$\pc$\rule[-12pt]{0pt}{30pt}&
  \protect$v^\frac{1}{2}$ ($v^\frac{3}{2}$)&
  \protect$v$ ($v^2$)&
  \protect$v^0$&
  \protect$v^\frac{1}{2}$ ($v^\frac{3}{2}$)&
  \protect$v$ ($v^2$)&
  \protect$v^0$
  \\
  \hline
\end{tabular}
\end{center}
\end{table}
\begin{table}[!htb]
  \caption{\label{twogluetwoquarkvertex} \sl Velocity power counting and
    vertices for the coupling of two gluons to a quark from the minimal
    coupling or the Fermi term in the NRQCD Lagrangean (\ref{nrqcdlagr}).} 
\begin{center}
  \footnotesize
\begin{tabular}{|r|c|c|c|c|c|c|}
  \hline
  Vertex\rule[-20pt]{0pt}{48pt}
  &
      \feynbox{50\unitlength}{
        \begin{fmfgraph*}(50,25)
          \fmfstraight
          \fmftop{i,o}
          \fmfbottom{u1,u2}
          \fmf{vanilla,width=thick}{i,v,o}
          \fmffreeze
          \fmf{dashes,width=thin,label=$\fs\Av$,label.side=left}{v,u1} 
          \fmf{dashes,width=thin,label=$\fs\Av$,label.side=right}{v,u2}
        \end{fmfgraph*}}
      &
      \feynbox{50\unitlength}{
        \begin{fmfgraph*}(50,25)
          \fmfstraight
          \fmftop{i,o}
          \fmfbottom{u1,u2}
          \fmf{vanilla,width=thick}{i,v,o}
          \fmffreeze
          \fmf{dashes,width=thin}{v,u1} 
          \fmf{photon,foreground=blue,width=thin}{v,u2}
        \end{fmfgraph*}}
      &
      \feynbox{50\unitlength}{
        \begin{fmfgraph*}(50,25)
          \fmfstraight
          \fmftop{i,o}
          \fmfbottom{u1,u2}
          \fmf{vanilla,width=thick}{i,v,o}
          \fmffreeze
          \fmf{photon,foreground=blue,width=thin}{v,u1}
          \fmf{photon,foreground=blue,width=thin}{v,u2}
        \end{fmfgraph*}}
      &
      \feynbox{50\unitlength}{
        \begin{fmfgraph*}(50,25)
          \fmfstraight
          \fmftop{i,o}
          \fmfbottom{u1,u2}
          \fmf{double,foreground=red,width=thin}{i,v,o}
          \fmffreeze
          \fmf{zigzag,foreground=red,width=thin}{v,u1} 
          \fmf{zigzag,foreground=red,width=thin}{v,u2}
        \end{fmfgraph*}}
      &
      \feynbox{50\unitlength}{
        \begin{fmfgraph*}(50,25)
          \fmfstraight
          \fmftop{i,o}
          \fmfbottom{u1,u2}
          \fmf{double,foreground=red,width=thin}{i,v}
          \fmf{vanilla,width=thick}{v,o}
          \fmffreeze
          \fmf{zigzag,foreground=red,width=thin}{v,u1} 
          \fmf{zigzag,foreground=red,width=thin}{v,u2}
        \end{fmfgraph*}}
      &
      \feynbox{50\unitlength}{
        \begin{fmfgraph*}(50,25)
          \fmfstraight
          \fmftop{i,o}
          \fmfbottom{u1,u2}
          \fmf{double,foreground=red,width=thin}{i,v,o}
          \fmffreeze
          \fmf{zigzag,foreground=red,width=thin}{v,u1} 
          \fmf{dashes,width=thin}{v,u2}
        \end{fmfgraph*}}
      \\[2ex]
  \hline
  \protect$\pc$\rule[-12pt]{0pt}{30pt}&
  \protect$v$&
  \protect$v^\frac{3}{2}$&
  \protect$v^2$&
  \protect$v$&
  \protect$v$&
  \protect$v^\frac{3}{2}$
  \\
  \hline
\end{tabular}
\\[8ex]
\begin{tabular}{|c|c|c|c|c|c|c|}
  \hline
  \rule[-20pt]{0pt}{48pt}
      \feynbox{50\unitlength}{
        \begin{fmfgraph*}(50,25)
          \fmfstraight
          \fmftop{i,o}
          \fmfbottom{u1,u2}
          \fmf{double,foreground=red,width=thin}{i,v}
          \fmf{vanilla,width=thick}{v,o}
          \fmffreeze
          \fmf{zigzag,foreground=red,width=thin}{v,u1}
          \fmf{dashes,width=thin}{v,u2} 
        \end{fmfgraph*}}
      &
      \feynbox{50\unitlength}{
        \begin{fmfgraph*}(50,25)
          \fmfstraight
          \fmftop{i,o}
          \fmfbottom{u1,u2}
          \fmf{double,foreground=red,width=thin}{i,v,o}
          \fmffreeze
          \fmf{dashes,width=thin}{v,u1} 
          \fmf{dashes,width=thin}{v,u2}
        \end{fmfgraph*}}
      &
      \feynbox{50\unitlength}{
        \begin{fmfgraph*}(50,25)
          \fmfstraight
          \fmftop{i,o}
          \fmfbottom{u1,u2}
          \fmf{vanilla,width=thick}{i,v,o}
          \fmffreeze
          \fmf{zigzag,foreground=red,width=thin}{v,u1} 
          \fmf{zigzag,foreground=red,width=thin}{v,u2}
        \end{fmfgraph*}}
      &
      \feynbox{50\unitlength}{
        \begin{fmfgraph*}(50,25)
          \fmfstraight
          \fmftop{i,o}
          \fmfbottom{u1,u2}
          \fmf{double,foreground=red,width=thin}{i,v,o}
          \fmffreeze
          \fmf{zigzag,foreground=red,width=thin}{v,u1} 
          \fmf{photon,foreground=blue,width=thin}{v,u2}
        \end{fmfgraph*}}
      &
      \feynbox{50\unitlength}{
        \begin{fmfgraph*}(50,25)
          \fmfstraight
          \fmftop{i,o}
          \fmfbottom{u1,u2}
          \fmf{double,foreground=red,width=thin}{i,v,o}
          \fmffreeze
          \fmf{photon,foreground=blue,width=thin}{v,u1}
          \fmf{photon,foreground=blue,width=thin}{v,u2}
        \end{fmfgraph*}}
      &
      \feynbox{50\unitlength}{
        \begin{fmfgraph*}(50,25)
          \fmfstraight
          \fmftop{i,o}
          \fmfbottom{u1,u2}
          \fmf{double,foreground=red,width=thin}{i,v}
          \fmf{vanilla,width=thick}{v,o}
          \fmffreeze
          \fmf{zigzag,foreground=red,width=thin}{v,u1} 
          \fmf{photon,foreground=blue,width=thin}{v,u2}
        \end{fmfgraph*}}
      &
      \feynbox{50\unitlength}{
        \begin{fmfgraph*}(50,25)
          \fmfstraight
          \fmftop{i,o}
          \fmfbottom{u1,u2}
          \fmf{double,foreground=red,width=thin}{i,v,o}
          \fmffreeze
          \fmf{dashes,width=thin}{v,u1} 
          \fmf{photon,foreground=blue,width=thin}{v,u2}
        \end{fmfgraph*}}
      \\[2ex]
  \hline
  \rule[-12pt]{0pt}{30pt}
  \protect$v^\frac{3}{2}$&
  \protect$v^2$&
  \protect$v$&
  \protect$v^2$&
  \protect$v^3$&
  \protect$v^2$&
  \protect$v^\frac{5}{2}$
  \\
  \hline
\end{tabular}
\end{center}
\end{table}

\subsection{Loop Rules}
\label{sec:loop}

As hinted upon above, the velocity power counting is not yet complete. The
propagators were constructed such that they count as order $v^0$ in each
r\'egime in a Feynman diagram. As one sees from the volume element used in
(\ref{ssplagr}), the vertex rules for the soft r\'egime count powers of $v$
with respect to the soft r\'egime. There, one hence retrieves the velocity
power counting of Heavy Quark Effective Theory~\cite{IsgurWise1,IsgurWise2}
(HQET), in which the interactions between one heavy (and hence static) and one
or several light quarks are described. HQET counts inverse powers of mass in
the Lagrangean, but because in the soft r\'egime $Mv\sim \mbox{const.}$, the
two approaches are actually equivalent. Now, HQET becomes a sub-set of NRQCD,
complemented by interactions between soft (HQET) and potential or ultrasoft
particles. This has already been used by Manohar~\cite{Manohar} to facilitate
and unify the matching of the NRQCD/HQET Lagrangean to QCD.

In NRQCD with on-shell quarks as initial and final states, the soft r\'egime
can occur only inside loops. Let us define as a ``soft blob'' each connected
graph containing (soft) loops which is obtained when all potential and
ultrasoft lines are cut. If the diagram contains even one soft particle, scale
conservation ensures that there is at least one loop which consists of only
soft particles, be they quarks or gluons, and that it is part of the soft
blob. Inside the soft blob, the power counting for the vertices is performed
in the soft r\'egime and has therefore to be transfered to the potential
r\'egime. Since soft loop momenta scale like $[\dd^{4}\!  k_\mathrm{s}]\sim
v^4$ while potential ones like $[\dd^{4}\!  k_\mathrm{p}]\sim v^5$, each soft
blob is enhanced by an additional factor $\frac{1}{v}$.

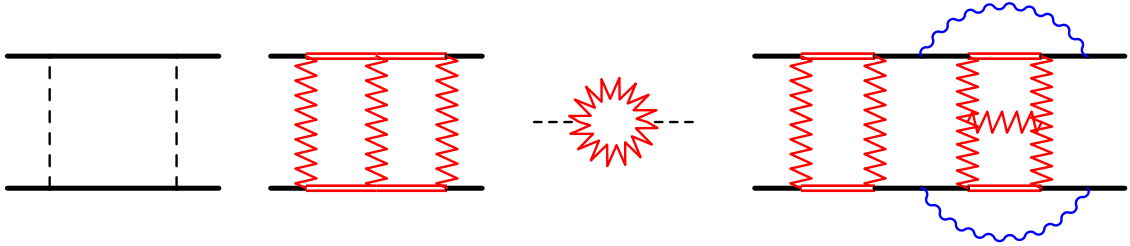
\begin{figure}[!htb]
\begin{center}
  \rule{0pt}{50pt}
  \feynbox{80\unitlength}{
  \begin{fmfgraph*}(80,50)
    \fmfstraight
    \fmftop{i1,o1}
    \fmfbottom{i2,o2}
    \fmf{vanilla,width=thick,tension=1.5}{i1,v1}
    \fmf{vanilla,width=thick,tension=0.5}{v1,v2}
    \fmf{vanilla,width=thick,tension=1.5}{v2,o1}
    \fmf{vanilla,width=thick,tension=1.5}{i2,v3}
    \fmf{vanilla,width=thick,tension=0.5}{v3,v4}
    \fmf{vanilla,width=thick,tension=1.5}{v4,o2}
    \fmffreeze
    \fmf{dashes,width=thin,tension=0.5}{v3,v1}
    \fmf{dashes,width=thin,tension=0.5}{v4,v2}
  \end{fmfgraph*}}
\hspace*{1em}
\feynbox{80\unitlength}{
  \begin{fmfgraph*}(80,50)
    \fmfstraight
    \fmftop{i1,o1}
    \fmfbottom{i2,o2}
    \fmf{vanilla,width=thick,tension=1}{i1,v1}
    \fmf{double,foreground=red,width=thin,tension=0.5}{v1,v5,v2}
    \fmf{vanilla,width=thick,tension=1}{v2,o1}
    \fmf{vanilla,width=thick,tension=1}{i2,v3}
    \fmf{double,foreground=red,width=thin,tension=0.5}{v3,v6,v4}
    \fmf{vanilla,width=thick,tension=1}{v4,o2}
    \fmffreeze
    \fmf{zigzag,foreground=red,width=thin,tension=0.5}{v3,v1}
    \fmf{zigzag,foreground=red,width=thin,tension=0.5}{v4,v2}
    \fmf{zigzag,foreground=red,width=thin,tension=0.5}{v5,v6}
  \end{fmfgraph*}}
\hspace*{1em}
\feynbox{60\unitlength}{
      \begin{fmfgraph*}(60,40)
        \fmfleft{o}
        \fmfright{u}
        \fmf{dashes,tension=3}{o,v1}
        \fmf{zigzag,foreground=red,left}{v1,v2,v1}
        \fmf{dashes,tension=3}{u,v2}
      \end{fmfgraph*}}
   \hspace*{1em}
   \rule[-30pt]{0pt}{20pt}
\feynbox{140\unitlength}{
  \begin{fmfgraph*}(140,50)
    \fmfstraight
    \fmftop{i1,o1}
    \fmfbottom{i2,o2}
    \fmf{vanilla,width=thick,tension=0.8}{i1,v1}
    \fmf{double,foreground=red,width=thin,tension=0.5}{v1,v5}
    \fmf{double,foreground=red,width=thin,tension=0.5}{v7,v2}
    \fmf{vanilla,width=thick,tension=0.8}{v2,v13}
    \fmf{vanilla,width=thick,tension=1}{v13,o1}
    \fmf{vanilla,width=thick,tension=0.8}{i2,v3}
    \fmf{double,foreground=red,width=thin,tension=0.5}{v3,v6}
    \fmf{double,foreground=red,width=thin,tension=0.5}{v8,v4}
    \fmf{vanilla,width=thick,tension=0.8}{v4,v14}
    \fmf{vanilla,width=thick,tension=1}{v14,o2}
    \fmf{vanilla,width=thick,tension=0.8}{v5,v11,v7}
    \fmf{vanilla,width=thick,tension=0.8}{v6,v12,v8}
    \fmffreeze
    \fmf{zigzag,foreground=red,width=thin,tension=0.5}{v3,v1}
    \fmf{zigzag,foreground=red,width=thin,tension=0.5}{v4,v9,v2}
    \fmf{zigzag,foreground=red,width=thin,tension=0.5}{v5,v6}
    \fmf{zigzag,foreground=red,width=thin,tension=0.5}{v7,v10,v8}
    \fmffreeze
    \fmf{zigzag,foreground=red,width=thin,tension=0.5}{v9,v10}
    \fmf{photon,foreground=blue,left=0.6}{v11,v13}
    \fmf{photon,foreground=blue,right=0.6}{v12,v14}
  \end{fmfgraph*}}
\end{center}
\caption{\label{loopfigs} \sl Power counting with soft loops. The loops in the
  second and third diagram obtain an inverse power of \protect$v$, the last
  diagram of \protect$v^2$ in addition to the power counting following from
  the vertex rules.}
\end{figure}

Consider for example the graphs of Fig.\ \ref{loopfigs}: Using the Lorentz
gauges, vertex power counting gives that the leading contribution is from the
exchange of two potential gluons, coupled via $Q^\dagger A_0 Q$. There are four
such vertices, so the diagram is $\calO(g^4)\,\pc(v^{-2})$ from table
\ref{scalarvertex}.
The next two diagrams are $\calO(g^6)\,\pc(v^0)$ and $\calO(g^2)\,\pc(v^1)$
from the vertex power counting in the soft r\'egime, but another factor
$\frac{1}{v}$ must be included because there is one soft blob in each diagram.
This way, the $v$ counting of the soft r\'egime is moved to the potential one.
The intermediate couplings in the second diagram take place in the soft
r\'egime and hence are counted in that r\'egime. After cutting all potential
and ultrasoft lines in the last diagram, two soft blobs are separated by the
propagation of two potential quarks. The graph is $\calO(g^{14})\,\pc(v^0)$
from the vertices, and the loop counting gives a factor $\frac{1}{v^2}$. Each
soft blob contributes at least four powers of $g$, but only one inverse power
of $v\sim g^2$. Power counting is preserved. These velocity power counting
rules in loops are also verified in explicit calculations of exemplary graphs,
\cite{hgpub3} and App.~\ref{sec:vertexcalc}.

There is no similar rule for ultrasoft loops: In the absence of ultrasoft
quarks (see Sect.\ \ref{sec:theorems}), the internal ultrasoft gluon couples
ultimately to a particle in the potential or soft r\'egime. Those vertices are
automatically counted in this stronger r\'egime, while couplings between
ultrasoft particles are counted in the weakest r\'egime. No ``ultrasoft blobs''
can therefore be isolated by cutting all potential and soft lines. It is hence
ultimately scale conservation which forbids a non-trivial loop counting rule
for the ultrasoft r\'egime.

Irrespective of the gauge chosen, there is only one relevant quark-gluon
coupling at tree level in the renormalisation group approach (i.e.\ only one
which dominates at zero velocity): As expected, it is the $\Qp^\dagger
A_{\mathrm{p},0}\Qp$ coupling providing the binding (table
\ref{scalarvertex}). The potential gluon ladders must therefore be re-summed
to all orders to yield the $\frac{1}{r}$ Coulomb potential. Diagrams higher
order in $v$ are corrections. In the Coulomb gauge, all other couplings and
insertions are irrelevant, while in general, there are three marginal
couplings: $\Qp^\dagger A_{\mathrm{u},0}\Qp,\Qs^\dagger A_{\mathrm{s},0}\Qs$
and $\Qs^\dagger A_{\mathrm{s},0}\Qp$. Because of the additional factor
$v^{-1}$ per soft blob, graphs containing the latter two couplings can indeed
be relevant and contribute as $v\to0$ (e.g.\ the second graph in Fig.\ 
\ref{loopfigs}). Eventually, retardation effects from $\Apmu$ will become
weaker than contributions from the soft r\'egime.

\absatz One finally turns to the inclusion of other relativistic particles. In
the same way as NRQCD replaces the physical gluon with one representative per
r\'egime, any light (relativistic) particle has to be tripled. There are
therefore three ghost fields $\eta_\mathrm{s},\eta_\mathrm{p},\eta_\mathrm{u}$
with the same rescaling rules as the gluon fields (\ref{gluonscaling}).
Fermions require more thought: In the real world, the kinetic energy of the
$\mathrm{b}$ quark in Bottomium is compatible to the strange quark mass,
$M_\mathrm{b}v^2\sim m_\mathrm{s}$. For the sake of simplicity, this article
assumes all light particles to have masses very much smaller than any other
scale, $m_\mathrm{q}\ll Mv^2$, so that the relativistic particles can be
treated as massless to lowest order and the denominators of the light particle
propagators are identical to the ones of the gluon in the respective
r\'egimes. The Dirac spinors representing light quarks scale in the three
r\'egimes as $\psi_\mathrm{s}\sim (Mv)^{\frac{3}{2}}\sim
\psi_\mathrm{p},\psi_\mathrm{u}\sim (Mv^2)^{\frac{3}{2}}$, i.e.\ similar to
the heavy quark but including its ultrasoft counterpart (see also Sect.\ 
\ref{sec:theorems}). The number of vertices per term in the rescaled
interaction Lagrangean increases as shown above. Quark-ghost couplings and
heavy-light quark couplings will appear in the Lagrangean (\ref{nrqcdlagr}) at
$\calO(g^4)$ to restore unitarity and can be obtained by matching NRQCD to
QCD.

\absatz With rescaling, multipole expansion and loop counting, the velocity
power counting rules are established. To witness, the rescaling rules
(\ref{xtscaling}/\ref{gluonscaling}/\ref{quarkscaling}) provide an efficient
and well-defined way to arrive at an NRQCD power counting: After rescaling, one
reads up the order in $v$ at which any term in the Lagrangean contributes in
each of the three kinematic r\'egimes and performs the appropriate multipole
expansions. This establishes the Feynman rules for NRQCD and HQET
simultaneously, and classifies the strength of the vertices in the r\'egime of
the particle with highest energy and momentum. Finally, the rescaling is
inverted, introducing one un-rescaled gluon and quark field as the
infrared-relevant degrees of freedom for each kinematic r\'egime. To obtain the
power counting for a certain graph, loop counting is taken into account to
transfer the strength of a soft blob into the potential r\'egime. Computations
are performed most naturally in the original, dimensionful variables. The
rescaling rules are only needed to establish the power counting, but in order
to maintain it, one is of course not allowed to re-sum the multipole expansions
in the un-rescaled variables. It is also interesting to note that there is no
choice but to assign one and the same coupling strength $g$ to each
interaction. Different couplings for one vertex in different r\'egimes are not
allowed. This is to be expected, as the fields in the various r\'egimes are
representatives of one and the same non-relativistic particle, whose
interactions are fixed by the non-relativistic Lagrangean (\ref{nrqcdlagr}).

\subsection{Gauge Invariance}
\label{sec:gaugeinvariance}

The Coulomb gauge $\dev\cdot\Av=0$ is a natural choice in NRQED because in it,
static charges do not radiate. In NRQCD, gluons carry colour and hence the
Coulomb gauge does not have this advantage. Luke and Savage~\cite{LukeSavage}
showed how to establish explicit velocity power counting in the Lorentz gauges
of NRQCD, too. The set of gauges applicable to NRQCD can be extended further:
The classification of the three kinematic r\'egimes (\ref{regimes}) itself
relied only on the typical excitation energy and momentum, and hence on gauge
invariant quantities, and on the form of the denominator in the propagators
(\ref{nonrelprop}) which is unchanged in any order of perturbation
theory\footnote{Non-perturbatively, the propagators are not of the form
  (\ref{nonrelprop}) because of confinement and the absence of coloured
  states, so that the power counting presented here may break down.}. The
perturbative quark propagator is gauge independent. Gauge fixing will
introduce gauge dependent denominators multiplying the gauge independent
denominators in the perturbative gluon propagators. As an example that this
will in general not change the identification of the soft and ultrasoft
r\'egimes from poles in the gluon propagators, consider the generalised axial
gauge ($n^2=-1$, $\alpha$ arbitrary), in which
\begin{equation}
  \label{axgaugeprop}
  \Amu\;:\;\frac{-\ii}{k^2+\ii\epsilon}\left[g_{\mu\nu}-
    \frac{k_\mu n_\nu+n_\mu k_\nu}{k\cdot n}+\frac{1+\alpha k^2}{(n\cdot k)^2}
    \right]\;\;.
\end{equation}
The additional denominators $n\cdot k$ introduce no new combinations of the
two infrared scales $Mv$ and $Mv^2$ for which the gluon propagator has a pole.
Therefore, the decomposition into the three r\'egimes (\ref{regimes}) remains
unchanged, as do the rescaling properties of the fields and interactions,
(\ref{xtscaling}/\ref{gluonscaling}/\ref{quarkscaling}) and tables
\ref{scalarvertex} to \ref{twogluetwoquarkvertex}.

The standard gauges (axial, Weyl, Lorentz, Coulomb) will therefore all show
the same power counting and vertex rules quoted above. Details of the gluon
propagator and its insertions look different in different r\'egimes and
gauges, and some gauges will not exhibit certain vertices, insertions and
representatives, e.g.\ the Coulomb gauge is unique in having $A_0$ contribute
only in the potential r\'egime. Only when the gauge dependent denominator
introduces a new r\'egime has the power counting to be modified. This is for
example the case in a gauge with denominator $k_0^4-M^2\kv^2$, in which an
exceptional r\'egime ($k_0\sim Mv\;,\;|\kv|\sim Mv^2$) enters. Rescaling is
then performed including the new r\'egime.

As is well known, the Weyl gauge $A_0=0$ wants a constraint quantisation:
Gau\3' law $\vec{D}\cdot \vec{E}=g Q^\dagger Q$, generating local gauge
transformations, is the equation of motion derived by varying the Lagrangean
with respect to $A_0$. When $A_0=0$, Gau\3' law will not be recovered as an
equation of motion. In order to restore local gauge invariance, a projector
onto states obeying Gau\3' law has therefore to be inserted into the path
integral. Resolving Gau\3' law explicitly, the Lagrangean of the Coulomb
interaction has the structure
\begin{equation}
  \label{weylcoulomb}
  \calL_\mathrm{Coulomb}= \int \dedreiy g^2\;Q^\dagger Q(\xv)\;\calG(\xv-\yv)
          \;Q^\dagger Q(\yv)\;\;,
\end{equation}
where $\calG(\xv-\yv)$ is an appropriate instantaneous Green's function to
Gau\3' law with dimension $[\mbox{mass}]^1$, e.g.\ in QED
$\frac{1}{|\xv-\yv|}$. Using the rescaling rules
(\ref{xtscaling}/\ref{gluonscaling}/\ref{quarkscaling}), one derives the
Coulomb interaction between two quarks to be $\pc(v^{-1})$ as in the other
gauges. Without constraint quantisation, the longitudinal component of the
vector gauge field mediating the Coulomb interaction in Weyl gauge QED,
$\Av^\mathit{l}$, has a static propagator $\frac{\ii}{k_0^2}$.

\section{The NRQCD $\beta$ Function}
\label{sec:betafunction}
\setcounter{equation}{0}

This chapter presents the computation of the perturbative part of the NRQCD
$\beta$ function to order $g^3,\; v^0$ in the MS scheme using the vertex and
rescaling rules derived in the previous section. Initially, the Lorentz gauges
are not only chosen because they allow a less cumbersome renormalisation than
e.g.\ the Coulomb gauge, but also to demonstrate that in the final result, the
gauge parameter $\alpha$ drops out. The Slavnov--Taylor identities will be
shown to be fulfilled, and the Lorentz gauges are a legitimate gauge choice in
NRQCD. As a by-product, Sect.\ \ref{sec:coulombbeta} will calculate the NRQCD
$\beta$ function to lowest order in the Coulomb gauge.

By construction, NRQCD and QCD must agree in the infrared limit, and
especially in the structure of collinear (infrared) divergences. Matching
proved that both the soft quark and the soft gluon are indispensable to
reproduce the correct structure of collinear divergences in a toy model given
by Beneke and Smirnov~\cite{BenekeSmirnov} and confirmed the proposed counting
rules~\cite{hgpub3}. The calculation of the NRQCD $\beta$ function will
manifest the relevance of the soft r\'egime beyond infrared divergences, and
it will endorse the power counting further. Deriving the lowest order $\beta$
function in the following, the relation between this power counting and
threshold expansion ~\cite{BenekeSmirnov} will be exemplified, and rules will
be developed which allow one to determine from the structure of a diagram
whether it is zero or not to all orders in $v$.

\absatz NRQCD with $N_\mathrm{F}$ light quarks must reproduce the $\beta$
function of QCD below the scale $M$,
\begin{equation}
  \label{qcdbetafunction}
  \beta_\mathrm{QCD}=-\;\frac{g_\mathrm{R}^3}{(4\pi)^2}\;
     \left[\frac{11}{3}\;N-\frac{2}{3}\;N_\mathrm{F}\right]
\end{equation}
to lowest order for the gauge group SU($N$) and renormalised coupling
$g_\mathrm{R}$. This means especially that the renormalised coupling strengths
of all interactions steming from expanding the same term in the Lagrangean in
the various r\'egimes are the same, except that they have to be taken at
different scales: For the interaction Lagrangean $-g Q^\dagger A_0 Q$, the six
vertices of table \ref{scalarvertex} all couple with the same un-renormalised
strength. Although the number of vertices is increased in the approach
presented here, the number of independent couplings is not. Indeed, one major
result of this chapter will be that renormalisation will not destroy this
symmetry because the renormalisation constants agree in all r\'egimes.

One can obtain the $\beta$ function from the ghost-gluon coupling, and the
ghost and gluon self energy. Because the relativistic sector of NRQCD is
identical to the one of QCD as seen in Sect.\ \ref{sec:vertex}, the
calculation proceeds in that case like in QCD and yields the same result. In
this article, the $\beta$ function is calculated by studying the heavy quark
sector, namely the renormalised coupling $Q^\dagger_\mathrm{R}
A_{0,\mathrm{R}} Q_\mathrm{R}$ between the heavy quark and the scalar gluon.
This term yields immediately the coupling renormalisation. All other vertices
are suppressed by at least one power of $v$.

Because the integrals to be performed are not Lorentz invariant, standard
formulae for dimensional regularisation often do not apply. Therefore,
Ref.~\cite{hgpub3} used split dimensional regularisation, which was introduced
by Leibbrandt and Williams~\cite{LeibbrandtWilliams} to cure the problems
arising from pinch singularities in non-covariant gauges. It treats the
temporal and spatial components of the loop integrations on an equal footing
by regularising the energy and momentum integration separately,
$\int\deintdim{d}{k}=\int\deintdim{\sigma}{k_0} \deintdim{d-\sigma}{\kv}$,
$\sigma\to1,\;d\to 4$~\cite[Chap.\ 4.1]{Collins}. Useful formulae for the
computation of the $\beta$ function are given in App.\ \ref{sec:splitdimreg}.

\subsection{NRQCD and Threshold Expansion: Potential Quark Self-Energy}
\label{sec:nrqcdandthreshold}

Threshold expansion~\cite{BenekeSmirnov} and NRQCD make use of very similar
basic techniques but different formulations, as this Section outlines. In
order to clarify the relation, consider the quark self-energy diagram lowest
order in $g$ which couples the scalar gluon and the quark in NRQCD. To
facilitate the presentation initially, the Feynman gauge $\alpha=1$ is chosen
in which scalar and vector gauge fields do not mix when propagating. Without
the scaling rules for the various r\'egimes, one obtains
\begin{equation}
  \rule{0pt}{34pt}
  \label{nrqcdselfenergy}
  \feynbox{60\unitlength}{
    \begin{fmfgraph*}(60,40)
      \fmfleft{i}
      \fmfright{o}
      \fmf{vanilla,width=thick,tension=5}{i,v1}
      \fmf{vanilla,width=thick,tension=5}{v2,o}
      \fmf{fermion,width=thick}{v1,v2}
      \fmffreeze
      \fmf{gluon,left,label=$\fs A_0,,q$,label.side=left}{v1,v2}
    \end{fmfgraph*}}
  \;:\; \ii\Sigma_\mathrm{p}(T,\pv)=(-\ii g)^2\; C_\mathrm{F}
  \;\int\deintdim{d}{q}\frac{-\ii}{q_0^2-\qv^2+\ii\epsilon}\; 
  \frac{\ii}{T+q_0-\frac{(\pv+\qv)^2}{2M}+\ii\epsilon}\;,
\end{equation}
with $C_\mathrm{F} \delta_{ij}=(t^a t^a)_{ij}=\frac{N^2-1}{2N} \delta_{ij}$
the Casimir operator of the fundamental representation of SU($N$). Threshold
expansion identifies the loop momentum of QCD to belong to a hard r\'egime
$q\sim M$ or to either of the three r\'egimes (\ref{regimes}) and expands the
integrand about the various saddle points, i.e.\ about the values of the
loop-momentum $q$ where poles occur, like the NRQCD classification in
Sect.~\ref{sec:regimes}. The hard r\'egime subsumes the relativistic effects
contained in the coefficients $c_i,d_i,e_i$ of the NRQCD Lagrangean
(\ref{nrqcdlagr}) and hence provides the matching between NRQCD and
QCD~\cite{BenekeSmirnov}. In deriving the NRQCD Lagrangean (\ref{nrqcdlagr}),
hard momenta have been integrated out, so that the ultraviolet behaviour of
the above loop integral is arbitrary and it is not necessary to take the poles
at $q\sim M$ in (\ref{nrqcdselfenergy}) into account. Indeed, to be
consistent, they should rather be discarded.

Let the incident quark be on shell, i.e.\ have a four-momentum
$(T_\mathrm{p},\pv)$ in the potential r\'egime. The integral decomposes then
into approximations about three saddle points: When $q$ is soft, the gluon
propagator has a pole, and one can expand the quark propagator in powers of
$T_\mathrm{p}/q_0\sim v$ and $\frac{(\pv+\qv)^2}{2M}/q_0\sim v$. The quark
becomes static:
\begin{eqnarray}
  \label{bsexprop}
  &&\!\!\!\!\!\!\!\!\!\!\!
   \frac{\ii}{{q_{0}}+{T_\mathrm{p}}-\frac{(\pv+\qv)^2}{2M}}
   \longrightarrow
   \frac{\ii}{{q_{0}}}+\frac{\ii}{{q_{0}}}\;\ii\bigg(
     {T_\mathrm{p}}-\frac{(\pv+\qv)^2}{2M}\bigg)\;
     \frac{\ii}{{q_{0}}}+\dots\\
  \label{thexpselfpottosoft}
  &&\!\!\!\!\!\!\!\!\!\!\!
  \Longrightarrow\;\ii\Sigma_\mathrm{p}(T,\pv)\Big|_\mathrm{soft}=
     (-\ii g)^2\;
     C_\mathrm{F} \;\int\deintdim{d}{q} \frac{-\ii}{q_0^2-\qv^2+\ii\epsilon}\;
     \frac{\ii}{q_0+\ii\epsilon}\;
     \sum\limits^\infty_{n=0}\left(\frac{
         \frac{(\pv+\qv)^2}{2M}-T_\mathrm{p}}{q_0+\ii\epsilon}\right)^n
\end{eqnarray}
The NRQCD power counting proceeds on the level of the Lagrangean instead of
the Feynman diagrams with corresponding diagrams for loop momenta in the soft
r\'egime,
\begin{eqnarray}
  &&
  \rule{0pt}{30pt}
  \feynbox{60\unitlength}{
    \begin{fmfgraph*}(60,40)
      \fmfleft{i}
      \fmfright{o}
      \fmf{vanilla,width=thick,tension=4}{i,v1}
      \fmf{vanilla,width=thick,tension=4}{v2,o}
      \fmf{double,foreground=red,width=thin}{v1,v2}
      \fmffreeze
      \fmf{zigzag,foreground=red,left,label=$\fs q$,
        label.side=left}{v1,v2}
    \end{fmfgraph*}}\;+\;
  \feynbox{60\unitlength}{
    \begin{fmfgraph*}(60,40)
      \fmfleft{i}
      \fmfright{o}
      \fmf{vanilla,width=thick,tension=2}{i,v1}
      \fmf{vanilla,width=thick,tension=2}{v2,o}
      \fmf{double,foreground=red,width=thin}{v1,v3,v2}
      \fmfv{decoration.shape=cross,foreground=red}{v3}
      \fmffreeze
      \fmf{zigzag,foreground=red,left}{v1,v2}
    \end{fmfgraph*}}\;+\;\dots\;:
  \label{nrqcdselfpottosoft}\\
  &&(-\ii g)^2\;
     C_\mathrm{F} \;\int\deintdim{d}{q} \frac{-\ii}{q_0^2-\qv^2+\ii\epsilon}\;
     \frac{\ii}{q_0+\ii\epsilon}\left(1-\ii\;\frac{(\pv+\qv)^2}{2M}\;
       \frac{\ii}{q_0+\ii\epsilon} + \ii T_\mathrm{p}
       \;\frac{\ii}{q_0+\ii\epsilon}+\dots
     \right)\;\;,\non
\end{eqnarray}
and recovers order by order in $v$ the result of threshold expansion. The
intermediate soft quark is static, and the higher order terms in threshold
expansion are interpreted as insertions into the soft quark propagator or as
resulting from the energy multipole expansion at the $\Qs\Asmu\Qp$ vertex. In
fact, using the equations of motion, a temporal multipole expansion may be
re-written such that the energy becomes conserved at the vertex. Now, both
soft and potential or ultrasoft energies are present in the propagators,
making it necessary to expand it in ultrasoft and potential energies. An
example would be to restate the vertex (\ref{sspvertex}) as
\begin{eqnarray} 
    \feynbox{40\unitlength}{
    \begin{fmfgraph*}(40,20)
      \fmfstraight
      \fmftop{i,o}
      \fmfbottom{u}
      \fmf{double,foreground=red,width=thin}{i,v}
      \fmf{vanilla,width=thick}{o,v}
      \fmffreeze
      \fmf{zigzag,foreground=red,width=thin}{u,v}
    \end{fmfgraph*}}
   &:&-\ii g\;(2\pi)^4\;\delta(T-T^\prime_\mathrm{p}+q_0)
   \;\delta^{(3)}(\pv-\pv^\prime+\qv)\;\;,
\end{eqnarray}
and the soft propagator to contain insertions $\pc(v)$ for potential
energies $T^\prime_\mathrm{p}$
\begin{equation}
    \feynbox{60\unitlength}{
    \begin{fmfgraph*}(60,30)
      \fmfleft{i}
      \fmfright{o}
      \fmf{heavy,foreground=red,width=thin,
        label=$\fs(T=q_0+T^\prime_\mathrm{p},,\pv)$,
        label.side=left}{i,o}
      \end{fmfgraph*}}
    \;:\;\frac{\ii}{q_0+\ii\epsilon}
    \sum\limits_{n=0}^\infty\left(\frac{-T^\prime_\mathrm{p}}{q_0}\right)^n
    \;\;.
\end{equation}
The same can be shown for the momentum-non-conserving vertices, too.

When the loop momentum $q$ is potential, threshold expansion picks the pole of
the quark propagator, and the gluon propagator is expanded as if $q_0\ll
|\qv|$, corresponding to NRQCD diagrams with insertions in the gluon
propagator:
\begin{eqnarray}
  \rule{0pt}{32pt}
  &&
  \feynbox{60\unitlength}{
    \begin{fmfgraph*}(60,40)
      \fmfleft{i}
      \fmfright{o}
      \fmf{vanilla,width=thick,tension=5}{i,v1}
      \fmf{vanilla,width=thick,tension=5}{v2,o}
      \fmf{fermion,width=thick}{v1,v2}
      \fmffreeze
      \fmf{dashes,left,label=$\fs q$,label.side=left}{v1,v2}
    \end{fmfgraph*}}\;+\;
  \feynbox{60\unitlength}{
    \begin{fmfgraph*}(60,40)
      \fmfleft{i}
      \fmfright{o}
      \fmf{vanilla,width=thick,tension=5}{i,v1}
      \fmf{vanilla,width=thick,tension=5}{v2,o}
      \fmf{fermion,width=thick}{v1,v2}
      \fmffreeze
      \fmf{phantom,left}{v1,v2}
      \fmfipath{p}
      \fmfiset{p}{vpath(__v1,__v2)}
      \fmfi{dashes}{subpath (0,length(p)/2) of p}
      \fmfi{dashes}{subpath (length(p)/2,length(p)) of p}
      \fmfiv{decor.shape=cross}{point length (p)/2 of p}
    \end{fmfgraph*}}\;+\;\dots\;:
  \label{thexpselfpottopot}\\
  &&\ii\Sigma_\mathrm{p}(T,\pv)\Big|_\mathrm{potential}=(-\ii g)^2\;
     C_\mathrm{F} \;\int\deintdim{d}{q}
     \frac{-\ii}{-\qv^2+\ii\epsilon}\;
     \sum\limits^\infty_{n=0}\left(\frac{q_0^2}{\qv^2}\right)^n\;
     \frac{\ii}{T+q_0-\frac{(\pv+\qv)^2}{2M}+\ii\epsilon}\non
\end{eqnarray}
The correspondence is again complete for ultrasoft $q$, and the higher order
terms of threshold expansion come from the spatial multipole expansion in
NRQCD:
\begin{eqnarray}
  &&
  \rule{0pt}{32pt}
  \feynbox{60\unitlength}{
    \begin{fmfgraph*}(60,40)
      \fmfleft{i}
      \fmfright{o}
      \fmf{vanilla,width=thick,tension=5}{i,v1}
      \fmf{vanilla,width=thick,tension=5}{v2,o}
      \fmf{fermion,width=thick}{v1,v2}
      \fmffreeze
      \fmf{photon,foreground=blue,left,label=$\fs q$,label.side=left}{v1,v2}
    \end{fmfgraph*}}\;:\;\ii\Sigma_\mathrm{p}(T,\pv)\Big|_\mathrm{ultrasoft}=
  \label{thexpselfpottoultra}\\
  &&=(-\ii g)^2\;
     C_\mathrm{F} \;\int\deintdim{d}{q}\frac{-\ii}{q_0^2-\qv^2+\ii\epsilon}\;
     \frac{\ii}{T+q_0-\frac{\pv^2}{2M}+\ii\epsilon}\;
     \sum\limits^\infty_{n=0}\left(\frac{
         \frac{2\pv\cdot\qv+\qv^2}{2M}}{T+q_0-\frac{\pv^2}{2M}+\ii\epsilon}
     \right)^n\;\;.\non
\end{eqnarray}
Threshold expansion and NRQCD concur indeed in all diagrams to be calculated
for the $\beta$ function. Further comparison of the two approaches is
postponed to the Conclusions.

\subsection{Some Diagrammatic Rules}
\label{sec:diagrammar}

In general, the number of possible diagrams in power counted NRQCD is
considerably larger than in the original theory because there are at least six
vertices per interaction allowed by scale conservation, cf.\ tables
\ref{scalarvertex} to \ref{twogluetwoquarkvertex} for interactions between
three and four particles. In the preceding sub-section, three diagrams were
drawn for the lowest order contribution to the potential quark self energy. The
analogue problem in threshold expansion is that each loop momentum can lie in
either of the three r\'egimes, so that the number of expansions for an $n$ loop
diagram grows like $3^n$.

The important step to cut down the computational effort in either approach is
to note that as an immediate consequence of the axioms of dimensional
regularisation (e.g.~\cite[Chap.\ 4.1 and 4.2]{Collins}), integrals without
scales set by external momenta or energies in the denominator vanish because
of homogeneity,
\begin{equation}
  \label{masterformula}
  \int\deintdim{d}{q} q^\alpha=0\;\;.
\end{equation}
For example, applying this theorem to the potential quark self energy
diagrams, contributions from soft (\ref{thexpselfpottosoft}) and potential
(\ref{thexpselfpottopot}) loop momenta are zero to all orders in the velocity
expansion. For the potential gluon contribution (\ref{thexpselfpottopot}),
this is seen by shifting the regularised loop integral
$T+q_0-\frac{(\pv+\qv)^2}{2M}\to q_0$ before using (\ref{masterformula}) for
the energy integral. The importance of (\ref{masterformula}) has already been
alluded to in threshold expansion~\cite{BenekeSmirnov}, and here a more
thorough and formal treatment is presented.

It is this theorem (\ref{masterformula}) which renders most of the diagrams in
NRQCD and threshold expansion calculations zero, and it is usually not even
necessary to consider the whole diagram but only a sub-set of it. The
following set of rules is helpful for reducing the numbers of diagrams to be
dealt with in the calculation of the $\beta$ function.

\absatz In order to establish these rules, recall that all graphs vanish which
contain a sub-graph zero by the rules developed. The routing of the loop
four-momentum inside a diagram is arbitrary as always in dimensional
regularisation so that having regularised the loop integration, all loop
four-momenta can be shifted at will like in ordinary integration. Since they
have identical denominators (Sect.\ \ref{sec:vertex}), gluon lines in any rule
can be replaced by any relativistic particle, e.g.\ ghosts and light quarks.
Finally and most importantly, numerators are unimportant to determine whether
an integral is scale-less following (\ref{masterformula}). As multipole
expansion and insertions do not change the denominators of the propagators, a
diagram is therefore zero to all orders when it is zero because of
(\ref{masterformula}) at leading order in $v$. This result is also insensitive
to the gauge chosen and to the specific vertex involved. For example, because
the numerators play no r\^ole in rendering
(\ref{thexpselfpottosoft}/\ref{thexpselfpottopot}) zero using
(\ref{masterformula}), the same will hold for any graph with the same
diagrammatic representation but different interactions, e.g.\ also for graphs
in which one or both vertices are replaced with the minimal coupled vector
fields $Q^\dagger \dev\cdot\Av Q$ or the Fermi interaction $Q^\dagger
\vec{\sigma}\cdot\vec{B} Q$.

\absatz As a first rule, consider a potential gluon between two soft
particles, represented by the overlay of double and zigzag lines in Fig.\ 
\ref{rule1} (a).
\begin{figure}[!htb]
  \vspace*{2ex}
  \begin{center}
    \feynbox{60\unitlength}{
      \begin{fmfgraph*}(60,30)
        \fmfleft{bl,tl}
        \fmfright{br,tr}
        \fmf{zigzag,foreground=red}{tl,v1,bl}
        \fmf{double,foreground=red}{tl,v1,bl}
        \fmf{zigzag,foreground=red}{tr,v2,br}
        \fmf{double,foreground=red}{tr,v2,br}
        \fmfv{label=${\fs(T,,\pv)\uparrow}$,label.angle=-90}{bl}
        \fmfv{label=${\fs(T,,\pv-\qv)\uparrow}$,label.angle=90}{tl}
        \fmfv{label=${\fs\uparrow(T^\prime,,\pv^\prime)
            }$,label.angle=-90}{br}
        \fmfv{label=${\fs\uparrow(T^\prime,,\pv^\prime+
            \qv)}$,label.angle=90}{tr}
        \fmfv{label=(a),label.angle=-60,label.dist=0.7w}{v1}
        \fmffreeze
        \fmf{dashes,label=${\rightarrow\atop\fs q}$}{v1,v2} 
      \end{fmfgraph*}}
    \hq$=0$
    \hq\hq\hq\hq\hq\hq\hq\hq\hq
    \feynbox{60\unitlength}{
      \begin{fmfgraph*}(60,30)
        \fmftop{t0,t1,t2,t3,t4}
        \fmfbottom{b1,b2}
        \fmf{dashes}{b1,v1,b2}
        \fmfv{label=${\rightarrow\atop\fs q+l}$,label.angle=-90}{b2}
        \fmfv{label=${\rightarrow\atop\fs q}$,label.angle=-90}{b1}
        \fmffreeze
        \fmf{dashes}{v2,v1}
        \fmf{dashes,tension=4}{t1,v2}
        \fmf{photon,foreground=blue}{v1,v3}
        \fmf{photon,foreground=blue,tension=4}{v3,t2}
        \fmf{vanilla,width=thick}{v1,v4}
        \fmf{vanilla,width=thick,tension=5}{v4,t3}
        \fmf{phantom,tension=5}{t1,t2,t3}
        \fmfv{label=$\fs\overbrace{\rule{40pt}{0pt}}^{\fs\downarrow l}$,
          label.angle=90}{t2}
        \fmfv{label=(b),label.angle=-90,label.dist=0.4w}{v1}
        \fmffreeze
        \fmf{dots}{v2,v3,v4}
      \end{fmfgraph*}}
    \hq\hq\hq\hq\hq\hq\hq\hq\hq
    \feynbox{70\unitlength}{
      \begin{fmfgraph*}(70,50)
        \fmftop{t0,t01,t1,t2,t3,t4,t41}
        \fmfleft{bl,tl}
        \fmfright{br,tr}
        \fmf{zigzag,foreground=red}{tl,vm1,bl}
        \fmf{double,foreground=red}{tl,vm1,bl}
        \fmf{zigzag,foreground=red}{tr,vm2,br}
        \fmf{double,foreground=red}{tr,vm2,br}
        \fmffreeze
        \fmf{dashes}{vm1,v1,vm2}
        \fmffreeze
        \fmf{dashes}{v2,v1}
        \fmf{dashes,tension=4}{t1,v2}
        \fmf{photon,foreground=blue}{v1,v3}
        \fmf{photon,foreground=blue,tension=4}{v3,t2}
        \fmf{vanilla,width=thick}{v1,v4}
        \fmf{vanilla,width=thick,tension=4}{v4,t3}
        \fmfv{label=(c),label.angle=-90,label.dist=0.5w}{v1}
        \fmffreeze
        \fmf{dots}{v2,v3,v4}
      \end{fmfgraph*}}
    \hq$=0$\\[8ex]
    \feynbox{200\unitlength}{
      \begin{fmfgraph*}(200,65)
        \fmfstraight
        \fmftopn{t}{26}
        \fmfbottomn{b}{25}
        \fmf{zigzag,foreground=red}{t1,v1,b1}
        \fmf{double,foreground=red}{t1,v1,b1}
        \fmf{zigzag,foreground=red}{t26,v2,b25}
        \fmf{double,foreground=red}{t26,v2,b25}
        \fmfblob{10}{v1}
        \fmfblob{10}{v2}
        \fmffreeze
        \fmf{dashes}{v1,v1a,v3,v3a}
        \fmf{dashes,tension=2}{v3a,v4}
        \fmf{dashes}{v2,v2a,v5}
        \fmf{dashes,tension=2}{v5,v6}
        \fmf{dots,tension=3}{v4a,v6a}
        \fmf{phantom,tension=2.5}{v4,v4a}
        \fmf{phantom,tension=2.5}{v6,v6a}
        \fmfblob{10}{v1a}
        \fmfblob{10}{v2a}
        \fmfblob{10}{v3a}
        \fmffreeze
        \fmf{dashes}{va1,v3}
        \fmf{photon,foreground=blue}{v3,va2}
        \fmf{vanilla,width=thick}{v3,va3}
        \fmf{dashes,tension=4}{t7,va1}
        \fmf{photon,foreground=blue,tension=4}{va2,t8}
        \fmf{vanilla,width=thick,tension=4}{va3,t9}
        \fmffreeze
        \fmf{dots}{va1,va2,va3}
        \fmfblob{10}{v3}
        \fmffreeze
        \fmf{dashes}{vb1,v5}
        \fmf{photon,foreground=blue}{v5,vb2}
        \fmf{vanilla,width=thick}{v5,vb3}
        \fmf{dashes,tension=4}{t18,vb1}
        \fmf{photon,foreground=blue,tension=4}{vb2,t19}
        \fmf{vanilla,width=thick,tension=4}{vb3,t20}
        \fmffreeze
        \fmf{dots}{vb1,vb2,vb3}
        \fmfblob{10}{v5}
        \fmffreeze
        \fmf{dashes}{vc1,v1}
        \fmf{photon,foreground=blue}{v1,vc2}
        \fmf{vanilla,width=thick}{v1,vc3}
        \fmf{dashes,tension=4}{b3,vc1}
        \fmf{photon,foreground=blue,tension=4}{vc2,b4}
        \fmf{vanilla,width=thick,tension=4}{vc3,b5}
        \fmffreeze
        \fmf{dots}{vc1,vc2,vc3}
        \fmf{dashes}{vd1,v2}
        \fmf{photon,foreground=blue}{v2,vd2}
        \fmf{vanilla,width=thick}{v2,vd3}
        \fmf{dashes,tension=4}{b23,vd1}
        \fmf{photon,foreground=blue,tension=4}{vd2,b22}
        \fmf{vanilla,width=thick,tension=4}{vd3,b21}
        \fmffreeze
        \fmf{dots}{vd1,vd2,vd3}
        \fmfv{label=(d),label.angle=-90,label.dist=0.1w}{b13}
      \end{fmfgraph*}}
    \hq$=0$\\[6ex]
  \end{center}
  \caption{\label{rule1}\sl The first rule as proven step by step in the
    text. (a) Simple case; (b) a more complicated sub-diagram; (c,d)
    generalisations. The overlay of double and zigzag line stands for any
    number of arbitrary soft particles, the triple line for any number of
    potential or ultrasoft particles, all entering at the same vertex. The
    blob represents vertex and propagator dressings. No time ordering is
    implied in the way the diagrams are drawn.}
\end{figure}
This sub-diagram is zero when the loop energy $q_0$ is integrated over: Assign
the potential loop momentum $q$ to the instantaneous gluon with a propagator
denominator $-\qv^2$ (\ref{gpprop}). As the coupling of soft particles to
potential ones is energy non-conserving, the denominators of the soft
particles coupling to $\Apmu$ do not contain $q_0\sim Mv^2$, either.
Therefore, $q_0$ does not occur in any denominator, and the dimensionally
regularised loop integral over $q_0$ is zero from (\ref{masterformula}). The
rule is extended by noting (Fig.\ \ref{rule1} (b)) that overall energy
conservation allows for the potential gluon line to be attached with an
arbitrary number of vertices into which an arbitrary number of potential or
ultrasoft gluons or quarks with collective (potential) four-momentum $l$
enters. The denominators of the potential gluon between two soft particles
still do not contain the loop energy, and the $q_0$ integral vanishes again,
Fig.\ \ref{rule1} (c). With this four-momentum routing, $l$ can be an external
or loop momentum.

One can finally dress the vertices and propagators without changing the
argument thanks to scale conservation. Also, when several other potential or
ultrasoft particles couple to the same soft particle in addition to the
potential gluon, a temporal multipole expansion is necessary for each of them.
Therefore, a potential gluon can never connect two soft particles,
irrespective of the number of particles coupling to the same vertices as the
potential gluon, and irrespective of other particles coupling to the potential
gluon in the intermediate state, Fig.\ \ref{rule1} (d): Diagrams with unbroken
potential gluon lines between soft particles vanish.

As predicted, the only vertex property which enters is that the coupling
between particles of different r\'egimes necessitates a multipole expansion
which can be read up from the diagram directly. The proof therefore extends to
massive, relativistic potential particles, like a light quark. It eliminates
for example the diagram corresponding to a non-Abelian vertex correction with
a denominator
\begin{equation}
  \rule{0pt}{36pt}
  \feynbox{60\unitlength}{
      \begin{fmfgraph*}(60,40)
        \fmftop{i,o}
        \fmfbottom{u}
        \fmf{double,foreground=red,tension=4,label=$\fs(T,,\pv)$,
          label.side=left}{i,v1}
        \fmf{double,foreground=red}{v1,v3}
        \fmf{double,foreground=red,tension=4,label=$\fs(T^\prime,,\pv^\prime)$,
          label.side=left}{v3,o}
        \fmffreeze
        \fmf{zigzag,foreground=red}{v1,v2}
        \fmf{dashes,label=$\fs q$}{v2,v3}
        \fmf{zigzag,foreground=red,tension=3}{v2,u}
      \end{fmfgraph*}}
    \;\;:\;\mathrm{Denom.}=T^\prime\;\qv^2\;
    \left((T-T^\prime)^2-(\pv-\pv^\prime+\qv)^2\right)\;\;.
\end{equation}

\absatz A similar rule can be proven for the potential quark between soft
particles, see Fig.\ \ref{rule2} (a) for its bare and (b) for its dressed
form.
\begin{figure}[!htb]
  \vspace*{2ex}
  \begin{center}
     \rule[-45pt]{0pt}{40pt}
    \feynbox{70\unitlength}{
      \begin{fmfgraph*}(70,50)
        \fmfleft{bl,tl}
        \fmfright{br,tr}
        \fmf{zigzag,foreground=red}{tl,vm1,bl}
        \fmf{double,foreground=red}{tl,vm1,bl}
        \fmf{zigzag,foreground=red}{tr,vm2,br}
        \fmf{double,foreground=red}{tr,vm2,br}
        \fmffreeze
        \fmf{vanilla,width=thick}{vm1,v1,vm2}
        \fmfblob{10}{v1}
        \fmfv{label=(a),label.angle=-90,label.dist=0.6w}{v1}
      \end{fmfgraph*}}
    \hq$=0$
    \hq\hq\hq\hq\hq\hq\hq
    \feynbox{100\unitlength}{
      \begin{fmfgraph*}(100,65)
        \fmfstraight
        \fmftopn{t}{26}
        \fmfbottomn{b}{13}
        \fmfleftn{l}{8}
        \fmfrightn{r}{8}
        \fmf{zigzag,foreground=red}{t1,v1,b1}
        \fmf{double,foreground=red}{t1,v1,b1}
        \fmf{zigzag,foreground=red}{t26,v2,b13}
        \fmf{double,foreground=red}{t26,v2,b13}
        \fmfblob{10}{v1}
        \fmfblob{10}{v2}
        \fmffreeze
        \fmf{vanilla,width=thick}{v1,v3,v2}
        \fmfblob{10}{v3}
        \fmffreeze
        \fmf{dashes}{vc1,v1}
        \fmf{photon,foreground=blue}{v1,vc2}
        \fmf{vanilla,width=thick}{v1,vc3}
        \fmf{dashes,tension=4}{b3,vc1}
        \fmf{photon,foreground=blue,tension=4}{vc2,b4}
        \fmf{vanilla,width=thick,tension=4}{vc3,b5}
        \fmffreeze
        \fmf{dots}{vc1,vc2,vc3}
        \fmf{dashes}{vd1,v2}
        \fmf{photon,foreground=blue}{v2,vd2}
        \fmf{vanilla,width=thick}{v2,vd3}
        \fmf{dashes,tension=4}{b11,vd1}
        \fmf{photon,foreground=blue,tension=4}{vd2,b10}
        \fmf{vanilla,width=thick,tension=4}{vd3,b9}
        \fmffreeze
        \fmf{dots}{vd1,vd2,vd3}
        \fmfv{label=(b),label.angle=-90,label.dist=0.1w}{b7}
      \end{fmfgraph*}}
    \hq$=0$
    \hq\hq\hq\hq\hq\hq\hq
    \feynbox{100\unitlength}{
      \begin{fmfgraph*}(100,50)
        \fmftop{t}
        \fmfleft{bl,tl}
        \fmfright{br,tr}
        \fmf{zigzag,foreground=red}{tl,vm1,bl}
        \fmf{double,foreground=red}{tl,vm1,bl}
        \fmf{zigzag,foreground=red}{tr,vm2,br}
        \fmf{double,foreground=red}{tr,vm2,br}
        \fmffreeze
        \fmf{vanilla,width=thick,label=$\fs q$}{vm1,v1}
        \fmf{vanilla,width=thick,label=$\fs(q_0+l_0,,\qv)$}{v1,vm2}
        \fmffreeze
        \fmf{photon,foreground=blue,label=$\fs l$}{v1,t}
        \fmfv{label=(c),label.angle=-90,label.dist=0.42w}{v1}
      \end{fmfgraph*}}
    \hq$\not=0$
  \end{center}
  \caption{\label{rule2}\sl The second rule in its bare (a) and dressed (b)
    version; (c) the generalisation analogous to Fig.\ \protect\ref{rule1} (c)
    fails. Conventions as in Fig.\ \ref{rule1}.}
\end{figure}
The vertices are again energy non-conserving, so that $q_0$ enters only in the
denominator of the potential quark in the combination $q_0-\frac{\qv^2}{2M}$.
Shifting $q_0-\frac{\qv^2}{2M}\to q_0$, the $q_0$ integral again does not
contain a scale in the denominator. In contradistinction to the previous case,
this rule cannot be extended to include potential or ultrasoft momenta
coupling to the blob, Fig.\ \ref{rule2} (c). For example, a bremsstrahlung
gluon with external, ultrasoft four-momentum $l$ renders the potential quark
denominators in the propagators as
$(q_0-\frac{\qv^2}{2M})(q_0+l_0-\frac{\qv^2}{2M})$. Now, $l_0$ provides the
necessary scale for the $q_0$ integration. As an example of the rule, the
Abelian vertex contribution with denominator
\begin{equation}
  \rule{0pt}{34pt}
   \feynbox{60\unitlength}{
      \begin{fmfgraph*}(60,20)
        \fmftop{i,o}
        \fmfbottom{u}
        \fmf{double,foreground=red,tension=2}{i,v1}
        \fmfv{label=$\fs(T,,\pv)$,label.angle=115}{i}
        \fmf{double,foreground=red}{v1,v2}
        \fmf{vanilla,width=thick}{v2,v3}
        \fmfv{label=$\fs q$,label.angle=-140}{v3}
        \fmf{double,foreground=red,tension=2}{v3,o}
        \fmfv{label=$\fs(T^\prime,,\pv^\prime)$,label.angle=65}{o}
        \fmffreeze
        \fmf{zigzag,foreground=red,left}{v1,v3}
        \fmf{zigzag,foreground=red}{v2,u}
      \end{fmfgraph*}}
    \;\;:\;\mathrm{Denom.}=\left(q_0-\frac{\qv^2}{2M}\right)
    \left(T-T^\prime\right)
    \left(T^{\prime 2}-(\pv^\prime-\qv)^2\right)
\end{equation}
is zero, but the last of the graphs in Fig.\ \ref{loopfigs} is not.

\absatz As the coupling of an ultrasoft gluon to a soft particle conserves
neither energy nor momentum, one may disconnect the ultrasoft gluon from such
a vertex, Fig.\ \ref{rule3} (a).
\begin{figure}[!htb]
  \vspace*{2ex}
  \begin{center}
    \rule[-20pt]{0pt}{40pt}
    \feynbox{50\unitlength}{
      \begin{fmfgraph*}(50,30)
        \fmftop{t}
        \fmfbottom{b}
        \fmfright{r}
        \fmf{zigzag,foreground=red}{t,v1,b}
        \fmf{double,foreground=red}{t,v1,b}
        \fmffreeze
        \fmf{photon,foreground=blue}{v1,r}
        \fmfv{label=(a),label.dist=0.45w,label.angle=-67}{r}
        \fmf{photon,foreground=blue}{v1,r}
      \end{fmfgraph*}}
    $\longrightarrow$\hspace*{-4ex}
    \feynbox{60\unitlength}{
      \begin{fmfgraph*}(60,30)
        \fmftop{t}
        \fmfbottom{b}
        \fmfright{r}
        \fmf{zigzag,foreground=red}{t,v1,b}
        \fmf{double,foreground=red}{t,v1,b}
        \fmffreeze
        \fmf{phantom,tension=2}{v1,v2}
        \fmf{photon,foreground=blue}{v2,r}
      \end{fmfgraph*}}
    \hq\hq\hq\hq
    \feynbox{60\unitlength}{
      \begin{fmfgraph*}(60,30)
        \fmfleft{i}
        \fmfright{o}
        \fmf{photon,foreground=blue,tension=3}{i,v1}
        \fmf{photon,foreground=blue,tension=3}{v2,o}
        \fmf{zigzag,foreground=red,left}{v1,v2,v1}
        \fmfv{label=(b),label.dist=0.4w,label.angle=-65}{o}
      \end{fmfgraph*}}
    $\longrightarrow$
    \feynbox{80\unitlength}{
      \begin{fmfgraph*}(80,30)
        \fmfleft{i}
        \fmfright{o}
        \fmf{photon,foreground=blue,tension=3}{i,vm1}
        \fmf{photon,foreground=blue,tension=3}{vm2,o}
        \fmf{phantom,tension=5}{vm1,v1}
        \fmf{phantom,tension=5}{vm2,v2}
        \fmf{zigzag,foreground=red,left}{v1,v2,v1}
      \end{fmfgraph*}}
    \hq$=0$
    \hq\hq\hq\hq
  \end{center}
  \caption{\label{rule3}\sl (a) Soft-to-ultrasoft vertices are cut off in a
    third rule; (b) an example. Conventions as in Fig.\ \ref{rule1}.}
\end{figure}
If after cutting, a line with loop momenta to be integrated over becomes
completely disconnected from the graph, the resulting tadpole makes the graph
vanish, Fig.\ \ref{rule3} (b). If only one of the $\Aumu$ legs becomes
dis-attached, the diagram can be non-zero, as the following diagram for the
Abelian vertex correction to the $\Qs A_{\mathrm{s}\,0}\Qp$ coupling
demonstrates:
\begin{equation}
  \label{sspvertex2}
  \rule{0pt}{40pt}
   \feynbox{60\unitlength}{
      \begin{fmfgraph*}(60,20)
        \fmftop{i,o}
        \fmfbottom{u}
        \fmf{double,foreground=red,tension=2}{i,v1}
        \fmfv{label=$\fs(T,,\pv)$,label.angle=115}{i}
        \fmf{double,foreground=red}{v1,v2}
        \fmf{vanilla,width=thick}{v2,v3}
        \fmf{vanilla,width=thick,tension=2}{v3,o}
        \fmfv{label=$\fs(T^\prime,,\pv^\prime)$,label.angle=65}{o}
        \fmffreeze
        \fmf{photon,foreground=blue,left,label=$\fs q$,label.side=left}{v1,v3}
        \fmf{zigzag,foreground=red}{v2,u}
      \end{fmfgraph*}}
    \hq\longrightarrow\;
    \feynbox{50\unitlength}{
      \begin{fmfgraph*}(50,40)
        \fmfleft{i}
        \fmfright{o}
        \fmftop{d1,d2,d4,d5,d6,p,d3}
        \fmfbottom{u}
        \fmf{double,foreground=red,tension=2}{i,v1}
        \fmf{double,foreground=red}{v1,v2}
        \fmf{vanilla,width=thick}{v2,v3}
        \fmf{vanilla,width=thick,tension=2}{v3,o}
        \fmffreeze
        \fmf{photon,foreground=blue}{v3,p}
        \fmf{zigzag,foreground=red}{v2,u}
      \end{fmfgraph*}}
    \;\;
    \not=0\;:\;\mathrm{Denom.}=T\;q^2\left(T^\prime-q_0-\frac{\pv^2}{2M}\right)
\end{equation}
See also Sect.\ \ref{sec:vertexcorrection} and the computation of this graph
in App.\ \ref{sec:vertexcalc}. 

When several ultrasoft gluons couple to a soft particle at the same point,
cutting does usually not result in a tadpole, either. For example, the
gluon-gluon vertex obtained by cutting the two-gluon two-quark vertex
\begin{equation}
  \label{nonzerocuttadpole}
  \feynbox{60\unitlength}{
      \begin{fmfgraph*}(60,40)
        \fmftop{i,o}
        \fmfbottom{u}
        \fmf{double,foreground=red}{i,v1,o}
        \fmffreeze
        \fmf{photon,foreground=blue,left,label=$\fs q$,label.side=left}{v1,v2}
        \fmf{photon,foreground=blue,left}{v2,v1}
        \fmf{photon,foreground=blue,tension=3,label=$\fs k$,
          label.side=right}{v2,u}
      \end{fmfgraph*}}
    \longrightarrow
    \feynbox{60\unitlength}{
      \begin{fmfgraph*}(60,50)
        \fmftop{i,o}
        \fmfbottom{u}
        \fmf{double,foreground=red}{i,v1,o}
        \fmfdot{v1}
        \fmffreeze
        \fmf{phantom,tension=3}{v1,v3}
        \fmf{photon,foreground=blue,left}{v3,v2,v3}
        \fmfdot{v3}
        \fmf{photon,foreground=blue,tension=3}{v2,u}
      \end{fmfgraph*}}
    \;\;\not=0\;:\;\mbox{Denom.}= (k-q)^2\; q^2 \not=0
\end{equation}
might be represented by a small blob as a reminder that it is neither energy
nor momentum conserving. In the ``tadpole'', a scale is set by the external
ultrasoft four-momentum $k$.

\absatz Loops made out of one soft quark and one soft gluon are restricted by
another rule.
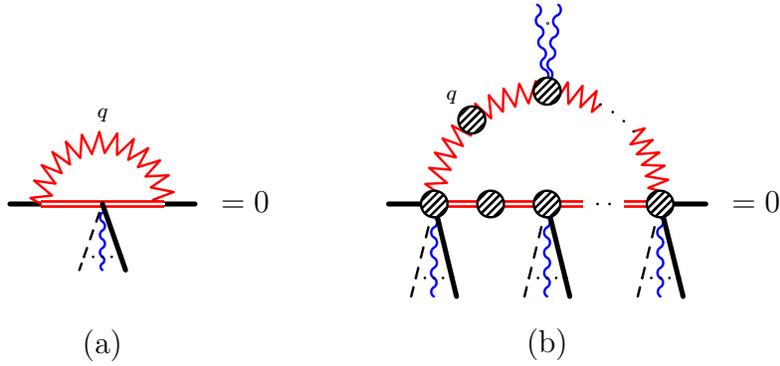
\begin{figure}[!htb]
  \vspace*{2ex}
  \begin{center}
    \rule[-45pt]{0pt}{110pt}
    \feynbox{70\unitlength}{
      \begin{fmfgraph*}(70,50)
        \fmfbottom{t0,t01,t02,t1,t2,t3,t4,t41,t42}
        \fmfleft{l}
        \fmfright{r}
        \fmf{vanilla,width=thick,tension=2}{l,vm2}
        \fmf{vanilla,width=thick,tension=2}{vm3,r}
        \fmf{double,foreground=red}{vm2,v1,vm3}
        \fmffreeze
        \fmf{zigzag,foreground=red,left,label=$\fs q$,
          label.side=left,label.dist=0.11w}{vm2,vm3}
        \fmffreeze
        \fmf{dashes}{v2,v1}
        \fmf{dashes,tension=4}{t1,v2}
        \fmf{photon,foreground=blue}{v1,v3}
        \fmf{photon,foreground=blue,tension=4}{v3,t2}
        \fmf{vanilla,width=thick}{v1,v4}
        \fmf{vanilla,width=thick,tension=4}{v4,t3}
        \fmffreeze
        \fmf{dots}{v2,v3,v4}
        \fmfv{label=(a),label.dist=0.32w}{t2}
      \end{fmfgraph*}}
    \hq$=0$
    \hq\hq\hq\hq\hq\hq\hq
    \feynbox{120\unitlength}{
      \begin{fmfgraph*}(120,70)
        \fmfstraight
        \fmftopn{t}{15}
        \fmfbottomn{b}{15}
        \fmfleft{l}
        \fmfright{r}
        \fmf{vanilla,width=thick,tension=0.79}{l,vm2}
        \fmf{vanilla,width=thick,tension=0.79}{vm3,r}
        \fmf{double,foreground=red,tension=0.64}{vm2,vm4,v1}
        \fmf{double,foreground=red}{v1,vm5}
        \fmf{double,foreground=red}{vm3,vm6}
        \fmf{dots,tension=3}{vm7,vm8}
        \fmf{phantom,tension=2.5}{vm6,vm7}
        \fmf{phantom,tension=2.5}{vm8,vm5}
        \fmfblob{10}{v1}
        \fmfblob{10}{vm2}
        \fmfblob{10}{vm3}
        \fmfblob{10}{vm4}
        \fmffreeze
        \fmf{phantom,left,tag=1}{vm2,vm3}
        \fmfipath{p[]}
        \fmfiset{p1}{vpath1(__vm2,__vm3)}
        \fmfi{zigzag,foreground=red,label=$\fs q$,label.side=left,
          label.dist=0.07w}{subpath (0,1/2 length(p1)) of p1}
        \fmfi{zigzag,foreground=red}
          {subpath (1/2 length(p1),2/3 length(p1)) of p1}
        \fmfi{zigzag,foreground=red}
          {subpath (4/5 length(p1),length(p1)) of p1}
        \fmfi{dots}{subpath (2/3 length(p1),4/5 length(p1)) of p1}
        \fmffreeze
        \fmfipair{minea}
        \fmfipair{mineb}
        \fmfipair{minec}
        \fmfiequ{minea}{(56,110)}
        \fmfiequ{mineb}{(64,110)}
        \fmfiequ{minec}{(60,103)}
        \fmfi{photon,foreground=blue}{point 1/2 length(p1) of p1 -- minea}
        \fmfi{photon,foreground=blue}{point 1/2 length(p1) of p1 -- mineb}
        \fmfiv{decoration.shape=circle,decoration.filled=shaded
          ,decoration.size=0.002w}{minec}
        \fmfiv{decoration.shape=circle,decoration.filled=shaded
          ,decoration.size=0.085w}{point 1/2 length(p1) of p1}
        \fmfiv{decoration.shape=circle,decoration.filled=shaded
          ,decoration.size=0.085w}{point 1/4 length(p1) of p1}
        \fmf{dashes}{va1,v1}
        \fmf{photon,foreground=blue}{v1,va2}
        \fmf{vanilla,width=thick}{v1,va3}
        \fmf{dashes,tension=4}{b7,va1}
        \fmf{photon,foreground=blue,tension=4}{va2,b8}
        \fmf{vanilla,width=thick,tension=4}{va3,b9}
        \fmffreeze
        \fmf{dots}{va1,va2,va3}
        \fmffreeze
        \fmf{dashes}{vc1,vm2}
        \fmf{photon,foreground=blue}{vm2,vc2}
        \fmf{vanilla,width=thick}{vm2,vc3}
        \fmf{dashes,tension=4}{b2,vc1}
        \fmf{photon,foreground=blue,tension=4}{vc2,b3}
        \fmf{vanilla,width=thick,tension=4}{vc3,b4}
        \fmffreeze
        \fmf{dots}{vc1,vc2,vc3}
        \fmf{dashes}{vd1,vm3}
        \fmf{photon,foreground=blue}{vm3,vd2}
        \fmf{vanilla,width=thick}{vm3,vd3}
        \fmf{dashes,tension=4}{b12,vd1}
        \fmf{photon,foreground=blue,tension=4}{vd2,b13}
        \fmf{vanilla,width=thick,tension=4}{vd3,b14}
        \fmffreeze
        \fmf{dots}{vd1,vd2,vd3}
        \fmfv{label=(b),label.angle=-90,label.dist=0.1w}{b8}
      \end{fmfgraph*}}
    \hq$=0$
  \end{center}
  \caption{\label{rule4}\sl The fourth rule in its (a) bare and (b) extended
    version, where the rule in Fig.\ \ref{rule3} (a) was used for attaching
    ultrasoft gluons to the soft gluon. Conventions as in Fig.\ \ref{rule1}.}
\end{figure}
In Fig.\ \ref{rule4} (a), the denominators are $q_0^2-\qv^2$ for the soft
gluon and $q_0$ for the soft quark. Therefore, the diagram is without scale.
The generalisation, Fig.\ \ref{rule4} (b), eliminates all soft contributions
to the potential quark self energy (\ref{nrqcdselfpottosoft}) and to Abelian
vertex corrections between potential and ultrasoft particles. Note that the
light particle exchanged is massless or has a mass considerably smaller than
$Mv^2$. A relativistic particle mass of order $Mv^2$ or bigger provides a
scale in the relativistic propagator, invalidating this rule. Also, as soon as
a potential particle couples to the soft gluon, the diagram is non-zero. An
example is the non-Abelian vertex correction to the $\Qp^\dagger \Apmu \Qp$
coupling with denominator
\begin{equation}
  \rule{0pt}{34pt}
  \feynbox{60\unitlength}{
      \begin{fmfgraph*}(60,40)
        \fmftop{i,o}
        \fmfbottom{u}
        \fmf{vanilla,width=thick,tension=4,label=$\fs(T,,\pv)$,
          label.side=left}{i,v1}
        \fmf{double,foreground=red}{v1,v3}
        \fmf{vanilla,width=thick,tension=4,label=$\fs(T^\prime,,\pv^\prime)$,
          label.side=left}{v3,o}
        \fmffreeze
        \fmf{zigzag,foreground=red}{v1,v2,v3}
        \fmf{dashes,tension=3}{v2,u}
      \end{fmfgraph*}}
    \;\;\not=0\;:\;
    \mathrm{Denom.}=q_0\;q^2\;\left(q_0^2-(\pv-\pv^\prime+\qv)^2\right)\;\;.
\end{equation}

\absatz To summarise, the principle underlying the rules is simple: One
identifies a potentially vanishing sub-set of the entire diagram and assigns
the external and loop momenta to it, taking into account multipole expansions.
Then, the denominators (i.e.\ inverse propagators) are written out at lowest
order in $v$. If by shifting integration variables, one arrives at an energy
or momentum integral without scale, the diagram vanishes to all orders. Along
this line, further rules are easily established.

\absatz Two more rules are more straightforwardly derived from the propagator
properties of NRQCD directly. The first one cuts down the number of diagrams
including potential gluons. For the heavy quarks, there is no distinction
between Feynman and retarded propagators in NRQCD (\ref{qsprop}/\ref{qpprop})
because antiparticle propagation has been eliminated by the field
transformation from the relativistic to the non-relativistic Lagrangean. The
fields $\Qs$ and $\Qp$ propagate hence only time-like into the forward light
cone. Still, Feynman's perturbation theory becomes more convenient than the
time-ordered formalism, as less diagrams have to be calculated. Soft and
ultrasoft gluons propagate light-like, and potential gluons instantaneously.
Therefore, any diagram which -- when time ordered -- involves quarks
propagating outside the forward light cone are zero, as are diagrams which
cannot be drawn with both ends of a potential gluon being space-like separated.
This ``forwardness'' excludes all non-zero contributions from the potential
gluon contribution to the quark self energy in \emph{any} r\'egime, e.g.\ 
(\ref{thexpselfpottopot}).

Another rule is that there must at least be one pole on each side of the real
$q_0$ axis for a diagram to be non-zero. This will e.g.\ not be the case when
the loop momentum $q$ only occurs in one quark propagator besides potential
gluon propagators. Likewise, the contribution to the Abelian vertex correction
\begin{equation}
  \rule{0pt}{34pt}
   \feynbox{60\unitlength}{
      \begin{fmfgraph*}(60,20)
        \fmftop{i,o}
        \fmfbottom{u}
        \fmf{double,foreground=red,tension=2}{i,v1}
        \fmfv{label=$\fs(T,,\pv)$,label.angle=115}{i}
        \fmf{vanilla,width=thick}{v1,v2,v3}
        \fmfv{label=$\fs q$,label.angle=-30,label.dist=0.13w}{v2}
        \fmf{double,foreground=red,tension=2}{v3,o}
        \fmfv{label=$\fs(T^\prime,,\pv^\prime)$,label.angle=65}{o}
        \fmffreeze
        \fmf{zigzag,foreground=red,left}{v1,v3}
        \fmf{dashes}{u,v2}
      \end{fmfgraph*}}
    \;\;:\;\mathrm{Denom.}=\left(q_0-\frac{\qv^2}{2M}+\ii\epsilon\right)
    \left(q_0-\frac{(\qv +\pv^\prime-\pv)^2}{2M}+\ii\epsilon\right)
    \left(T^2-(\pv-\qv)^2\right)
\end{equation}
has only poles below the real axis. Closing the contour of the $q_0$
integration above it, the diagram is zero. This can also be proven using
dimensional regularisation.

\absatz In order to demonstrate that the rules established above are useful in
reducing the number of graphs, Fig.\ \ref{softselffilter} shows that only one
of the four scale-conserving graphs contributing to the one loop soft quark
self-energy survives their filter.
\begin{figure}[!htb]
  \vspace*{2ex}
  \begin{center}
    \feynbox{60\unitlength}{
      \begin{fmfgraph*}(60,40)
        \fmfleft{i}
        \fmfright{o}
        \fmf{double,foreground=red,tension=4}{i,v1}
        \fmf{double,foreground=red,tension=4}{v2,o}
        \fmf{heavy,foreground=red,width=thin,label=(a),label.dist=0.25w,
          label.side=right}{v1,v2}
        \fmffreeze
        \fmf{zigzag,foreground=red,left}{v1,v2}
      \end{fmfgraph*}}
    $\;\not=0$
    \hq\hq\hq
    \feynbox{60\unitlength}{
      \begin{fmfgraph*}(60,40)
        \fmfleft{i}
        \fmfright{o}
        \fmf{double,foreground=red,tension=4}{i,v1}
        \fmf{double,foreground=red,tension=4}{v2,o}
        \fmf{fermion,width=thick,label=(b),label.dist=0.25w,
          label.side=right}{v1,v2}
        \fmffreeze
        \fmf{zigzag,foreground=red,left}{v1,v2}
      \end{fmfgraph*}}
    $\;=0$
    \hq\hq\hq
    \feynbox{60\unitlength}{
      \begin{fmfgraph*}(60,40)
        \fmfleft{i}
        \fmfright{o}
        \fmf{double,foreground=red,tension=4}{i,v1}
        \fmf{double,foreground=red,tension=4}{v2,o}
        \fmf{heavy,foreground=red,label=(c),label.dist=0.25w,
          label.side=right}{v1,v2}
        \fmffreeze
        \fmf{dashes,left}{v1,v2}
      \end{fmfgraph*}}
    $\;=0$
    \hq\hq\hq
    \feynbox{60\unitlength}{
      \begin{fmfgraph*}(60,40)
        \fmfleft{i}
        \fmfright{o}
        \fmf{double,foreground=red,tension=4}{i,v1}
        \fmf{double,foreground=red,tension=4}{v2,o}
        \fmf{heavy,foreground=red,label=(d),label.dist=0.25w,
          label.side=right}{v1,v2}
        \fmffreeze
        \fmf{photon,foreground=blue,left}{v1,v2}
      \end{fmfgraph*}}
    $\;=0$
  \end{center}
  \caption{\label{softselffilter}\sl Out of the four scale-conserving
    one loop contributions to the soft quark self energy, only the first one,
    (a), survives. (b) is zero because it contains a sub-graph zero by Fig.\ 
    \protect\ref{rule2} (a); (c) vanishes because of Fig.\ \protect\ref{rule1}
    (c) or because of forwardness; (d) is eliminated by cutting off the
    ultrasoft gluons (Fig.\ \protect\ref{rule3} (a)). The resulting tadpole
    is zero.}
\end{figure}
The classification of the only non-zero diagrams for Abelian and non-Abelian
one-loop corrections to one gluon vertices addressed later,
Figs.~\ref{abelianvertex}, \ref{nonabelianvertex} and \ref{nonabelianvertexa2},
may serve as another example.

\absatz In conclusion, the homogene{\ia}ty of dimensional regularisation
(\ref{masterformula}) can be translated into diagrammatic rules to
systematically identify graphs which vanish to all orders in threshold
expansion or velocity expansion in NRQCD. The concept is sensitive only to the
multipole expansion of the vertices involved, not to their precise nature. It
is also independent of insertions and gauge chosen. Only the denominators of
the graph lowest order in $v$ have to be looked at.

\subsection{Theorems from Diagrammatic Rules}
\label{sec:theorems}

Diagrammatic rules are quite effective to prove decoupling theorems to all
orders in perturbation theory. ``Scale conservation'' entered in Sect.\ 
\ref{sec:vertex} as a constraint on both the overall energy and momentum of a
vertex because both have to be conserved order by order in $v$. Now, one may
resort to a more formal argument: If one rescales the fields in the Lagrangean
in all possible combinations, diagrams containing scale non-conserving
vertices are eliminated because they correspond to temporal or spatial
tadpoles. As an example, consider a soft particle coupling to two potential
ones. The dashed-solid line representing any potential particle, the multipole
expansion is
\begin{equation}
  \rule{0pt}{32pt}
  \label{sppvertex}
    \feynbox{75\unitlength}{
    \begin{fmfgraph*}(75,30)
      \fmfstraight
      \fmftop{i,o}
      \fmfbottom{u}
      \fmf{double_dashsolid_lr,label=$\fs(T,,\pv)$,label.side=left,
        label.dist=0.1w}{i,v}
      \fmf{double_dashsolid_lr,label=$\fs(T^\prime,,\pv^\prime)$,
        label.side=left,label.dist=0.1w}{v,o}
      \fmffreeze
      \fmf{zigzag,foreground=red,width=thin,label=$\uparrow\fs q,,A_0$}{u,v}
      \fmf{double,foreground=red}{u,v}
    \end{fmfgraph*}}
          \;\propto \;(2\pi)^4\;
          \delta^{(3)}(\pv-\pv^\prime+\qv)\;
          \Big[\exp\Big({(T-T^\prime)\;\frac{\partial}{\partial
          q_0}}\Big)\;\delta(q_0)\Big]\;\;.
\end{equation}
The temporal tadpole of the $q_0$ integration makes all diagrams vanish in
which this vertex is embedded such that $q$ is a loop momentum. If $q_0$ is an
external energy, its being zero contradicts the assumption that it is of order
$Mv$. Therefore, diagrams containing scale non-conserving vertices are zero.
In threshold expansion, this follows automatically from the observation that
not the individual vertices but the loop momentum as a whole is expanded, see
Sect.\ \ref{sec:nrqcdandthreshold}.

\absatz Another application is the decoupling of an ultrasoft heavy quark from
NRQCD. When the particle content of NRQCD in the three different r\'egimes was
outlined in Sect.\ \ref{sec:regimes}, the bremsstrahlung gluon was assumed to
be the only ultrasoft field, and an ultrasoft quark $\Qu$ was not necessary to
make the theory self-consistent. An ultrasoft quark might still be produced by
the radiation of a potential gluon off a potential quark, but it is not
unavoidable. In contradistinction, a soft quark is indispensable because a
soft, on shell gluon excites a potential quark to a kinetic energy of order
$Mv$: The potential quark coupling to a soft gluon becomes necessarily soft.
Taking advantage of the homogene{\ia}ty (\ref{masterformula}) of dimensional
regularisation, one can show that a hypothetical ultrasoft heavy quark
decouples completely from the theory: All graphs containing it are zero. This
fact was already remarked upon by Beneke and Smirnov \cite{BenekeSmirnov}, and
the formal proof using the NRQCD techniques proceeds as follows.

Of all properties of the ultrasoft quark, the only ones needed are that its
propagator (denoted by a dotted line) is static because $T\sim
Mv^2,\;|\pv|\sim Mv^2$,
\begin{equation}
  \label{quprop}
  \Qu:\;
       \feynbox{60\unitlength}{
       \begin{fmfgraph*}(60,30)
         \fmfleft{i}
         \fmfright{o}
         \fmf{dbl_dots_arrow,foreground=blue,width=thin,label=$\fs(T,,\pv)$,
           label.side=left}{i,o}
       \end{fmfgraph*}}
       \;:\;\frac{\ii}{T+\ii\epsilon}\;\;;\,
\end{equation}
and that it couples momentum non-conserving to all but ultrasoft particles.
These features can of course be derived from the rescaling rules of an
ultrasoft quark, $\Qu(\xv,t)=(Mv^2)^\frac{3}{2} \calQu(\xu,\tu)$, in full
analogy to Sect.\ \ref{sec:props}. If it exists, $\Qu$ can only enter in
internal lines. Fermion number conservation dictates that it is produced and
annihilated in a vertex into which at least one soft or potential quark
enters.

\begin{figure}[!htb]
  \vspace*{3ex}
  \begin{center}
     \feynbox{80\unitlength}{
      \begin{fmfgraph*}(80,40)
        \fmfstraight
        \fmfleft{bl,tl}
        \fmfright{br,tr}
        \fmf{vanilla,width=thick}{tl,v1}
        \fmf{double_dashsolid_ud}{bl,v1}
        \fmf{vanilla,width=thick}{tr,v2}
        \fmf{double_dashsolid_ud}{br,v2}
        \fmfv{label=${\fs(T-q_0,,\pv)\uparrow}$,label.angle=90}{tl}
        \fmfv{label=${\fs(T,,\pv)\uparrow}$,label.angle=-90}{bl}
        \fmfv{label=${\fs\uparrow(T^\prime+q_0,,\pv^\prime)}$,
          label.angle=90}{tr}
        \fmfv{label=${\fs\uparrow(T^\prime,,\pv^\prime)}$,label.angle=-90}{br}
        \fmfv{label=$\!\!\!\!$(a),label.angle=-45,label.dist=0.7w}{v1}
        \fmffreeze
        \fmf{dbl_dots,foreground=blue,
          label=${\rightarrow\atop\fs q}$,label.side=right}{v1,v2} 
      \end{fmfgraph*}}
      \hq$=0$
      \hq\hq\hq\hq\hq\hq
      \feynbox{80\unitlength}{
      \begin{fmfgraph*}(80,50)
        \fmfstraight
        \fmftopn{t}{9}
        \fmfleft{bl,tl}
        \fmfright{br,tr}
        \fmf{vanilla,width=thick}{tl,v1}
        \fmf{double_dashsolid_ud}{bl,v1}
        \fmf{vanilla,width=thick}{tr,v2}
        \fmf{double_dashsolid_ud}{br,v2}
        \fmffreeze
        \fmf{dbl_dots,foreground=blue,label=$\fs q$,label.side=right}{v1,v3}
        \fmf{dbl_dots,foreground=blue,label=$\fs q+l$,label.side=right}{v3,v2}
        \fmffreeze
        \fmf{photon,foreground=blue}{va1,v3}
        \fmf{dbl_dots,foreground=blue}{v3,va2}
        \fmf{photon,foreground=blue}{v3,va3}
        \fmf{photon,foreground=blue,tension=4}{t4,va1}
        \fmf{dbl_dots,foreground=blue,tension=4}{va2,t5}
        \fmf{photon,foreground=blue,tension=4}{va3,t6}
        \fmfv{label=$\fs\overbrace{\rule{25pt}{0pt}}^{\fs\downarrow l}$,
          label.angle=90}{t5}
        \fmffreeze
        \fmf{dots}{va1,va2,va3}
        \fmfv{label=(b),label.angle=-90,label.dist=0.5w}{v3}
      \end{fmfgraph*}}
      \hq$=0$
      \hq\hq\hq\hq\hq\hq
    \feynbox{60\unitlength}{
      \begin{fmfgraph*}(60,50)
        \fmfstraight
        \fmftopn{t}{6}
        \fmfbottomn{b}{6}
        \fmfleft{l}
        \fmfright{r}
        \fmf{dbl_dots,foreground=blue,label=$\fs q$,label.side=right}{l,v1}
        \fmf{dbl_dots,foreground=blue,label=$\fs l$,label.side=right}{v1,r}
        \fmffreeze
        \fmf{dashes}{va1,v1}
        \fmf{dashes,tension=4}{t3,va1}
        \fmf{vanilla,width=thick}{v1,va2}
        \fmf{vanilla,width=thick,tension=5}{va2,t4}
        \fmf{photon,foreground=blue}{v1,vb1}
        \fmf{photon,foreground=blue,tension=4}{vb1,b3}
        \fmf{dbl_dots,foreground=blue}{v1,vb2}
        \fmf{dbl_dots,foreground=blue,tension=4}{vb2,b4}
        \fmf{phantom,tension=5}{t3,t4}
        \fmf{phantom,tension=5}{b3,b4}
        \fmfv{label=(c),label.angle=-90,label.dist=0.67w}{v1}
        \fmffreeze
        \fmf{dots}{va1,va2}
        \fmf{dots}{vb1,vb2}
      \end{fmfgraph*}}
    \\[10ex]
    \feynbox{200\unitlength}{
      \begin{fmfgraph*}(200,75)
        \fmfstraight
        \fmftopn{t}{26}
        \fmfbottomn{b}{26}
        \fmf{vanilla,width=thick}{t1,v1}
        \fmf{double_dashsolid_ud}{b1,v1}
        \fmf{vanilla,width=thick}{t26,v2}
        \fmf{double_dashsolid_ud}{b26,v2}
        \fmffreeze
        \fmf{dbl_dots,foreground=blue}{v1,v1a,v3}
        \fmf{dbl_dots,foreground=blue,tension=2}{v3,v4}
        \fmf{dbl_dots,foreground=blue}{v2,v2a,v5}
        \fmf{dbl_dots,foreground=blue,tension=2}{v5,v6}
        \fmf{dots,tension=3}{v4a,v6a}
        \fmf{phantom,tension=2.2}{v4,v4a}
        \fmf{phantom,tension=2.2}{v6,v6a}
        \fmffreeze
        \fmf{dashes}{va1,v3}
        \fmf{vanilla,width=thick}{v3,va3}
        \fmf{dashes,tension=4}{t8,va1}
        \fmf{vanilla,width=thick,tension=4}{va3,t10}
        \fmffreeze
        \fmf{dots}{va1,va3}
        \fmf{photon,foreground=blue}{v3,va2}
        \fmf{dbl_dots,foreground=blue}{v3,va4}
        \fmf{photon,foreground=blue,tension=7}{va2,b8}
        \fmf{dbl_dots,foreground=blue,tension=7}{va4,b10}
        \fmffreeze
        \fmf{dots}{va2,va4}
        \fmffreeze
        \fmf{dashes}{vb1,v5}
        \fmf{vanilla,width=thick}{v5,vb3}
        \fmf{dashes,tension=4}{t17,vb1}
        \fmf{vanilla,width=thick,tension=4}{vb3,t19}
        \fmffreeze
        \fmf{dots}{vb1,vb3}
        \fmf{photon,foreground=blue}{v5,vb2}
        \fmf{dbl_dots,foreground=blue}{v5,vb4}
        \fmf{photon,foreground=blue,tension=7}{vb2,b17}
        \fmf{dbl_dots,foreground=blue,tension=7}{vb4,b19}
        \fmffreeze
        \fmf{dots}{vb2,vb4}
        \fmffreeze
        \fmf{dashes}{vc1,v1}
        \fmf{photon,foreground=blue}{v1,vc2}
        \fmf{vanilla,width=thick}{v1,vc3}
        \fmf{dashes,tension=4}{b3,vc1}
        \fmf{photon,foreground=blue,tension=4}{vc2,b4}
        \fmf{vanilla,width=thick,tension=4}{vc3,b5}
        \fmffreeze
        \fmf{dots}{vc1,vc2,vc3}
        \fmf{dashes}{vd1,v2}
        \fmf{photon,foreground=blue}{v2,vd2}
        \fmf{vanilla,width=thick}{v2,vd3}
        \fmf{dashes,tension=4}{b24,vd1}
        \fmf{photon,foreground=blue,tension=4}{vd2,b23}
        \fmf{vanilla,width=thick,tension=4}{vd3,b22}
        \fmffreeze
        \fmf{dots}{vd1,vd2,vd3} 
        \fmfv{label=$\;\;\,,$(d),label.angle=-90,label.dist=0.1w}{b13}
      \end{fmfgraph*}}
    \hq$=0$\\[6ex]
  \end{center}
  \caption{\label{qudecoupling}\sl Decoupling of the heavy, ultrasoft 
    quark. (a) primitive diagram; (b) extension to the case of ultrasoft
    particles coupling to \protect$\Qu$; (c) general vertex involving
    potential and ultrasoft particles; (d) ultrasoft quarks decouple from any
    graph containing potential and ultrasoft particles. Conventions as in
    Fig.\ \ref{rule1}.}
\end{figure}
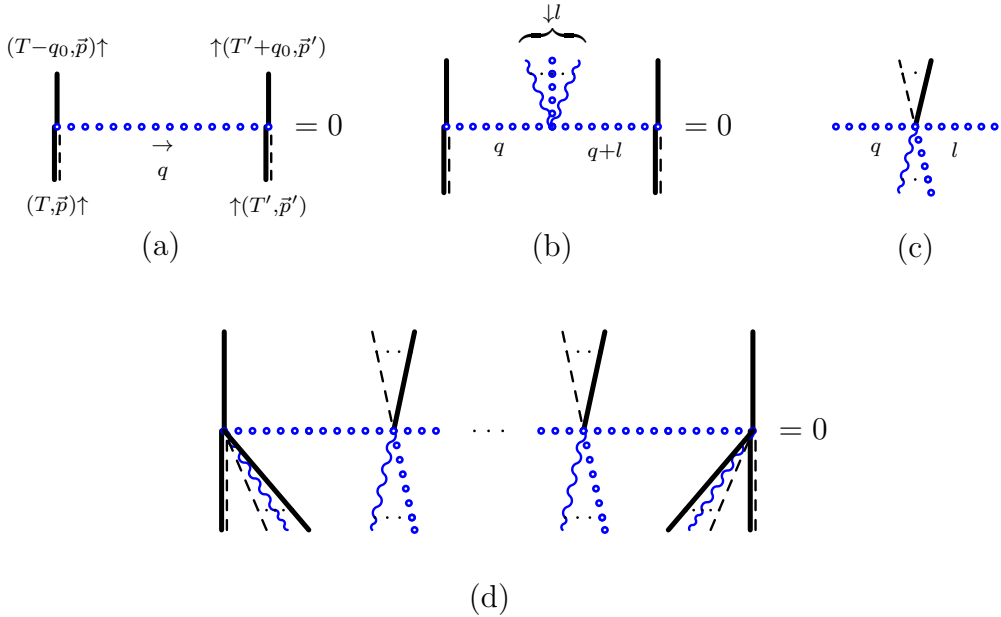

The simplest sub-diagram containing an ultrasoft quark is depicted in Fig.\ 
\ref{qudecoupling} (a) and vanishes because there is no loop momentum $\qv$ in
any denominator.
This is still true when an arbitrary number of ultrasoft particles couple to
$\Qu$, Fig.\ \ref{qudecoupling} (b), and when such a coupling occurs
repetitively. Now one considers couplings to potential particles like in Fig.\ 
\ref{qudecoupling} (c). The momentum multipole expansion necessary for all
ultrasoft particles separately in such a vertex disallows $\qv$ to be
transferred to a denominator. Therefore, any diagram containing an ultrasoft
quark coupled to potential or ultrasoft particles is zero because of the
spatial tadpole in the $q$ integration, Fig.\ \ref{qudecoupling} (d). Because
couplings of ultrasoft particles to soft ones involve the same momentum
multipole expansion as couplings to potential ones, the proof extends to all
graphs containing ultrasoft quarks: Every diagram with ultrasoft quarks is
zero. Recursive application dresses all vertices and propagators.

\subsection{The Quark Self-Energy}
\label{sec:selfenergy}

The rules cover the lowest order contributions to the potential quark self
energy as discussed in Sect.\ \ref{sec:nrqcdandthreshold}. The graph in
(\ref{nrqcdselfpottosoft}) is zero because of Fig.\ \ref{rule4}, the one in
(\ref{thexpselfpottopot}) because of forwardness. The only remaining graph is
(\ref{thexpselfpottoultra}), whose lowest order contribution is $\pc(v^0)$ by
the power counting of table \ref{scalarvertex}. Indeed, corrections to particle
propagators, quark or gluon, must count as $v^0$ since the power counting was
constructed such that free particle propagators in each r\'egime scale like
$v^0$ (\ref{gluonscaling}/\ref{quarkscaling}) and renormalisation corrections
to propagators must be of the same order. Using the integral
(\ref{firstmaster}) of App.\ \ref{sec:splitdimreg}, the pole in four dimensions
is extracted as
\begin{equation}
  \label{selfpot}
    \feynbox{60\unitlength}{
    \begin{fmfgraph*}(60,40)
      \fmfleft{i}
      \fmfright{o}
      \fmf{vanilla,width=thick,tension=5}{i,v1}
      \fmf{vanilla,width=thick,tension=5}{v2,o}
      \fmf{fermion,width=thick}{v1,v2}
      \fmffreeze
      \fmf{photon,foreground=blue,left}{v1,v2}
    \end{fmfgraph*}}
  \;\;:\;\;
  \ii\Sigma_\mathrm{p}(T,\pv)|_\mathrm{pole}=\frac{-\ii
    g^2_\mathrm{R}\;\mu^{2\varepsilon}}{8 \pi^2}\;
  C_\mathrm{F}\;\frac{3-\alpha}{2}\;\Gamma[\varepsilon]\;
  \left(T-\frac{\pv^2}{2M}\right)+\pc(v)\;\;,
\end{equation}
where $\varepsilon=2-\frac{d}{2}$ and $\mu$ is the renormalisation parameter.
The potential quark propagator is recovered, confirming that different
r\'egimes do not mix under renormalisation. The non-relativistic quark
propagator does not need renormalisation of the heavy quark mass. The quark
wave function renormalisation is hence in the MS scheme
\begin{equation}
  \label{z2}
  Z_2=1+\frac{g^2_\mathrm{R}\;\mu^{2\varepsilon}}{8\pi^2}
    \;C_\mathrm{F}\;\frac{3-\alpha}{2}\;\Gamma[\varepsilon]\;\;. 
\end{equation}
For the soft quark self energy, the only diagram surviving the diagrammatic
filter (Fig.\ \ref{softselffilter}) is at lowest order in the power counting
using the Feynman rules and (\ref{firstmaster})
\begin{equation}
  \label{nrqcdselfsoft}
  \feynbox{60\unitlength}{
    \begin{fmfgraph*}(60,40)
      \fmfleft{i}
      \fmfright{o}
      \fmf{double,foreground=red,tension=4}{i,v1}
      \fmf{double,foreground=red,tension=4}{v2,o}
      \fmf{heavy,foreground=red,width=thin}{v1,v2}
      \fmffreeze
      \fmf{zigzag,foreground=red,left}{v1,v2}
    \end{fmfgraph*}}
  \;:\;
  \ii\Sigma_\mathrm{s}(T,\pv)|_\mathrm{pole}=\frac{-\ii g^2_\mathrm{R}\;
    \mu^{2\varepsilon}}{8\pi^2}\;C_\mathrm{F}\;\frac{3-\alpha}{2}
  \;\Gamma[\varepsilon]\;T\;\;.
\end{equation}
That the soft quark renormalisation is the same as for the potential quark is
not surprising as both stem from the same un-expanded NRQCD diagram
(\ref{nrqcdselfenergy}).

\subsection{The Vacuum Polarisation}
\label{sec:vacpol}

In contradistinction to the self energy for $\Qp$ which does not receive
contributions from soft particles, the vacuum polarisation of the potential
gluon is non-zero only because of the presence of soft gluons, independent of
the gauge chosen. The scale-conserving graphs with gluon loops are
\begin{equation}
  \label{vacpolpot}
  \feynbox{60\unitlength}
  {\begin{fmfgraph*}(60,40)
    \fmfleft{i}
    \fmfright{o}
    \fmf{dashes,tension=4}{i,v1}
    \fmf{dashes,tension=4}{v2,o}
    \fmf{zigzag,foreground=red,left}{v1,v2,v1}
  \end{fmfgraph*}}
  \hq\hq\hq\hq\hq\hq\hq\hq
  \feynbox{60\unitlength}
  {\begin{fmfgraph*}(60,40)
    \fmfleft{i}
    \fmfright{o}
    \fmf{dashes,tension=4}{i,v1}
    \fmf{dashes,tension=4}{v2,o}
    \fmf{dashes,left}{v1,v2,v1}
  \end{fmfgraph*}}
  \hq\hq\hq\hq\hq\hq\hq\hq
  \feynbox{60\unitlength}
  {\begin{fmfgraph*}(60,40)
    \fmfleft{i}
    \fmfright{o}
    \fmf{dashes,tension=4}{i,v1}
    \fmf{dashes,tension=4}{v2,o}
    \fmf{dashes,left}{v1,v2}
    \fmf{photon,foreground=blue,left}{v2,v1}
  \end{fmfgraph*}}\;\;.
\end{equation}
The integral in the second graph does not contain $q_0$ in the denominator and
hence is zero; the third diagram vanishes because it couples a potential gluon
to a light-like particle twice. Soft gluons are therefore indispensable to
provide the gluons mediating the Coulomb interaction with a non-zero vacuum
polarisation, and hence to have a running $\beta$ function. As the gluons in
the soft and ultrasoft r\'egimes are on-shell particles and must run in NRQCD
as in QCD, there is no reason to expect the potential gluons to freeze out in
perturbation theory. The only non-zero ghost and light quark contributions
come analogously from soft ghost and soft light quark propagation in the loop.

The rescaling rules of table \ref{threegluonvertex} give the three gluon vertex
to count as $v^\frac{1}{2}$, and an additional $v^{-1}$ from the loop power
counting makes the graph count as $v^0$ at leading order. Again, this is
expected not only because gluons are relativistic particles, but also because
one wants to renormalise a propagator which is $\pc(v^0)$ in the power
counting. Without the loop rule of Sect.\ \ref{sec:loop}, the graph would be of
order $v^1$, and the power counting would predict that its contribution would
vanish as $v\to 0$ so that the potential gluon propagator at rest would reduce
to the bare one, a clearly unacceptable conclusion. The explicit computation
(\ref{vacpol}) will also show that this graph is $\pc(v^0)$.

\absatz The soft and ultrasoft gluon receive their vacuum polarisations from
loops with on shell gluons, $\pc(v^0)$:
\begin{equation}
  \label{vacpolsoftultra}
  \feynbox{60\unitlength}{
    \begin{fmfgraph*}(60,40)
      \fmfleft{i}
      \fmfright{o}
      \fmf{zigzag,foreground=red,tension=4}{i,v1}
      \fmf{zigzag,foreground=red,tension=4}{v2,o}
      \fmf{zigzag,foreground=red,left}{v1,v2,v1}
    \end{fmfgraph*}}
  \hq\hq\hq\hq\hq\hq\hq\hq
  \feynbox{60\unitlength}{
    \begin{fmfgraph*}(60,40)
      \fmfleft{i}
      \fmfright{o}
      \fmf{photon,foreground=blue,tension=4}{i,v1}
      \fmf{photon,foreground=blue,tension=4}{v2,o}
      \fmf{photon,foreground=blue,left}{v1,v2}
      \fmf{photon,foreground=blue,left}{v2,v1}
    \end{fmfgraph*}}
\end{equation}
All other diagrams vanish by the rules developed above. Both contributions are
identical to the ordinary QCD result, as no insertions or expansions enter.
The gluon vacuum polarisation from gluon, ghost and light fermion
contributions is therefore in the soft and ultrasoft r\'egime~\cite[eq.\ 
(2.5.132)]{Muta}
\begin{eqnarray}
  \label{vacpol}
  &\dis\Pi_{\mu\nu}^{ab}(k)=\delta^{ab}\;(k_\mu k_\nu- k^2 g_{\mu\nu})
  \;\Pi(k^2)\;\;\mbox{with}&\\
  &\dis\Pi(k^2) = \frac{g^2_\mathrm{R}\;\mu^{2\varepsilon}}{(4\pi)^2}\;
    \left[\frac{2}{3}\; N_\mathrm{F}
    -\frac{1}{2}\;N\left(\frac{13}{3}-\alpha\right)\right]
  \Gamma[\varepsilon]+\mbox{ finite.}&\non
\end{eqnarray}
Because the potential gluon vacuum polarisation does not contain insertions or
multipole expansions in the internal lines, either, the QCD result can be
taken and expanded in powers of the external energy $k_0\ll|\kv|$. As the
infinite part of $\Pi(k^2)$ does not contain $k$, the only change will be that
to lowest order in $v$, the part guaranteeing transversality of the gluon
becomes
\begin{equation}
  \label{ginvforpotglue}
  (k_\mu k_\nu- k^2 g_{\mu\nu})\;\to\;(\delta_{\mu i}\delta_{\nu j}
  k_i k_j+\kv^2 g_{\mu\nu})+\pc(v)\;\;.
\end{equation}
Renormalisation therefore keeps the contributions from the three r\'egimes
separate and the potential gluon propagator transversal up to higher order in
$v$. For all three r\'egimes, the gluon wave function renormalisation is in the
end the one of QCD~\cite[eq.\ (2.5.135)]{Muta},
\begin{equation}
  \label{z3}
  Z_3=1-\frac{g_\mathrm{R}^2\;\mu^{2\varepsilon}}{(4\pi)^2}\left[\frac{2}{3}\;
    N_\mathrm{F}-\frac{1}{2}\;N\left(\frac{13}{3}-\alpha\right)\right]
  \Gamma[\varepsilon]=Z_3^\mathrm{QCD}\;\;.
\end{equation}

\subsection{The Vertex Correction and NRQCD $\beta$ Function}
\label{sec:vertexcorrection}

Since they probe only the relativistic sector of the theory, the
renormalisation constants for gluons, ghosts and other light particles are the
same in QCD and NRQCD. The quark wave function renormalisation is computed in
the non-relativistic sector and for a non-relativistic bi-spinor rather than a
relativistic Dirac spinor, so that it is not surprising that the result
(\ref{z2}) differs from its QCD counterpart
$Z_2^\mathrm{QCD}=1-\frac{g_\mathrm{R}^2\;\mu^{2\varepsilon}}{(4\pi)^2}\;
C_\mathrm{F}\;\alpha\;\Gamma[\varepsilon]$~\cite[eq.\ (2.5.139)]{Muta} even in
the dependence on $\alpha$, although the MS scheme was used in both cases. As
both Lagrangeans are gauge invariant and agree in the light particle sector,
the Slavnov--Taylor identities of QCD must also hold in NRQCD. Therefore, the
renormalisation $Z_1$ of the quark-gluon vertex $Q^\dagger A_0 Q$ can be
inferred from $Z_2, Z_3$ and the three-gluon renormalisation $Z_{1,\,g}$ of
QCD (and NRQCD) not to be the same as in QCD where
$Z_1^\mathrm{QCD}=1-\frac{g_\mathrm{R}^2\;
  \mu^{2\varepsilon}}{8\pi^2}\;(C_\mathrm{F}\;\alpha+N\;\frac{3+\alpha}{4})
\;\Gamma[\varepsilon]$~\cite[eq.\ (2.5.145)]{Muta}:
\begin{equation}
  \label{z1}
  Z_1=\frac{Z_2\;Z_{1,\,g}}{Z_3}=1+\frac{g_\mathrm{R}^2\;
    \mu^{2\varepsilon}}{(4\pi)^2}\left(C_\mathrm{F}\left(3-\alpha\right)-
    \;N\;\frac{3+\alpha}{4}\right)\;\Gamma[\varepsilon]\;
  \not=Z_1^\mathrm{QCD}\;\;. 
\end{equation}
In the following, one will identify as the cause that all non-zero
contributions from both Abelian and non-Abelian vertex corrections to the
scalar gluon vertex in NRQCD are different from their QCD values: Most
notably, the non-Abelian vertex only provides gauge parameter corrections. As
a by-product, the topologies of all one loop corrections to interactions of
one gluon with a quark will be found and their leading order velocity power
counting determined.

Which vertices may be encountered in the graphs leading in $v$? For every
vertex, the $Q^\dagger A_0Q$ coupling is at least one order stronger in $v$
than any other coupling (tables \ref{scalarvertex} and \ref{vectorvertex}).
Any contribution is therefore suppressed in which scalar gluons in a vertex
are replaced with vector gluons, coupled either minimally or via the Fermi
term $\frac{gc_3}{2M}Q^\dagger\vec{\sigma}\cdot\vec{B}Q$ in (\ref{nrqcdlagr}).
Moreover, the Fermi term couples one or two gluons to the quark spin. When
only one Fermi interaction is found in any vertex correction, parity
conservation requires the strongest possible correction involving $A_0$ as
outgoing particle to be of the form of the spin-orbit term, i.e.\ proportional
to $Q^\dagger\vec{\sigma}\cdot(\vec{D}\times\vec{E})Q$. This cannot correct
the leading scalar gluon vertex. Two Fermi interactions are even more
suppressed because vector couplings are weaker than scalar couplings.

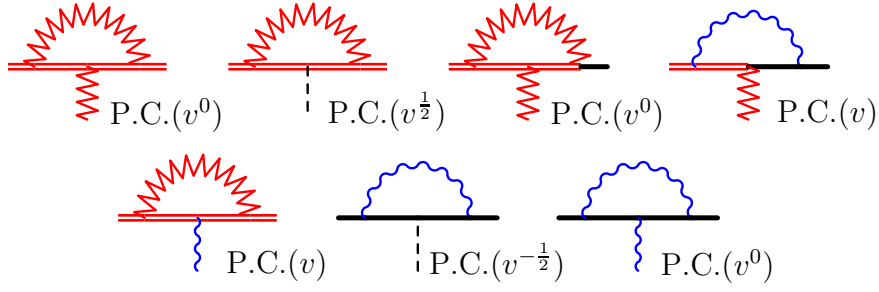
\begin{figure}[!htb]
  \vspace*{6ex}
  \begin{center}
    \feynbox{60\unitlength}{
      \begin{fmfgraph*}(60,20)
        \fmftop{i,o}
        \fmfbottom{u}
        \fmf{double,foreground=red,tension=2}{i,v1}
        \fmf{double,foreground=red}{v1,v2,v3}
        \fmf{double,foreground=red,tension=2}{v3,o}
        \fmffreeze
        \fmf{zigzag,foreground=red,left
          }{v1,v3}
        \fmf{zigzag,foreground=red
          }{v2,u}
        \fmfv{label=$\pc(v^0)$,label.angle=-90,label.dist=0.2w}{o}
      \end{fmfgraph*}}
    \hspace{3ex}
    \feynbox{60\unitlength}{
      \begin{fmfgraph*}(60,20)
        \fmftop{i,o}
        \fmfbottom{u}
        \fmf{double,foreground=red,tension=2}{i,v1}
        \fmf{double,foreground=red}{v1,v2,v3}
        \fmf{double,foreground=red,tension=2}{v3,o}
        \fmffreeze
        \fmf{zigzag,foreground=red,left}{v1,v3}
        \fmf{dashes}{v2,u}
        \fmfv{label=$\pc(v^\frac{1}{2})$,label.angle=-90,label.dist=0.16w}{o}
      \end{fmfgraph*}}
    \hspace{3ex}
    \feynbox{60\unitlength}{
      \begin{fmfgraph*}(60,20)
        \fmftop{i,o}
        \fmfbottom{u}
        \fmf{double,foreground=red,tension=2}{i,v1}
        \fmf{double,foreground=red}{v1,v2,v3}
        \fmf{vanilla,width=thick,tension=2}{v3,o}
        \fmffreeze
        \fmf{zigzag,foreground=red,left}{v1,v3}
        \fmf{zigzag,foreground=red}{v2,u}
        \fmfv{label=$\pc(v^0)$,label.angle=-90,label.dist=0.2w}{o}
      \end{fmfgraph*}}
    \hspace{3ex}
    \feynbox{60\unitlength}{
      \begin{fmfgraph*}(60,20)
        \fmftop{i,o}
        \fmfbottom{u}
        \fmf{double,foreground=red,tension=2}{i,v1}
        \fmf{double,foreground=red}{v1,v2}
        \fmf{vanilla,width=thick}{v2,v3}
        \fmf{vanilla,width=thick,tension=2}{v3,o}
        \fmffreeze
        \fmf{photon,foreground=blue,left}{v1,v3}
        \fmf{zigzag,foreground=red}{v2,u}
        \fmfv{label=$\pc(v)$,label.angle=-90,label.dist=0.2w}{o}
      \end{fmfgraph*}}
    \\[7ex]  
    \feynbox{60\unitlength}{
      \begin{fmfgraph*}(60,20)
        \fmftop{i,o}
        \fmfbottom{u}
        \fmf{double,foreground=red,tension=2}{i,v1}
        \fmf{double,foreground=red}{v1,v2,v3}
        \fmf{double,foreground=red,tension=2}{v3,o}
        \fmffreeze
        \fmf{zigzag,foreground=red,left}{v1,v3}
        \fmf{photon,foreground=blue}{v2,u}
        \fmfv{label=$\pc(v)$,label.angle=-90,label.dist=0.2w}{o}
      \end{fmfgraph*}}
    \hspace{3ex}
    \feynbox{60\unitlength}{
      \begin{fmfgraph*}(60,20)
        \fmftop{i,o}
        \fmfbottom{u}
        \fmf{vanilla,width=thick,tension=2}{i,v1}
        \fmf{vanilla,width=thick}{v1,v2,v3}
        \fmf{vanilla,width=thick,tension=2}{v3,o}
        \fmffreeze
        \fmf{photon,foreground=blue,left}{v1,v3}
        \fmf{dashes}{u,v2}
        \fmfv{label=$\pc(v^{-\frac{1}{2}})$,label.angle=-90,
          label.dist=0.16w}{o}
      \end{fmfgraph*}}
    \hspace{3ex}
    \feynbox{60\unitlength}{
      \begin{fmfgraph*}(60,20)
        \fmftop{i,o}
        \fmfbottom{u}
        \fmf{vanilla,width=thick,tension=2}{i,v1}
        \fmf{vanilla,width=thick}{v1,v2,v3}
        \fmf{vanilla,width=thick,tension=2}{v3,o}
        \fmffreeze
        \fmf{photon,foreground=blue,left}{v1,v3}
        \fmf{photon,foreground=blue}{v2,u}
        \fmfv{label=$\pc(v^0)$,label.angle=-90,label.dist=0.2w}{o}
      \end{fmfgraph*}}
    \end{center}
  \caption{\label{abelianvertex}\sl The Abelian vertex corrections of
    order \protect$g^3$ from the interactions $Q^\dagger A_0 Q$ which survive
    the diagrammatic filter. The leading order power counting is obtained when
    all vertices are the leading order scalar gluon interactions. After
    applying the rules of Sect.\ \ref{sec:diagrammar}, the seven diagrams
    drawn are the only survivors of 22 scale conserving graphs.}
\end{figure}

Each Abelian correction graph (Fig.\ \ref{abelianvertex}) starts off at the
same order as the bare graph whose vertex correction it presents (table
\ref{scalarvertex}) when only scalar gluons couple to the quark.
The only exception is the last diagram in the first row of Fig.\ 
\ref{abelianvertex}, which by the power counting is $\pc(v^1)$ while the bare
$\Qs^\dagger A_{\mathrm{s}\,0} \Qp$ vertex is $\pc(v^0)$ from table
\ref{scalarvertex}, as confirmed by an explicit computation of this graph in
App.\ \ref{sec:vertexcalc}. This diagram does therefore not enter in the
computation of the vertex correction at lowest order.

\begin{figure}[!htb]
  \vspace*{3ex}
  \begin{center}
    \feynbox{60\unitlength}{
      \begin{fmfgraph*}(60,40)
        \fmftop{i,o}
        \fmfbottom{u}
        \fmf{double,foreground=red,tension=4}{i,v1}
        \fmf{double,foreground=red}{v1,v3}
        \fmf{double,foreground=red,tension=4}{v3,o}
        \fmffreeze
        \fmf{zigzag,foreground=red
          }{v1,v2}
        \fmf{zigzag,foreground=red
          }{v2,v3}
        \fmf{zigzag,foreground=red,tension=3
          }{v2,u}
        \fmfv{label=$\pc(v)$,label.angle=-90,label.dist=0.5w}{o}
      \end{fmfgraph*}}
    \hspace{3ex}
    \feynbox{60\unitlength}{
      \begin{fmfgraph*}(60,40)
        \fmftop{i,o}
        \fmfbottom{u}
        \fmf{double,foreground=red,tension=4}{i,v1}
        \fmf{double,foreground=red}{v1,v3}
        \fmf{double,foreground=red,tension=4}{v3,o}
        \fmffreeze
        \fmf{zigzag,foreground=red}{v1,v2,v3}
        \fmf{dashes,tension=3}{u,v2}
        \fmfv{label=$\pc(v^\frac{3}{2})$,label.angle=-90,label.dist=0.47w}{o}
      \end{fmfgraph*}}
    \hspace{3ex}
    \feynbox{60\unitlength}{
      \begin{fmfgraph*}(60,40)
        \fmftop{i,o}
        \fmfbottom{u}
        \fmf{double,foreground=red,tension=4}{i,v1}
        \fmf{double,foreground=red}{v1,v3}
        \fmf{vanilla,width=thick,tension=4}{v3,o}
        \fmffreeze
        \fmf{zigzag,foreground=red}{v1,v2,v3}
        \fmf{zigzag,foreground=red,tension=3}{v2,u}
        \fmfv{label=$\pc(v)$,label.angle=-90,label.dist=0.5w}{o}
      \end{fmfgraph*}}
    \hspace{3ex}
    \feynbox{60\unitlength}{
      \begin{fmfgraph*}(60,40)
        \fmftop{i,o}
        \fmfbottom{u}
        \fmf{double,foreground=red,tension=4}{i,v1}
        \fmf{vanilla,width=thick}{v1,v3}
        \fmf{vanilla,width=thick,tension=4}{v3,o}
        \fmffreeze
        \fmf{zigzag,foreground=red}{v1,v2}
        \fmf{photon,foreground=blue}{v2,v3}
        \fmf{zigzag,foreground=red,tension=3}{v2,u}
        \fmfv{label=$\pc(v^2)$,label.angle=-90,label.dist=0.5w}{o}
      \end{fmfgraph*}}
    \\[5ex]
    \feynbox{60\unitlength}{
      \begin{fmfgraph*}(60,40)
        \fmftop{i,o}
        \fmfbottom{u}
        \fmf{double,foreground=red,tension=4}{i,v1}
        \fmf{double,foreground=red}{v1,v3}
        \fmf{double,foreground=red,tension=4}{v3,o}
        \fmffreeze
        \fmf{zigzag,foreground=red}{v1,v2,v3}
        \fmf{photon,foreground=blue,tension=3}{v2,u}
        \fmfv{label=$\pc(v^2)$,label.angle=-90,label.dist=0.5w}{o}
      \end{fmfgraph*}}
    \hspace{3ex}
    \feynbox{60\unitlength}{
      \begin{fmfgraph*}(60,40)
        \fmftop{i,o}
        \fmfbottom{u}
        \fmf{vanilla,width=thick,tension=4}{i,v1}
        \fmf{double,foreground=red}{v1,v3}
        \fmf{vanilla,width=thick,tension=4}{v3,o}
        \fmffreeze
        \fmf{zigzag,foreground=red}{v1,v2,v3}
        \fmf{dashes,tension=3}{u,v2}
        \fmfv{label=$\pc(v^\frac{1}{2})$,label.angle=-90,label.dist=0.46w}{o}
      \end{fmfgraph*}}
    \hspace{3ex}
    \feynbox{60\unitlength}{
      \begin{fmfgraph*}(60,40)
        \fmftop{i,o}
        \fmfbottom{u}
        \fmf{vanilla,width=thick,tension=4}{i,v1}
        \fmf{vanilla,width=thick}{v1,v3}
        \fmf{vanilla,width=thick,tension=4}{v3,o}
        \fmffreeze
        \fmf{dashes}{v1,v2}
        \fmf{photon,foreground=blue}{v2,v3}
        \fmf{dashes,tension=3}{u,v2}
        \fmfv{label=$\pc(v^\frac{3}{2})$,label.angle=-90,label.dist=0.46w}{o}
      \end{fmfgraph*}}
    \hspace{3ex}
    \feynbox{60\unitlength}{
      \begin{fmfgraph*}(60,40)
        \fmftop{i,o}
        \fmfbottom{u}
        \fmf{vanilla,width=thick,tension=4}{i,v1}
        \fmf{vanilla,width=thick}{v1,v3}
        \fmf{vanilla,width=thick,tension=4}{v3,o}
        \fmffreeze
        \fmf{photon,foreground=blue}{v1,v2,v3}
        \fmf{photon,foreground=blue,tension=3}{v2,u}
        \fmfv{label=$\pc(v)$,label.angle=-90,label.dist=0.5w}{o}
      \end{fmfgraph*}}
  \end{center}
  \caption{\label{nonabelianvertex}\sl The non-Abelian vertex corrections of
    order \protect$g^3$ which survive the diagrammatic filter. The soft blob in
    the second diagram in the second row provides an additional factor
    $\frac{1}{v}$ by loop power counting. The power counting indicated is the
    leading contribution in the Feynman gauge, when the three gluon vertex is
    the standard QCD one and both one vector and one scalar gluon couples
    minimally to the quark. In the Lorentz gauges, two scalar gluons couple to
    the quark, making all diagrams a factor $\frac{1}{v}$ stronger and hence
    contribute at the same order as the Abelian corrections in Fig.\ 
    \ref{abelianvertex}. Out of 25 scale-conserving diagrams, only eight
    survive the diagrammatic filter.}
\end{figure}
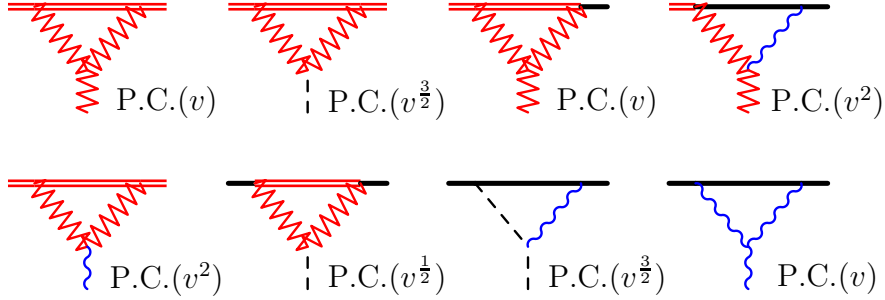

The non-Abelian corrections to the $Q^\dagger A_0 Q$ vertex at order $g^3$
fall into two categories: The topology of the first class of diagrams, Fig.\ 
\ref{nonabelianvertex}, is analogous to the one of the non-Abelian vertex
correction in QCD. In the Feynman gauge $\alpha=1$, its leading order
contribution involves two scalar gluons and one vector gluon because in that
gauge, different components of the gluon field do not mix when propagating,
(\ref{gpprop}), and the coupling of three scalar gluons is forbidden at order
$g$ by the Lorentz structure.
Again because for every vertex, the $Q^\dagger A_0Q$ coupling is stronger in
$v$ than any other coupling, the NRQCD analogue to the non-Abelian QCD vertex
correction to the scalar gluon vertex is sub-leading in the Feynman gauge. The
resulting power counting is reported in Fig.\ \ref{nonabelianvertex}. In the
generalised Lorentz gauges, the scalar and vector components of the gauge field
mix in the gluon propagation, with the amount of mixing proportional to
$(1-\alpha)$ (\ref{gpprop}). A scalar gluon emitted from a quark will therefore
partially turn into a vector gluon, which then enters a three gluon vertex. The
non-Abelian vertex corrections now have a non-trivial Lorentz structure and
evaluate to be proportional to $(1-\alpha)$ and $(1-\alpha)^2$, as is confirmed
in App.\ \ref{sec:vertexcalc}, (\ref{lastexample}). A factor $\frac{1}{v}$ is
gained with respect to the power counting with one vector gluon coupling as
given in Fig.\ \ref{nonabelianvertex}, making the diagrams of the same, leading
order as the Abelian vertex corrections. The non-Abelian graphs can
consequently be seen as merely providing gauge corrections to the Abelian
vertex result. The last diagram in the first row and the next-to-last in the
second row of Fig.\ \ref{nonabelianvertex} are sub-leading and vanish using the
equations of motion at lowest order for the outgoing potential quark, see
App.\ \ref{sec:vertexcalc}.

\begin{figure}[!htb]
  \vspace*{3ex}
  \begin{center}
    \feynbox{60\unitlength}{
      \begin{fmfgraph*}(60,40)
        \fmftop{i,o}
        \fmfbottom{u}
        \fmf{double,foreground=red}{i,v1,o}
        \fmffreeze
        \fmf{zigzag,foreground=red,left,label=$\fs\Av$,
          label.side=left}{v1,v2,v1}
        \fmf{zigzag,foreground=red,tension=3,label=$\fs A_\mu$,
          label.side=right}{v2,u}
        \fmfv{label=$\pc(v)$,label.angle=-90,label.dist=0.5w}{o}
      \end{fmfgraph*}}
    \hspace{3ex}
    \feynbox{60\unitlength}{
      \begin{fmfgraph*}(60,40)
        \fmftop{i,o}
        \fmfbottom{u}
        \fmf{double,foreground=red}{i,v1,o}
        \fmffreeze
        \fmf{zigzag,foreground=red,left}{v1,v2,v1}
        \fmf{dashes,tension=3}{u,v2}
        \fmfv{label=$\pc(v^\frac{3}{2})$,label.angle=-90,label.dist=0.46w}{o}
      \end{fmfgraph*}}
    \hspace{3ex}
    \feynbox{60\unitlength}{
      \begin{fmfgraph*}(60,40)
        \fmftop{i,o}
        \fmfbottom{u}
        \fmf{double,foreground=red}{i,v1}
        \fmf{vanilla,width=thick}{v1,o}
        \fmffreeze
        \fmf{zigzag,foreground=red,left}{v1,v2,v1}
        \fmf{zigzag,foreground=red,tension=3}{v2,u}
        \fmfv{label=$\pc(v)$,label.angle=-90,label.dist=0.5w}{o}
      \end{fmfgraph*}}
    \\[5ex]
    \feynbox{60\unitlength}{
      \begin{fmfgraph*}(60,40)
        \fmftop{i,o}
        \fmfbottom{u}
        \fmf{double,foreground=red}{i,v1,o}
        \fmffreeze
        \fmf{photon,foreground=blue,left}{v1,v2,v1}
        \fmf{photon,foreground=blue,tension=3}{v2,u}
        \fmfv{label=$\pc(v^3)$,label.angle=-90,label.dist=0.5w}{o}
      \end{fmfgraph*}}
    \hspace{3ex}
    \feynbox{60\unitlength}{
      \begin{fmfgraph*}(60,40)
        \fmftop{i,o}
        \fmfbottom{u}
       \fmf{vanilla,width=thick}{i,v1,o}
        \fmffreeze
        \fmf{zigzag,foreground=red,left}{v1,v2,v1}
        \fmf{dashes,tension=3}{u,v2}
        \fmfv{label=$\pc(v^\frac{1}{2})$,label.angle=-90,label.dist=0.46w}{o}
      \end{fmfgraph*}}
    \hspace{3ex}
    \feynbox{60\unitlength}{
      \begin{fmfgraph*}(60,40)
        \fmftop{i,o}
        \fmfbottom{u}
        \fmf{vanilla,width=thick}{i,v1,o}
        \fmffreeze
        \fmf{photon,foreground=blue,left}{v1,v2,v1}
        \fmf{photon,foreground=blue,tension=3}{v2,u}
        \fmfv{label=$\pc(v^2)$,label.angle=-90,label.dist=0.5w}{o}
      \end{fmfgraph*}}
  \end{center}
  \caption{\label{nonabelianvertexa2}\sl The only non-zero diagrams of the
    vertex corrections of order \protect$g^3$ from the Fermi term proportional
    to $c_3$ in the NRQCD Lagrangean (\ref{nrqcdlagr}). The soft blob in the
    next-to-last diagram provides an additional factor $\frac{1}{v}$ by loop
    power counting. The leading order power counting is obtained when the
    three gluon vertex is the standard QCD one and the gluons couple to the
    quark via the Fermi term. Out of 24 scale-conserving diagrams, only six
    survive the diagrammatic filter.}
\end{figure}
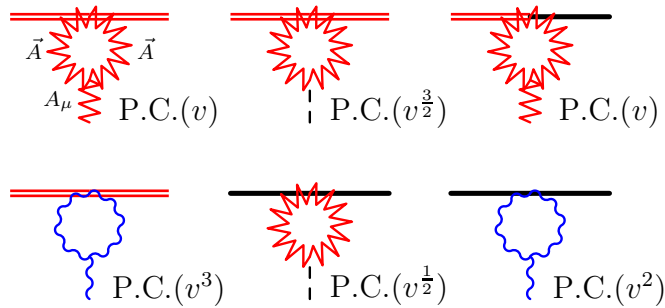

There is no comparable graph in QCD to the second category of non-Abelian
vertex corrections in NRQCD at order $g^3$. Figure \ref{nonabelianvertexa2}
lists the topologies of all non-zero corrections to one gluon vertices
involving one gluon loop.
Again, one convinces oneself easily from tables \ref{threegluonvertex} and
\ref{twogluetwoquarkvertex} that these diagrams are at least one order in $v$
weaker than the Abelian vertex corrections, even before the precise vertex
structure is specified. In NRQED, this type of diagrams occurs first at order
$g^5$ from the $e_2$ term in the Lagrangean (\ref{nrqcdlagr}) because a three
photon coupling does not occur earlier. In QCD, the lowest order three gluon
coupling is totally antisymmetric in colour space, so that its contraction with
the totally symmetric expression in colour space from the vertex
$\frac{g^2}{2M}Q^\dagger\Av^2Q$ vanishes. The Fermi interaction has a colour
antisymmetric two gluon part because
$B^a_i=\frac{1}{2}\epsilon^{ijk}(\de_jA^a_k-\de_kA^a_j+gf^{abc}A^b_jA^c_k)$,
but is not a possible source of an order $g^3$ vertex correction in
Fig.~\ref{nonabelianvertexa2} to the minimal coupling of scalar gluons and
quarks because the spin structures do not match.

Hence, the non-Abelian nature of the gauge field enters into the NRQCD $\beta$
function only via the gluon vacuum polarisation and as what may be called
gauge correction represented by non-Abelian vertex corrections. Each of the
six quark gluon couplings between the various r\'egimes is renormalised by one
and only one non-zero Abelian and non-Abelian vertex correction.

Since the classification of diagrams following the rules of Sect.\ 
\ref{sec:diagrammar} is insensitive to the precise nature of the vertex
coupling, the list leads as a by-product to a classification of all
non-zero one loop graphs in which two quarks and one gluon couple. The same is
true for the quark self energy and vacuum polarisation graphs. The $v$ power
counting of the diagrams in Figs.\ \ref{abelianvertex}, \ref{nonabelianvertex}
and \ref{nonabelianvertexa2} is specific to the interactions chosen, but the
fact that all other graphs with the same topology are zero is not.

All Abelian graphs lead to the same vertex vertex correction, as
demonstrated in App.\ \ref{sec:vertexcalc} at exemplary graphs:
\begin{equation}
  \label{abeliangamma}
  -\ii\;g_\mathrm{R}\;\mu^\varepsilon\;t^a\;
  \Gamma^\mathrm{Abelian}|_\mathrm{pole}=
  \frac{\ii\;g^3_\mathrm{R}\;\mu^{3\varepsilon}}{8\pi^2}
  \left(C_\mathrm{F}-\frac{N}{2}\right)\;t^a\;\frac{3-\alpha}{2}\;
  \Gamma[\varepsilon]
\end{equation}
Likewise for the non-Abelian vertex corrections:
\begin{equation}
  \label{nonabeliangamma}
  -\ii\;g_\mathrm{R}\;\mu^\varepsilon\;t^a\;
  \Gamma^\mathrm{non-Abelian}|_\mathrm{pole}=
  \frac{-\ii\;g^3_\mathrm{R}\;\mu^{3\varepsilon}}{(4\pi)^2}\;\frac{N}{2}\;t^a\;
  \frac{3(1-\alpha)}{2}\;\Gamma[\varepsilon]
\end{equation}
The vertex renormalisation in the MS scheme, 
\begin{equation}
  Z_1=1-\Gamma^\mathrm{Abelian}|_\mathrm{pole}-
  \Gamma^\mathrm{non-Abelian}|_\mathrm{pole}\;\;,
\end{equation}
coincides therefore with the one expected from the Slavnov--Taylor identities
(\ref{z1}).

The NRQCD $\beta$ function is computed from the scale invariance of the bare
coupling
\begin{equation}
  \label{couplingren}
  g=g_\mathrm{R}\;\mu^\varepsilon\;\frac{Z_1}{Z_2\;\sqrt{Z_3}}
\end{equation}
and agrees with the anticipated result (\ref{qcdbetafunction}),
\begin{equation}
  \label{nrqcdbetafunction}
  \beta_\mathrm{NRQCD}=-\;\frac{g_\mathrm{R}^3}{(4\pi)^2}\;
  \left[\frac{11}{3}\;N-\frac{2}{3}\;N_\mathrm{F}\right]=\beta_\mathrm{QCD}
  \;\;.
\end{equation}
To the order calculated, one may ``match'' the relativistic and
non-relativistic $\beta$ functions to infer that indeed to lowest order in $g$
and $v$,
\begin{equation}
  \label{trivia}
  g^\mathrm{QCD}_\mathrm{R}=g^\mathrm{NRQCD}_\mathrm{R}\;\;.
\end{equation}
To match NRQCD to QCD via the renormalisation group behaviour of its couplings
and operators is of course considerably more complicated than to match matrix
elements. Nonetheless, it demonstrates that matching relations survive
renormalisation. The renormalised NRQCD coupling is as expected the same in all
r\'egimes, gauge invariant, and runs simultaneously in all r\'egimes. The NRQCD
$\beta$ function is independent of the gauge parameter $\alpha$. The Lorentz
gauges are a legitimate gauge choice in NRQCD, obeying the Slavnov--Taylor
identities. Wave function and vertex renormalisations differ in QCD and NRQCD,
while the $\beta$ functions do not.

In dimensionally regularised NRQCD, the vacuum polarisation receives its sole
contribution from the propagation of on-shell relativistic particles, i.e.\ of
physical soft or ultrasoft gluons and soft or ultrasoft ghosts and light
quarks. Potential gluons, i.e.\ such mediating the Coulomb interaction, do not
give rise to a gluon renormalisation, nor do they contribute to \emph{any}
other renormalisation at lowest order in $v$, (\ref{selfpot}) and Figs.\ 
\ref{softselffilter}, \ref{abelianvertex} and \ref{nonabelianvertex}. This
observation applies to all standard gauges, including physical ones like the
Coulomb gauge. It is traced back to the homogene{\ia}ty of dimensional
regularisation (\ref{masterformula}), which in turn is well known to be
connected to a cancellation of ultraviolet and infrared divergences in
scale-less integrals, e.g.~\cite[p.\ 172 f.]{Muta}. In a cut-off
regularisation, the situation is different as massless tadpoles require both
infrared and ultraviolet regularisation resulting in logarithms, invalidating
(\ref{masterformula}) and all diagrammatic rules, e.g.~\cite[Chap.\ 18.5]{Lee}.
In such a regularisation, the Coulomb gluon vacuum polarisation can therefore
indeed receive its main contribution from Coulomb gluons in the intermediate
state.

To close this section, mind that this derivation of the NRQCD $\beta$ function
is slightly simpler than its QCD counterpart but may be applied with equal
generality. Except in the determination of the gluon vacuum polarisation, the
Lorentz structure of the vertices and propagators did not enter. The Dirac
algebra is not needed except when light fermions are included, and then the
manipulations are straightforward. The vertex corrections in the Feynman gauge
reduce to the computation of one graph: the Abelian vertex which involves
neither Lorentz nor Dirac indices. Non-Abelian vertex corrections are easily
extracted, see App.\ \ref{sec:vertexcalc}. The presentation demonstrates also
intuitively that quarks with mass $M$ higher than the scale at which the
$\beta$ function is computed freeze out and are not to be included in the light
quark number $N_\mathrm{F}$ of (\ref{nrqcdbetafunction}). In a world with
$N_\mathrm{f}$ quarks, the QCD $\beta$ function can hence be found from NRQCD
with $N_\mathrm{f}=N_\mathrm{F}$ ``light'' quarks and one fictitious quark with
a mass much bigger than any real quark mass.

\subsection{The Coulomb Gauge $\beta$ Function}
\label{sec:coulombbeta}

With the gauge independent classification of all one-loop self-energy, vacuum
polarisation and vertex correction diagrams completed, it is straightforward to
calculate the $\beta$ function in the Coulomb gauge from the renormalisation of
the scalar gluon field. In order to classify the contributing diagrams, one
combines the notion that in the Coulomb gauge, $A_0$ is only potential (Sect.\ 
\ref{sec:props}) with the knowledge of the topology of the non-zero one loop
diagrams as reported in (\ref{selfpot}/\ref{nrqcdselfsoft}/\ref{vacpolpot}) and
Figs.\ \ref{abelianvertex} to \ref{nonabelianvertexa2}. As noted in the
previous section, the leading order contribution in each self energy, vacuum
polarisation or vertex correction graph is obtained when only scalar gluons
couple to quarks, and when all gluons are on shell, i.e.\ either soft or
ultrasoft. Because for the Coulomb gauge, this eliminates any correction which
is of the same order as the quark propagator or the $Q^\dagger A_0Q$ vertex
(table \ref{scalarvertex}), the soft and potential quark self energy and all
vertex corrections, Abelian or non-Abelian, are therefore zero \emph{to all
  orders} in $g$ and $v$:
\begin{equation}
  \label{coulombz1z2}
  Z_1^\mathrm{Coulomb}=1\;\;,\;\;Z_2^\mathrm{Coulomb}=1\;\;.
\end{equation}
The coupling constant is hence renormalised only because the potential gluon
vacuum polarisation in the Coulomb gauge is non-zero. A computation of the
gluonic part which makes again use of split dimensional regularisation can be
found in \cite[eq.\ (41)]{LeibbrandtWilliams}:
\begin{equation}
  \label{coulombvacpol}
  \feynbox{60\unitlength}
  {\begin{fmfgraph*}(60,40)
    \fmfleft{i}
    \fmfright{o}
    \fmfv{label=$\fs A_0$,label.angle=90}{i}
    \fmfv{label=$\fs A_0$,label.angle=90}{o}
    \fmf{dashes,tension=4}{i,v1}
    \fmf{dashes,tension=4}{v2,o}
    \fmf{zigzag,foreground=red,left,label=$\fs \Av$,label.side=left}{v1,v2,v1}
  \end{fmfgraph*}}
  \;\;:\;\Pi_{00}^{ab,\,\mathrm{Coulomb}}(k)=
  \frac{\ii\;g_\mathrm{R}^2\;\mu^{2\varepsilon}}{(4\pi)^2}\;\delta^{ab}\;
  \kv^2\;\frac{11}{3}\;N\;\Gamma[\varepsilon]+\mbox{ finite}
\end{equation}
The ghost part is zero because ghosts do not propagate. The contribution for
light fermions is gauge independent and can hence be extracted from the Lorentz
gauge result (\ref{z3}). One arrives finally at
\begin{equation}
  \label{coulombz3}
  Z_3^\mathrm{Coulomb}=1-\frac{g_\mathrm{R}^2\;\mu^{2\varepsilon}}{(4\pi)^2}
  \left[\frac{2}{3}\;N_\mathrm{F}-\frac{11}{3}\;N\right]
  \Gamma[\varepsilon]\;\;.
\end{equation}
Not surprisingly, the NRQCD $\beta$ function is extracted as
(\ref{nrqcdbetafunction}) using (\ref{couplingren}). This particular part of
renormalising NRQCD in the Coulomb gauge may appear simpler than in the Lorentz
gauges, but other parts suffer severely from technical problems like the
non-transversality of the gluon propagator.

\section{Conclusions and Outlook}
\label{sec:conclusions}
\setcounter{equation}{0}

After presenting the extension to NRQCD of a recently proposed explicit
velocity power counting scheme for a non-relativistic toy field
theory~\cite{hgpub3} following Beneke and Smirnov's threshold
expansion~\cite{BenekeSmirnov}, this article presented the computation of the
NRQCD $\beta$ function to lowest order in $g$ and $v$ in the Lorentz gauges and
in the Coulomb gauge. It endorsed the relevance of a new quark representative
and a new gluon representative in the soft scaling r\'egime $E\sim |\pv|\sim
Mv$ in which quarks are static and gluons on shell. HQET becomes a sub-set of
NRQCD. The identification of three different r\'egimes of scale for on-shell
particles from the poles of the NRQCD propagators leads in a natural way to
this r\'egime, in addition to the well known potential one with on-shell quarks
and instantaneous gluons mediating the Coulomb binding~\cite{LukeManohar} and
the ultrasoft one with bremsstrahlung gluons~\cite{GrinsteinRothstein}. Neither
of the five fields in the three r\'egimes should be thought of as ``physical
particles''. Rather, they represent the ``true'' quark and gluon in the
respective r\'egimes as the infrared-relevant degrees of freedom. None of the
r\'egimes overlap. Using a rescaling
technique~\cite{LukeManohar,LukeSavage,GrinsteinRothstein}, Sect.\ 
\ref{sec:velocitypowercounting} proposed an NRQCD Lagrangean which leads to the
correct behaviour of scattering and production amplitudes~\cite{hgpub3}. It
establishes explicit velocity power counting, preserved to all orders in
perturbation theory once dimensional regularisation is chosen to complete the
theory. The reason is the non-commutativity of the expansion in small
parameters with dimensionally regularised integrals and the homogene{\ia}ty of
dimensional regularisation.

There is an intimate relation between the recently discovered threshold
expansion~\cite{BenekeSmirnov} and this version of NRQCD which uses dimensional
regularisation and has been developed over the past
years~\cite{LukeManohar,hgpub3,LukeSavage,GrinsteinRothstein}. This was
demonstrated at the outset of the computation of the $\beta$ function in
Sect.~\ref{sec:nrqcdandthreshold} and has already led to fruitful exchange. As
noted by Beneke and Smirnov~\cite{BenekeSmirnov}, threshold expansion provides
an efficient way to derive the NRQCD Lagrangean (\ref{nrqcdlagr}) from QCD,
deepening our understanding of effective field theories: The coefficients
$c_i,d_i,e_i$ are obtained by computing QCD diagrams and sub-diagrams in the
hard r\'egime, $q\sim M$. Threshold expansion also shows how to systematise
identifying the different kinematic r\'egimes of NRQCD. Indeed, NRQCD was
incomplete without the soft r\'egime and its corresponding degrees of freedom
as pointed out by Beneke and Smirnov~\cite{BenekeSmirnov}. One might therefore
see threshold expansion and NRQCD as two sides of the same coin. The one
expands loop integrals to establish a power counting for Feynman diagrams,
while the other uses rescaling properties to classify the relative strengths of
all vertices, allowing for intuitive interpretations of the derived vertex
power counting rules on the level of the Lagrangean (see the end of Sect.\ 
\ref{sec:vertex}). An effective field theory formulation can also be applied
more easily to bound state problems. Finally, while threshold expansion has not
yet been proven to be valid on a formal level, the existence of NRQCD as an
effective field theory is guaranteed by the renormalisation group,
e.g.~\cite[Chap.\ 8]{Collins}.

The calculation of the NRQCD $\beta$ function in the Lorentz gauges to one loop
order in Sect.\ \ref{sec:betafunction} -- albeit its result was easily
anticipated -- was non-trivial as it demonstrated that the power counting
proposed is consistent also after renormalisation. Because of the splitting of
quark and gluon fields in different representatives in the various r\'egimes,
this might be thought not to be straightforward. It also resolved a puzzle
about the running of the coupling of gluons mediating the Coulomb interaction.
The rescaling and power counting rules are gauge independent for a wide class
of gauges including all standard gauges. Bare graphs and their corrections are
of the same power in $v$. The soft r\'egime is necessary not only to render the
same $\beta$ function as in QCD, but even to attain a result which is
independent of the gauge parameter $\alpha$ at all and in which the
renormalisation constants obey the Slavnov--Taylor identities. In the Coulomb
gauge calculation of the $\beta$ function, the quark self energy and the vertex
correction of the vertex coupling a quark minimally to a scalar gluon is easily
shown to be zero to all orders in $g$ and $v$. In the Coulomb gauge, the vacuum
polarisation is therefore the only non-trivial renormalisation necessary for
the $\beta$ function.

The derivation was greatly simplified by diagrammatic rules (Sect.\ 
\ref{sec:diagrammar}) which follow from the homogene{\ia}ty property of
dimensional regularisation and allow one to systematically recognise the
majority of NRQCD graphs as zero to all orders in $v$ just by drawing them,
independently of the gauge and of details of the vertices involved. The rules
apply equally well to threshold expansion and will prove fruitful in more
rigorous investigations, also into NRQCD. Here, they were already used to
prove scale conservation and the decoupling of the ultrasoft, heavy quark in
Sect.\ \ref{sec:theorems}. All topologies of one loop corrections to vertices
involving one gluon and one quark were also classified, and the power counting
of the leading order contributions determined.

Dimensional regularisation in NRQCD has many advantages over other
regularisation schemes in which scale-less integrals are non-zero. It preserves
the symmetries and guarantees the tree level power counting to be corrected at
most by logarithms. It is well known that cut-off regularisation violates gauge
symmetry and wants removal of power divergences in the renormalisation process
in order to conserve the tree level power counting to all orders. When massless
tadpoles do not vanish, the number of diagrams to be computed in order to
obtain even a simple vertex correction was seen to increase from one in
dimensional regularisation to three or more. In general, it is not clear
whether scale conservation holds in cut-off regularisation: The three different
r\'egimes require the introduction of at least three artificial cut-off scales
to separate them and to cure ultraviolent divergences. Furthermore, the
ultrasoft quark does not decouple, adding further diagrams. In the end, the
answers of all well-defined regularisation schemes do agree, but dimensional
regularisation provides an elegant way which furthermore often leads to
analytic expressions. A disadvantage of dimensional regularisation is that in
its standard formulation, it cannot be used to explore the non-perturbative
sector of NRQCD.

Finally, NRQCD with dimensional regularisation shows how to establish a power
counting in any effective field theory with several low energy scales: The
theory is first divided into one sector containing the particles relativistic
at a given heavy scale and another sector containing the non-relativistic
particles. In NRQCD, the former are light quarks, gluons and ghosts, the latter
the heavy quark. The effect of all physics at larger scales is absorbed into
the coefficients of the Lagrangean. Then, one identifies the combinations of
scales in which particles become on shell by looking at the denominators of the
various propagators. This gives the scaling r\'egimes. Finally, the Lagrangean
is rescaled to dimensionless fields in each r\'egime to exhibit the vertex and
loop power counting rules. A priori, all couplings obeying scale conservation
are allowed. The diagrammatic rules developed in Sect.\ \ref{sec:diagrammar}
still hold. Non-relativistic effective nuclear field theory is one obvious
application with three low energy scales: the pion mass and production
threshold, and the anomalously large scattering length of the two body
system~\footnote{Following this article, Mehen and Stewart investigated the
  soft r\'egime in effective nuclear theory~\cite{MehenStewartradiation}.}.

Among the points to be addressed further is an extension of the diagrammatic
rules. An investigation of the influence of soft quarks and gluons on bound
state calculations in NRQED and NRQCD is also important because -- as seen at
the end of Sect.\ \ref{sec:loop} -- their contribution at $\calO(g^4)$ and
higher becomes stronger than retardation effects. To investigate the
non-perturbative sector of NRQCD and to establish its equivalence to QCD in
the continuum is another formidable task.


\section*{Acknowledgements} 
I am indebted to M.\ Luke for initial discussions about the problem of the
$\beta$ function in power counted NRQCD out of which this article grew, and to
M.\ Burkardt, J.\ Gasser, G.\ P.\ Lepage, H.\ Reinhardt, A.\ Pineda, M.\ J.\ 
Savage and J.\ Soto for valuable further exchanges and discussions. The work
was supported in part by a Department of Energy grant DE-FG03-97ER41014.

\newpage


\begin{appendix}

\section{Some Details on Split Dimensional Regularisation}
\label{app:splitdimreg}
\setcounter{equation}{0}

\subsection{Useful Split Dimensionally Regularised Integrals}
\label{sec:splitdimreg}

This appendix presents some useful formulae for non-covariant loop integrals
using split dimensional regularisation as introduced by Leibbrandt and
Williams~\cite{LeibbrandtWilliams}, see also the Appendix in~\cite{hgpub3}. In
its results, split dimensional regularisation agrees with other methods to
compute loop integrals in non-covariant gauges, such as the non-principal
value prescription~\cite{LeeNyeo}, but two features make it especially
attractive: It treats the temporal and spatial components of the loop
integrations on an equal footing, and no recipes are necessary to deal with
e.g.\ pinch singularities. Rather, it uses the fact that, like in ordinary
integration, the axioms of dimensional regularisation~\cite[Chap.\ 
4.1]{Collins} allow to split the integration into two separate integrals:
\begin{equation}
  \label{splitdimreg}
  \int\deintdim{d}{k}=\int\deintdim{\sigma}{k_0}\deintdim{d-\sigma}{\kv}
\end{equation}
Where applicable, split and ordinary dimensional regularisation of covariant
integrals must hence agree once the limit $\sigma\to 1$ is taken with $d$
still arbitrary.

For the quark self energy, the integral
\begin{equation}
  \int\deintdim{d}{q}
  \frac{1}{(q^2+\ii\epsilon)^\alpha} \;\frac{1}{q_0+a+\ii\epsilon}
\end{equation}
is most easily computed by combining denominators as in HQET,
\begin{equation}
  2 \alpha\int\limits_0^\infty \deint{}{\lambda}\int\deintdim{d}{q} \left[
    q^2 + 2\lambda(q_0+a)+\ii\epsilon\right]^{-(\alpha+1)}\;\;.
\end{equation}
Now, the integrand is in a standard covariant form and can be computed using
Minkowski space methods with the result
\begin{equation}
  \label{firstmaster}
  \int\deintdim{d}{q}
  \frac{1}{(q^2+\ii\epsilon)^\alpha}\;\frac{1}{q_0+a+\ii\epsilon}
  =\frac{2\;\ii\;\Gamma[\frac{d}{2}-\alpha]\;
    \Gamma[2\alpha+1-d]}{(4\pi)^\frac{d}{2}\;\Gamma[\alpha]} 
  \left(2a+\ii\epsilon\right)^{d-(2\alpha+1)}\;
  \e^{\ii\pi(\frac{d}{2}-\alpha)}\;\;.
\end{equation}
By differentiating with respect to $a$, the following formula useful for the
computation of the vertex corrections is derived:
\begin{equation}
  \label{secondmaster}
  \int\deintdim{d}{q}
  \frac{1}{(q^2+\ii\epsilon)^\alpha}\;\frac{1}{(q_0+a+\ii\epsilon)^\beta}=
  \frac{2\;\ii\;\Gamma[\frac{d}{2}-\alpha]\;\Gamma[2\alpha+\beta-d]}{
    (4\pi)^\frac{d}{2}\;\Gamma[\alpha]\;\Gamma[\beta]}
  \left(2a+\ii\epsilon\right)^{d-2\alpha-\beta}\;
  \e^{\ii\pi(\frac{d}{2}-\alpha)}
\end{equation}
Another generalisation is easiest computed by first combining the denominators
linear in $q_0$ using Feynman parameters,
\begin{equation}
  \frac{1}{q_0+a+\ii\epsilon}\;\frac{1}{q_0+b+\ii\epsilon}=
  \int\limits_0^1\deint{}{x}\frac{1}{\left[q_0+x a+(1-x)b+\ii\epsilon
    \right]^2}\;\;,
\end{equation}
and employing (\ref{secondmaster}):
\begin{eqnarray}
  \label{thirdmaster}
  \lefteqn{\int\deintdim{d}{q}
  \frac{1}{(q^2+\ii\epsilon)^\alpha}\;
  \frac{1}{q_0+a+\ii\epsilon}\;\frac{1}{q_0+b+\ii\epsilon}=}\\
  &&=\frac{-2\;\ii\;\Gamma[\frac{d}{2}-\alpha]\;\Gamma[2\alpha+1-d]}{
    (4\pi)^\frac{d}{2}\;\Gamma[\alpha]}\;
  \frac{(2a+\ii\epsilon)^{d-2\alpha-1}-(2b+\ii\epsilon)^{d-2\alpha-1}}{a-b}\;
  \e^{\ii\pi(\frac{d}{2}-\alpha)}\non
\end{eqnarray}
Note that in all three integrals
(\ref{firstmaster}/\ref{secondmaster}/\ref{thirdmaster}), a shift of $\qv$ by
an arbitrary value leaves the result unaffected. A scale in the momentum
integration is only induced by the scales $a,b$ of the energy integration.

\subsection{Computation of Exemplary Vertex Corrections}
\label{sec:vertexcalc}

In order to illustrate the NRQCD power counting scheme further, this appendix
outlines the computation of characteristic Abelian and non-Abelian vertex
corrections in Sect.\ \ref{sec:vertexcorrection}.

The Feynman rules give for the Abelian correction to the $\Qs^\dagger
A_{\mathrm{s},0}\Qs$
\begin{equation}
  \rule[-15pt]{0pt}{58pt}
  \feynbox{60\unitlength}{
      \begin{fmfgraph*}(60,20)
        \fmftop{i,o}
        \fmfv{label=$\fs (T,,\pv)$}{i}
        \fmfv{label=$\fs (T^\prime,,\pv^\prime)$}{o}
        \fmfbottom{u}
        \fmfv{label=$\fs {(T-T^\prime,,\pv-\pv^\prime)}$}{u}
        \fmf{double,foreground=red,tension=2}{i,v1}
        \fmf{double,foreground=red}{v1,v2,v3}
        \fmf{double,foreground=red,tension=2}{v3,o}
        \fmffreeze
        \fmf{zigzag,foreground=red,left,label=$\fs q$,label.side=left}{v1,v3}
        \fmf{zigzag,foreground=red}{v2,u}
      \end{fmfgraph*}}
    \;\;:\;
    (-\ii\;g_\mathrm{R})^3\;\mu^{3\varepsilon}\;t^b t^a t^c\;
    \int\deintdim{d}{q}\frac{-\ii\delta^{bc}}{q^2}
      \left[1-(1-\alpha)\frac{q^2_0}{q^2}\right]
    \frac{\ii}{T+q_0}\;\frac{\ii}{T^\prime+q_0}\;\;.
\end{equation}
As usual, $t^b t^a t^b=(C_\mathrm{F}-\frac{N}{2})t^a$. The integral containing
$q_0^2$ in the numerator is simplified by re-writing it as
\begin{equation}
  \label{numeratortreatment}
  \frac{q^2_0}{(T+q_0)(T^\prime+q_0)}=1+\;
  \frac{T\,T^\prime}{(T+q_0)(T^\prime+q_0)}\;-\;\frac{T}{(T+q_0)}\;-
  \;\frac{T^\prime}{(T^\prime+q_0)}
\end{equation}
and noting that the first integral is zero as it is without scale
(\ref{masterformula}). The remaining integrals are using
(\ref{firstmaster}/\ref{thirdmaster})
\begin{equation}
  \label{sssvertex}
  \frac{\ii\;g^3_\mathrm{R}\mu^{3\varepsilon}}{8\pi^2}
  \left(C_\mathrm{F}-\frac{N}{2}\right)\;t^a\;\frac{3-\alpha}{2}\;
  \Gamma[\varepsilon]+\mbox{ finite,}
\end{equation}
from which (\ref{abeliangamma}) is read up. The Abelian correction to
$\Qp^\dagger A_{\mathrm{p},0}\Qp$,
\begin{eqnarray}
  \lefteqn{
  \rule[-20pt]{0pt}{60pt}
  \feynbox{60\unitlength}{
      \begin{fmfgraph*}(60,20)
        \fmftop{i,o}
        \fmfv{label=$\fs (T,,\pv)$}{i}
        \fmfv{label=$\fs (T^\prime,,\pv^\prime)$}{o}
        \fmfbottom{u}
        \fmfv{label=$\fs {(T-T^\prime,,\pv-\pv^\prime)}$}{u}
        \fmf{vanilla,width=thick,tension=2}{i,v1}
        \fmf{vanilla,width=thick}{v1,v2,v3}
        \fmf{vanilla,width=thick,tension=2}{v3,o}
        \fmffreeze
        \fmf{photon,foreground=blue,left,label=$\fs q$,label.side=left}{v1,v3}
        \fmf{dashes}{u,v2}
      \end{fmfgraph*}}
     \;\;:}\\
   &&\;\;(-\ii\;g_\mathrm{R})^3\;\mu^{3\varepsilon}\;t^b t^a t^c\;
    \int\deintdim{d}{q}\frac{-\ii\delta^{bc}}{q^2}
      \left[1-(1-\alpha)\frac{q^2_0}{q^2}\right]
    \frac{\ii}{T+q_0-\frac{\pv^2}{2M}}\;
    \frac{\ii}{T^\prime+q_0-\frac{\pv^{\prime\,2}}{2M}}\;\;,
    \non
\end{eqnarray}
is inferred by replacing $T\to T-\frac{\pv^2}{2M}$ and $T^\prime\to
T^\prime-\frac{\pv^{2\,\prime}}{2M}$ in (\ref{sssvertex}). Again,
$\Gamma^\mathrm{Abelian}$ is extracted as quoted in (\ref{abeliangamma}). Note
that -- if one were to employ the equations of motion for the outgoing
particles -- the integral would appear to vanish as no scale is present. This
would mean that there is no Abelian vertex correction to the potential (or
ultrasoft) scalar gluon vertex. In that case, the result for the NRQCD $\beta$
function is not even gauge invariant. Since the diagram is the leading order
contribution to the correction of the $\Qp^\dagger A_{\mathrm{p},0}\Qp$ vertex
(Fig.\ \ref{abelianvertex}), the application of the equations of motion is not
legitimate.

\absatz As noted in Sect.\ \ref{sec:vertexcorrection}, there are two Abelian
corrections to $\Qs^\dagger A_{\mathrm{s},0}\Qp$. The leading one is
\begin{equation}
  \rule[-22pt]{0pt}{60pt}
  \feynbox{60\unitlength}{
      \begin{fmfgraph*}(60,20)
        \fmftop{i,o}
        \fmfv{label=$\fs (T,,\pv)$}{i}
        \fmfv{label=$\fs (T^\prime,,\pv^\prime)$}{o}
        \fmfbottom{u}
        \fmfv{label=$\fs {(T-T^\prime\to T,,\pv-\pv^\prime)}$}{u}
        \fmf{double,foreground=red,tension=2}{i,v1}
        \fmf{double,foreground=red}{v1,v2,v3}
        \fmf{vanilla,width=thick,tension=2}{v3,o}
        \fmffreeze
        \fmf{zigzag,foreground=red,left,label=$\fs q$,label.side=left}{v1,v3}
        \fmf{zigzag,foreground=red}{v2,u}
      \end{fmfgraph*}}
    \;\;:\;\;(-\ii\;g_\mathrm{R})^3\;\mu^{3\varepsilon}\;t^b t^a t^c\;
    \int\deintdim{d}{q}\frac{-\ii\delta^{bc}}{q^2}
      \left[1-(1-\alpha)\frac{q^2_0}{q^2}\right]\frac{\ii}{q_0}\;
    \frac{\ii}{T+q_0}\;\;.
\end{equation}
The integration (\ref{firstmaster}/\ref{thirdmaster}) gives the expected
Abelian contribution (\ref{abeliangamma}) to the vertex normalisation
(\ref{z1}).

The vertex correction already discussed in connection with the cutting rule,
(\ref{sspvertex2}), is according to Sec.\ \ref{sec:vertexcorrection} of order
$v^1$ and hence not the leading contribution. Because
$T^\prime\sim\frac{\pv^{\prime\,2}}{2M}\sim Mv^2$ and $T\sim|\pv|\sim Mv$, one
indeed finds
\begin{eqnarray}
  \rule{0pt}{40pt}
  \feynbox{60\unitlength}{
      \begin{fmfgraph*}(60,20)
        \fmftop{i,o}
        \fmfv{label=$\fs (T,,\pv)$}{i}
        \fmfv{label=$\fs (T^\prime,,\pv^\prime)$}{o}
        \fmfbottom{u}
        \fmfv{label=$\fs {(T,,\pv-\pv^\prime)}$}{u}
        \fmf{double,foreground=red,tension=2}{i,v1}
        \fmf{double,foreground=red}{v1,v2}
        \fmf{vanilla,width=thick}{v2,v3}
        \fmf{vanilla,width=thick,tension=2}{v3,o}
        \fmffreeze
        \fmf{photon,foreground=blue,left,label=$\fs q$,label.side=left}{v1,v3}
        \fmf{zigzag,foreground=red}{v2,u}
      \end{fmfgraph*}}
    &:&(-\ii\;g_\mathrm{R})^3\;\mu^{3\varepsilon}\;t^b t^a t^c\;
    \int\deintdim{d}{q}\frac{-\ii\delta^{bc}}{q^2}
    \left[1-(1-\alpha)\frac{q_0^2}{q^2}\right]\frac{\ii}{T}\;
    \frac{\ii}{T^\prime+q_0-\frac{\pv^{\prime\,2}}{2M}}=\non\\
    &&\!\!\!\!\!\!=\frac{\ii\;g^3_\mathrm{R}\mu^{3\varepsilon}}{8\pi^2}
  \left(C_\mathrm{F}-\frac{N}{2}\right)\;t^a\;\frac{3-\alpha}{2}\;
  \Gamma[\varepsilon]\;\frac{T^\prime-\frac{\pv^{\prime\,2}}{2M}}{T}
  +\mbox{ finite }=\pc(v)!
  \label{sspvertex2calculation}
\end{eqnarray}
The diagram will therefore only contribute to the renormalisation of the
multipole expansion. Moreover, because it is sub-leading, one may use inside
the integral the equations of motion at leading order in the power counting,
$T=\frac{\pv^2}{2M}$, to set the outgoing potential quark on its mass-shell.
The integral vanishes then, as no scale is present, see (\ref{sspvertex2}) and
(\ref{sspvertex2calculation}).

The cutting rule suggests that the $\Qs^\dagger A_{\mathrm{u},0}\Qs$ correction
should be related to the soft quark self energy:
\begin{eqnarray}
  \rule{0pt}{40pt}
  \feynbox{60\unitlength}{
      \begin{fmfgraph*}(60,20)
        \fmftop{i,o}
        \fmfv{label=$\fs (T,,\pv)$}{i}
        \fmfv{label=$\fs (T^\prime,,\pv^\prime)$}{o}
        \fmfbottom{u}
        \fmfv{label=$\fs {(T-T^\prime\to 0,,\pv-\pv^\prime\to 0)}$}{u}
        \fmf{double,foreground=red,tension=2}{i,v1}
        \fmf{double,foreground=red}{v1,v2,v3}
        \fmf{double,foreground=red,tension=2}{v3,o}
        \fmffreeze
        \fmf{zigzag,foreground=red,left,label=$\fs q$,label.side=left}{v1,v3}
        \fmf{photon,foreground=blue}{v2,u}
      \end{fmfgraph*}}
    \hq\hq&&
    \rightarrow
    \feynbox{60\unitlength}{
      \begin{fmfgraph*}(60,20)
        \fmftop{i,o}
        \fmfbottom{u}
        \fmf{double,foreground=red,tension=2}{i,v1}
        \fmf{double,foreground=red}{v1,v2,v3}
        \fmf{double,foreground=red,tension=2}{v3,o}
        \fmfdot{v2}
        \fmfdot{v4}
        \fmffreeze
        \fmf{zigzag,foreground=red,left}{v1,v3}
        \fmf{phantom,tension=2}{v2,v4}
        \fmf{photon,foreground=blue}{v4,u}
      \end{fmfgraph*}}
    \;:\;\\
    &&(-\ii\;g_\mathrm{R})^3\;\mu^{3\varepsilon}\;t^b t^a t^c\;
    \int\deintdim{d}{q}\frac{-\ii\delta^{bc}}{q^2}
      \left[1-(1-\alpha)\frac{q^2_0}{q^2}\right]
    \frac{\ii}{T+q_0}\;\frac{\ii}{T+q_0}\non 
\end{eqnarray}
The dot at the cut-off vertex stands again as a reminder that the structure of
the vertex will enter here, and one finds the vertex correction to be
proportional to $\frac{\de}{\de T} \Sigma_\mathrm{s}(T,\pv)$
(\ref{nrqcdselfsoft}). Using (\ref{secondmaster}), the vertex correction is
again (\ref{abeliangamma}).

\absatz Finally, two examples for the straightforward but tedious non-Abelian
vertex corrections of Fig.\ \ref{nonabelianvertex} are given. First, one may
calculate the non-Abelian correction to $\Qs^\dagger A_{\mathrm{u},0}\Qs$
\begin{eqnarray}
  \label{lastexample}
  \rule{0pt}{35pt}
  \feynbox{60\unitlength}{
      \begin{fmfgraph*}(60,40)
        \fmftop{i,o}
        \fmfv{label=$\fs (T,,\pv)$}{i}
        \fmfv{label=$\fs (T^\prime,,\pv^\prime)$}{o}
        \fmfbottom{u}
        \fmfv{label=$\fs {(T-T^\prime\to 0,,\pv-\pv^\prime\to 0)}$}{u}
        \fmf{double,foreground=red,tension=4}{i,v1}
        \fmf{double,foreground=red}{v1,v3}
        \fmf{double,foreground=red,tension=4}{v3,o}
        \fmffreeze
        \fmf{zigzag,foreground=red}{v1,v2,v3}
        \fmf{photon,foreground=blue,tension=3}{v2,u}
      \end{fmfgraph*}}
    &:&(-\ii\;g_\mathrm{R})^2\;g_\mathrm{R}\;\mu^{3\varepsilon}\;f^{abc}t^b
    t^c\;\int\deintdim{d}{q}
    \frac{\ii}{T+q_0}\;\frac{\ii\calG^{0\mu}(q)}{q^2}
    \;\frac{\ii\calG^{0\nu}(q)}{q^2} \times\non\\
  &&\hspace*{20ex}\times\;\;\left[-2q_0 g_{\mu\nu}+q_\mu
      g_{\nu0}+q_\nu g_{\mu0}\right]\;\;.
\end{eqnarray}
The Feynman gauge term, being independent of $(1-\alpha)$, is zero as expected.
For this diagram, even the terms proportional to $(1-\alpha)^2$ vanish, which
can be shown to be finite in general and hence not to contribute to the vertex
normalisation. In (\ref{lastexample}), only the contribution leading in $v$ is
considered. Using the well known relation $\ii f^{abc}t^at^b=-\frac{N}{2}\;t^a$
for SU($N$), this can be collapsed down by re-writing numerators analogously to
(\ref{numeratortreatment}) to
\begin{equation}
  g_\mathrm{R}^3\;\mu^{3\varepsilon}\;N \;t^a\;(1-\alpha)\;T\;
  \int\deintdim{d}{q}
  \frac{1}{T+q_0}\;\frac{1}{q^4}\;\left[\frac{T^2}{q^2}\;-1\right]\;\;.
\end{equation}
One extracts the pole part of the non-Abelian vertex correction as
(\ref{nonabeliangamma}) from the result obtained using (\ref{firstmaster}):
\begin{equation}
  \frac{-\ii\;g^3_\mathrm{R}\;\mu^{3\varepsilon}}{(4\pi)^2}\;\frac{N}{2}\;t^a\;
  \frac{3(1-\alpha)}{2}\;\Gamma[\varepsilon]+\mbox{ finite.}
\end{equation}
The leading piece of the non-Abelian correction to $\Qp^\dagger
A_{\mathrm{p},0}\Qp$ is extracted as
\begin{eqnarray}
  \rule{0pt}{35pt}
  \feynbox{60\unitlength}{
      \begin{fmfgraph*}(60,40)
        \fmftop{i,o}
        \fmfv{label=$\fs (T,,\pv)$}{i}
        \fmfv{label=$\fs (T^\prime,,\pv^\prime)$}{o}
        \fmfbottom{u}
        \fmfv{label=$\fs {(T-T^\prime,,\pv-\pv^\prime)}$}{u}
        \fmf{vanilla,width=thick,tension=4}{i,v1}
        \fmf{double,foreground=red}{v1,v3}
        \fmf{vanilla,width=thick,tension=4}{v3,o}
        \fmffreeze
        \fmf{zigzag,foreground=red}{v1,v2,v3}
        \fmf{dashes,tension=3}{v2,u}
      \end{fmfgraph*}}
    &:&(-\ii\;g_\mathrm{R})^2\;g_\mathrm{R}\;\mu^{3\varepsilon}\;f^{abc}t^b
    t^c\;\int\deintdim{d}{q}
    \frac{\ii}{q_0}\;\frac{\ii\calG^{0\mu}(q)}{q^2}
    \;\frac{\ii\calG^{0\nu}(q+\Delta E)}{(q+\Delta E)^2}
    \;\times\non\\
  &&\hspace*{10ex}\times\;\left[2q_0 g_{\mu\nu}+(\Delta E-q)_\mu
      g_{\nu0}-(q+2\Delta E)_\nu g_{\mu0}\right]\;\;,
\end{eqnarray}
where $\Delta E=(0,\pv^\prime-\pv)$. The term proportional to $(1-\alpha)$
contains the ultraviolet singularity and can be written as
\begin{equation}
  -g_\mathrm{R}^3\;\mu^{3\varepsilon}\;N\; t^a\;(1-\alpha)\;
  \int\deintdim{d}{q}\frac{1}{q^4}\;\frac{1}{(q+\Delta E)^2}\;\left(q_0^2-q^2-2
    E\cdot q\right)\;\;.
\end{equation}
Standard formulae for dimensional regularisation yield as divergent part
(\ref{nonabeliangamma}). The term proportional to $(1-\alpha)^2$ is
ultraviolet finite and can be reduced to the integral
\begin{equation}
  g_\mathrm{R}^3\;\mu^{3\varepsilon}\;\frac{N}{2} \;t^a\;(1-\alpha)^2\;
  \Delta E^2\; \int\deintdim{d}{q}\frac{q_0^2}{q^4}\; \frac{1}{(q+\Delta E)^4}=
  \frac{\ii\;g_\mathrm{R}^3\;\mu^{3\varepsilon}}{(4\pi)^2}\;
  \frac{N}{2} \;t^a\;\frac{(1-\alpha)^2}{2}\;\;.
\end{equation}

\end{appendix}


\end{fmffile}
\end{document}